%% file: MagnusHamiltonianSimulation.tex
\documentclass[a4paper,onecolumn,11pt]{quantumarticle}
\pdfoutput=1

\input{def_papersetup}

\begin{document}

\title{Quantum simulation of highly-oscillatory many-body Hamiltonians %
for near-term devices}

\author{Guannan Chen}
\affiliation{Department of Mathematical Sciences, University of Bath, Bath BA2 7AY, United Kingdom}

\author{Mohammadali Foroozandeh}
\affiliation{Chemistry Research Laboratory, University of Oxford, Mansfield Road, OX1 3TA Oxford, United Kingdom}

\author{Chris Budd}
\affiliation{Department of Mathematical Sciences, University of Bath, Bath BA2 7AY, United Kingdom}

\author{Pranav Singh}
\email{ps2106@bath.ac.uk}
\affiliation{Department of Mathematical Sciences, University of Bath, Bath BA2 7AY, United Kingdom}

\maketitle

\begin{abstract}
We develop a fourth-order Magnus expansion based quantum algorithm for the simulation of many-body problems involving two-level quantum systems with time-dependent Hamiltonians, $\ha(t)$. A major hurdle in the utilization of the Magnus expansion is the appearance of a commutator term which leads to prohibitively long circuits. We present a technique for eliminating this commutator and find that a single time-step of the resulting algorithm is only marginally costlier than that required for time-stepping with a time-independent Hamiltonian, requiring only three additional single-qubit layers. For a large class of Hamiltonians appearing in liquid-state nuclear magnetic resonance (NMR) applications, we further exploit symmetries of the Hamiltonian and achieve a surprising reduction in the expansion, whereby a single time-step of our fourth-order method has a circuit structure and cost that is identical to that required for a fourth-order Trotterized time-stepping procedure for time-independent Hamiltonians. Moreover, our algorithms are able to take time-steps that are larger than the wavelength of oscillation of the time-dependent Hamiltonian, making them particularly suited for highly-oscillatory controls. The resulting quantum circuits have shorter depths for all levels of accuracy when compared to first and second-order Trotterized methods, as well as other fourth-order Trotterized methods, making the proposed algorithm a suitable candidate for simulation of time-dependent Hamiltonians on near-term quantum devices.
\end{abstract}

\input{def_toggles}

\input{1_introduction}

\input{2_theory}

\input{3_approx_matrix_exp}

\input{4_trotterized_circuit}

\input{5_results}

\input{6_conclusion}

\input{7_acknowledgements}

\bibliographystyle{apsrevcustom}

\bibliography{MagnusHamiltonianSimulation}

\input{A_appendix}

\end{document}

%% file: def_papersetup.tex
\usepackage[sort&compress,square,numbers]{natbib}

\makeatother
\usepackage[dvipsnames]{xcolor}
\usepackage[utf8]{inputenc}
\usepackage[english]{babel}
\usepackage[T1]{fontenc}
\usepackage{amsmath}
\usepackage{tikz}
\usepackage{lipsum}

\usepackage[utf8]{inputenc} %
\usepackage[T1]{fontenc}    %
\usepackage{url}            %
\usepackage{booktabs}       %
\usepackage{amsmath}
\usepackage[english]{babel}
\usepackage{amsthm}
\usepackage{amsfonts}       %
\usepackage{nicefrac}       %
\usepackage{microtype}      %
\usepackage{lipsum}
\usepackage{fancyhdr}       %
\usepackage{graphicx}       %
\usepackage{subcaption}
\usepackage{mathtools}
\usepackage{multirow}
\usepackage{bbold}

\usepackage{scalerel}[2016-12-29]
\def\stretchint#1{\vcenter{\hbox{\stretchto[440]{\displaystyle\int}{#1}}}}

\usepackage{enumitem}
\newlist{customenum}{enumerate}{1}
\setlist[customenum,1]{label=(\roman*), ref=(\roman*)}

\usepackage{tikz}
\usepackage{pgfplots}

\usepackage{diagbox}

\usepgfplotslibrary{groupplots}

\usepackage{pgfplotstable}
\pgfplotsset{compat=newest}

\usepackage{qcircuit}
\usepackage{mathrsfs}
\usepackage{makecell}
\usepackage{bbm}
\usepackage[colorlinks,allcolors=red!50!black]{hyperref}
\usepackage{cleveref}
\usepackage{comment}
\usepackage{float}

\usepackage{graphicx}
\newlength{\subcolumnwidth}
\newenvironment{subcolumns}[1][0.45\columnwidth]
 {\valign\bgroup\hsize=#1\setlength{\subcolumnwidth}{\hsize}\vfil##\vfil\cr}
 {\crcr\egroup}
\newcommand{\nextsubcolumn}[1][]{%
  \cr\noalign{}
  \if\relax\detokenize{#1}\relax\else\hsize=#1\setlength{\subcolumnwidth}{\hsize}\fi
}

\pgfplotsset{compat = 1.3}
\pgfplotsset{improved scaled ticks/.style 2 args={
    scaled #1 ticks=#2,
    extra #1 tick style={
        scaled #1 ticks=#2,
        #1tick scale label code/.code=,
    },
}}

\definecolor{darkgray176}{RGB}{176,176,176}
\definecolor{darkorange25512714}{RGB}{255,127,14}
\definecolor{forestgreen4416044}{RGB}{44,160,44}
\definecolor{lightgray204}{RGB}{204,204,204}
\definecolor{steelblue31119180}{RGB}{31,119,180}
\definecolor{mediumpurple148103189}{RGB}{148,103,189}
\definecolor{sienna1408675}{RGB}{140,86,75}
\definecolor{crimson2143940}{RGB}{214,39,40}
\definecolor{yellowgreen152204112}{RGB}{152,204,112}
\definecolor{orchid227119194}{RGB}{227,119,194}
\definecolor{lightgray230}{RGB}{230,230,230}

\definecolor{darkgray176}{RGB}{176,176,176}
\definecolor{steelblue31119180}{RGB}{31,119,180}
\definecolor{darkorange25512714}{RGB}{255,127,14}
\definecolor{lightgray204}{RGB}{204,204,204}
\definecolor{forestgreen4416044}{RGB}{44,160,44}
\definecolor{lightgray230}{RGB}{230,230,230}
\definecolor{steelblue31119180}{RGB}{31,119,180}
\definecolor{mediumpurple148103189}{RGB}{148,103,189}
\definecolor{sienna1408675}{RGB}{140,86,75}
\definecolor{crimson2143940}{RGB}{214,39,40}
\definecolor{orchid227119194}{RGB}{227,119,194}
\definecolor{yellowgreen152204112}{RGB}{152,204,112}

\definecolor{sienna1408675}{RGB}{214,39,40}
\definecolor{darkgray176}{RGB}{176,176,176}
\definecolor{lightgray204}{RGB}{204,204,204}
\definecolor{orchid227119194}{RGB}{227,119,194}
\definecolor{mediumpurple148103189}{RGB}{148,103,189}
\definecolor{sienna1408675}{RGB}{140,86,75}
\definecolor{darkgray176}{RGB}{176,176,176}
\definecolor{orchid227119194}{RGB}{44,160,44}
\definecolor{gray}{RGB}{128,128,128}
\definecolor{darkorange25512714}{RGB}{255,127,14}

\definecolor{crimson2143940}{RGB}{214,39,40}
\definecolor{darkgray176}{RGB}{176,176,176}
\definecolor{lightgray204}{RGB}{204,204,204}
\definecolor{227119194}{RGB}{227,119,194}
\definecolor{steelblue31119180}{RGB}{31,119,180}
\definecolor{mediumpurple148103189}{RGB}{148,103,189}
\definecolor{sienna1408675}{RGB}{140,86,75}
\definecolor{darkgray176}{RGB}{176,176,176}
\definecolor{forestgreen4416044}{RGB}{44,160,44}
\definecolor{gray}{RGB}{128,128,128}
\definecolor{lightgray204}{RGB}{204,204,204}
\definecolor{darkorange25512714}{RGB}{255,127,14}

\definecolor{crimson2143940}{RGB}{214,39,40}
\definecolor{darkgray176}{RGB}{176,176,176}
\definecolor{lightgray204}{RGB}{204,204,204}
\definecolor{forestgreen4416044}{RGB}{44,160,44}
\definecolor{steelblue31119180}{RGB}{31,119,180}
\definecolor{darkorange25512714}{RGB}{255,127,14}

\definecolor{darkgray176}{RGB}{176,176,176}
\definecolor{steelblue31119180}{RGB}{31,119,180}
\definecolor{darkorange25512714}{RGB}{255,127,14}
\definecolor{lightgray204}{RGB}{204,204,204}
\definecolor{forestgreen4416044}{RGB}{44,160,44}
\definecolor{mediumpurple148103189}{RGB}{148,103,189}
\definecolor{sienna1408675}{RGB}{140,86,75}
\definecolor{crimson2143940}{RGB}{214,39,40}

\definecolor{darkgray176}{RGB}{176,176,176}
\definecolor{darkorange25512714}{RGB}{255,127,14}
\definecolor{lightgray204}{RGB}{204,204,204}

\definecolor{darkgray176}{RGB}{176,176,176}
\definecolor{darkorange25512714}{RGB}{255,127,14}
\definecolor{lightgray204}{RGB}{204,204,204}

\definecolor{darkgray176}{RGB}{176,176,176}
\definecolor{steelblue31119180}{RGB}{31,119,180}
\definecolor{darkorange25512714}{RGB}{255,127,14}
\definecolor{lightgray204}{RGB}{204,204,204}
\definecolor{lightgray230}{RGB}{230,230,230}
\definecolor{mediumpurple148103189}{RGB}{148,103,189}
\definecolor{sienna1408675}{RGB}{140,86,75}
\definecolor{crimson2143940}{RGB}{214,39,40}
\definecolor{orchid227119194}{RGB}{227,119,194}

\definecolor{darkgray176}{RGB}{176,176,176}
\definecolor{darkorange25512714}{RGB}{255,127,14}
\definecolor{lightgray204}{RGB}{204,204,204}
\definecolor{lightgray230}{RGB}{230,230,230}
\definecolor{steelblue31119180}{RGB}{31,119,180}
\definecolor{orchid227119194}{RGB}{227,119,194}

\definecolor{crimson2143940}{RGB}{214,39,40}
\definecolor{darkgray176}{RGB}{176,176,176}
\definecolor{lightgray204}{RGB}{204,204,204}
\definecolor{orchid227119194}{RGB}{227,119,194}
\definecolor{steelblue31119180}{RGB}{31,119,180}
\definecolor{darkorange25512714}{RGB}{255,127,14}
\definecolor{emeraldgreen}{RGB}{46, 139, 87}
\definecolor{royalblue}{RGB}{0, 65, 255}
\definecolor{crimsonred}{RGB}{255, 80, 0}
\definecolor{lightgray}{RGB}{230,225,225}
\definecolor{midgray}{RGB}{185,190,190}
\definecolor{darkergray}{RGB}{95,95,100}
\definecolor{forestgreen4416044}{RGB}{44,160,44}

\definecolor{darkgray176}{RGB}{176,176,176}
\definecolor{steelblue31119180}{RGB}{31,119,180}

\definecolor{darkgray176}{RGB}{176,176,176}
\definecolor{steelblue31119180}{RGB}{31,119,180}
\definecolor{darkorange25512714}{RGB}{255,127,14}
\definecolor{steelblue31119180}{RGB}{31,119,180}
\definecolor{lightgray204}{RGB}{204,204,204}

\graphicspath{{media/}}     %

\newtheorem{Theorem}{Theorem}
\newtheorem{Property}{Property}
\newtheorem{Lemma}{Lemma}
\newtheorem{Remark}{Remark}
\newtheorem{Definition}{Definition}
\newtheorem{Example}{Example}

\DeclarePairedDelimiter\bra{\langle}{\rvert}
\DeclarePairedDelimiter\ket{\lvert}{\rangle}
\DeclarePairedDelimiterX\braket[2]{\langle}{\rangle}{#1 \delimsize\vert #2}

\newcommand{\px}{X}
\newcommand{\py}{Y}
\newcommand{\pz}{Z}

\newcommand{\vone }{\mathbf{1}}
\newcommand{\va}{\mathbf{a}}
\newcommand{\vb}{\mathbf{b}}
\newcommand{\vc}{\mathbf{c}}

\newcommand{\ve}{\mathbf{e}}
\newcommand{\vex}{\ve^X}
\newcommand{\vey}{\ve^Y}
\newcommand{\vez}{\ve^Z}
\newcommand{\vo}{\boldsymbol{\Omega}}
\newcommand{\vmu}{\boldsymbol{\mu}}

\newcommand{\vr}{\mathbf{r}}

\newcommand{\para}[1]{{\bf #1}}

\newcommand{\jg}{C_{\mathrm{in}}}
\newcommand{\jhomo}{C_{\mathrm{iso}}}

\newcommand{\vu}{\mathbf{u}}

\newcommand{\vcs}{\mathbb{S}}
\newcommand{\vca}{\mathbb{A}}
\newcommand{\vcb}{\mathbb{B}}
\newcommand{\vcc}{\mathbb{C}}
\newcommand{\tr}{\text{Tr}}
\newcommand{\ha}{\mathcal{H}}

\newcommand{\bch}{\mathrm{BCH}}

\newcommand{\CC}{\mathbb{C}}

\newcommand{\RR}{\mathbb{R}}

\newcommand{\oneb}{\mathbb{1}}
\newcommand{\customI}[2]{I_{#1}^{#2}}
\newcommand{\XYZ}[2]{{#1}_{#2}}

\newcommand{\ee}{\mathrm{e}}
\newcommand{\dd}{\mathrm{d}}
\newcommand{\ii}{\text{i}}

\newcommand{\A}{\mathcal{A}}
\newcommand{\order}{\mathcal{O}}

\newcommand{\lb}{\left(}
\newcommand{\rb}{\right)}
\newcommand{\lbb}{\left[}
\newcommand{\rbb}{\right]}

\newcommand{\tp}{\otimes}

\newcommand{\hs}{\ha_{\text{ss}}} %
\newcommand{\hi}{\ha_{\text{in}}} %

\newcommand{\eli}{\text{E}} %

\newcommand{\ie}{\emph{i.e.}\,}
\newcommand{\eg}{\emph{e.g.}\,}

\definecolor{LimeGreen}{RGB}{50, 205, 50}

\renewcommand{\cite}{\citep}

%% file: def_toggles.tex
\newtoggle{rotationgate}
\newtoggle{couplinggates}
\newtoggle{circuit}
\newtoggle{Gaussian_pulse}
\newtoggle{chirped_pulse}
\newtoggle{observables}
\newtoggle{indep_result}
\newtoggle{chirped_result}
\newtoggle{bandwidthTheta2}
\newtoggle{couplingstrengthTheta2}
\newtoggle{chirped_CFTO_result}
\newtoggle{bandwidth_methods}
\newtoggle{bandwidth_integrals}
\newtoggle{Hin_methods}
\newtoggle{Hin_integrals}
\newtoggle{general_no_mixed_result}
\newtoggle{general_mixed_result}

\toggletrue{rotationgate}
\toggletrue{couplinggates}
\toggletrue{circuit}
\toggletrue{Gaussian_pulse}
\toggletrue{chirped_pulse}
\toggletrue{observables}
\toggletrue{indep_result}
\toggletrue{chirped_result}
\toggletrue{bandwidthTheta2}
\toggletrue{couplingstrengthTheta2}
\toggletrue{chirped_CFTO_result}
\toggletrue{bandwidth_methods}
\toggletrue{bandwidth_integrals}
\toggletrue{Hin_methods}
\toggletrue{Hin_integrals}
\toggletrue{general_no_mixed_result}
\toggletrue{general_mixed_result}


%% file: 1_introduction.tex
\section{Introduction}
In this paper we develop effective methods for the Hamiltonian simulation for many-body two-level systems with time-dependent Hamiltonians, $\ha(t)$, $\ie$, for solving the ordinary differential equation,
\begin{equation}
    \ii \frac{\partial}{\partial t} \ket{\psi(t)} = \ha(t) \ket{\psi(t)}, \qquad \ket{\psi(0)} = \ket{\psi_0}\in \mathscr{H} = \mathbb{C}^{2^M}, \qquad t \in \mathbb{R},
    \label{eq: Schrodinger}
\end{equation}
where $\ket{\psi(t)}$ describes the state of a quantum system at time $t$.

The quantum state $\ket{\psi(t)}$ resides in the Hilbert space $\mathscr{H}$, whose dimension increases exponentially for a many-body quantum system \cite{feynman1982simulating}. This {\em curse of dimensionality} potentially renders the solution of \cref{eq: Schrodinger} using classical computers infeasible beyond a small number of particles. On the other hand, due to a similarly exponential increase in the Hilbert space of a quantum computer with the number of qubits, the Hamiltonian simulation problem can be solved very effectively on a fault-tolerant quantum computer in the case of time-independent Hamiltonians, $\ha(t) = \ha_0$.

The significance of this problem to computational quantum chemistry and physics, the suspected exponential difficulty for classical computers, and the demonstrably linear growth in terms of number of qubits required, makes Hamiltonian simulation problems for many-body systems an important class of candidates for demonstrating `{\em quantum advantage}' \citep{childs2018toward,seetharam2021digital,IBM23nisq}.
Hamiltonian simulation is also a fundamental subroutine of other sophisticated quantum algorithms such as quantum phase estimation (QPE) \cite{kitaev1995quantum}  and the Harrow-Hassidim-Lloyd (HHL) algorithm \cite{harrow2009quantum} for solving linear system of equations.

Along with rapid development of quantum hardware \cite{Castelvecchi2023,moses2023race,bluvstein2023logical}, this has fuelled a resurgence of interest in the development of quantum algorithms for Hamiltonian simulation \cite{childs2012hamiltonian,berry2014exponential,berry2015hamiltonian,BCCKS_berry2015simulating,low2017hamiltonian,low2017optimal,low2019hamiltonian}.
These algorithms typically solve \cref{eq: Schrodinger} by utilizing techniques for approximating the matrix exponential. Arguably the oldest and the most widely utilized among these techniques are {\em Trotterization} techniques \cite{trotter1959product}, also known as {\em splitting} methods \cite{mclachlan2002splitting,blanes2008splitting}, which lead to simpler circuit designs that are easier to implement \cite{nam2019low}, and that can be optimized \cite{tepaske} for near-term applications. Trotterized circuits have been utilized in recent attempts at demonstrating the utility of existing {\em noisy intermediate scale quantum} (NISQ) devices for Hamiltonian simulation \cite{smith2019simulating,IBM23nisq}. Another notable approach is the qubitization procedure of \cite{low2019hamiltonian}, which provides a substantially improved {\em additive complexity} for simulation of time-independent Hamiltonians, and is expected to prove superior in the context of large-scale fault-tolerant quantum computers.

Hamiltonian simulation for time-dependent Hamiltonians, which is required for describing the interaction of quantum systems with external time-dependent fields such as lasers, magnetic fields and light, involves solving a non-autonomous differential equation \cref{eq: Schrodinger} whose solution is no longer given by a direct exponential of the Hamiltonian. Thus, new techniques are required for the simulation of time-dependent Hamiltonians,   and this has become an area of intense investigation with recent approaches including the interaction picture approach \cite{low2019interaction}, approximation of the Dyson series \cite{kieferova2019simulating}, L1 norm scaling \cite{berry2020time}, the treatment of slowly varying time-dependent for lattice Hamiltonians
 \cite{haah2021quantum}, permutation expansion \cite{chen2021quantum}, approaches based on
time-ordered operators \cite{watkins2022timedependent}, a Floquet based approach for periodic time-dependent Hamiltonians \cite{mizuta2023optimal}, and the autonomization approach of \cite{cao2023quantum}.

\para{Time-dependent Hamiltonians in many-body two-level systems.} Two-level quantum systems are ubiquitous in quantum technologies, and among other applications, they form the physical realization of qubits in a quantum computer \cite{divincenzo2000physical,devoret2004superconducting} and describe the phenomenon of nuclear magnetic resonance (NMR) \cite{ernst1990principles}. They appear either naturally in systems that have only two energy levels (and eigenstates), $\eg$, spin-$\nicefrac{1}{2}$ systems, or as anharmonic quantum systems such as superconducting qubits, where the lowest two energy levels stay sufficiently separated from the rest during operation to effectively behave like a two-level system.

The Hamiltonians of interest in this paper are time-dependent Hamiltonians involving multiple interacting two-level systems. Such Hamiltonians appear prominently in the control of quantum technologies, including gate design and control of superconducting qubits \cite{blais2004cavity,sporl2007optimal,krantz2019quantum}, trapped ion control \cite{Zhao2008}, NV-centers in diamond \cite{Chou2015,Tian2019}, quantum error-correction and quantum information registers \cite{Zhang2012,Waldherr2014,Dolde2014}, pulse design in magnetic resonance spectroscopy (NMR) and imaging (MRI) \cite{Skinner2003,Tosner2006,Reeth2019}, electron paramagnetic resonance (EPR) \cite{Spindler2012,Doll2013,Kaufmann2013}, cold atom interferometry \cite{Saywell2018,Saywell2020a,Saywell2020}, and terahertz technologies \cite{Coudert2018,Rasanen2007}, among others. A wide range of general purpose algorithms for the optimal control of such two-level systems have been developed, including GRAPE \cite{khaneja2005optimal}, CRAB \cite{caneva2011chopped}, GOAT \cite{machnes2018tunable}, and QOALA
\cite{goodwin2022adaptive}, and have been employed widely in systems of practical interest.

An additional difficulty is posed by the highly-oscillatory {\em chirped} pulses that are often encountered in the control of quantum systems \cite{AmstrupChirped,garwood2001return, FOROOZANDEH2020106768}. A highly-oscillatory
Hamiltonian conventionally forces the use of small time-steps. For classical computers, this is undesirable due to
longer computational times, but is otherwise achievable. On a quantum computer, where small time-steps correspond to larger circuit depths, the accumulation of errors in the quantum circuit can significantly restrict the feasibility of simulating such Hamiltonians on near-term quantum devices. This makes the Hamiltonian simulation problem for time-dependent Hamiltonians with highly-oscillatory pulses significantly harder than for time-independent Hamiltonians of the same structure.

\para{Main contributions. }
In this manuscript we develop a fourth-order method for Hamiltonian simulation of many-body two-level system under time-dependent Hamiltonians, which can take long time-steps and achieve high accuracy. Our approach is independent of the frequency of the Hamiltonian oscillation and can handle a continuum of frequencies. A unique feature is that our method can take time-steps that are larger than the wavelength of oscillation of the Hamiltonian, allowing us to handle chirped pulses very effectively. The ability of our method to be used with long time-steps corresponds to shorter circuit depths. Moreover, all of this can be achieved using minor modifications of a Trotterized circuit for time-independent Hamiltonians, or a qubitized circuit for piecewise-constant Hamiltonians,
and involves only a marginal increase in circuit depth.

Our approach is based on the use of the Magnus expansion \cite{magnus1954exponential,iserles2000lie}, which
amounts to finding an effective Hamiltonian as an infinite series of nested integrals of nested commutators. Magnus based methods are utilised widely in classical algorithms for quantum dynamics \cite{hochbruck2003magnus,  kormann2008accurate, blanes2009magnus,iserles2018magnus,iserles2019solving}.

{\em Mean Hamiltonian.} The first term of the effective Hamiltonian is simply the mean Hamiltonian, which has the same structure as a time-independent Hamiltonian, and techniques for exponentiating time-independent Hamiltonians such as Trotterization \cite{trotter1959product} and qubitization \cite{low2019hamiltonian} can be applied without altering the circuit structure. However, as we demonstrate with numerical evidence, the second-order accuracy of this method is inadequate for simulating time-dependent Hamiltonians, requiring
very small time-steps or long circuit depths for reasonable accuracy. This necessitates the development of higher-order schemes. In the context of chirped pulses, this is more immediately evident since they involve a continuous range of frequencies and the Hamiltonian averaging techniques is particularly poorly suited here.

{\em Issues with commutators.} The immediate problem in the use of higher-order Magnus expansions for quantum algorithms is the appearance of a commutator term. The time-independent and mean Hamiltonians have a common structure and can be expressed as a linear combination of $3M+\frac12 |\jg|$ unitaries, where $M$ is the number of two-level systems and $|\jg|\leq 9 M(M-1)$ is the number of pairwise couplings. On the other hand, the commutator term appearing in the fourth-order Magnus expansion involves a linear combination of $\order \lb M^3 \rb$ unitaries. This makes direct application of Trotterization or qubitization for exponentiating the fourth-order Magnus expansion infeasible due to a prohibitive increase in the circuit depth, and is arguably the biggest restriction in the application of Magnus-based methods for Hamiltonian simulation on a quantum computer.

{\em Eliminating commutators.} We present an approach that eliminates the commutator term in the fourth-order Magnus expansion for a general spin Hamiltonian, leaving us with an effective Hamiltonian that has the same structure as the time-independent and mean Hamiltonians. Consequently, it can be expressed as a linear combination of $3 M+\frac12|\jg|$ unitaries and efficiently exponentiated via qubitization, or combined with a Trotterized approach with very minor modification of the circuit structure. Specifically the minor overhead is the appearance of two additional terms that can be computed with $\order \lb M \rb$ single-qubit gates. In the special case of {\em isotropic} coupling with identical controls, which appears frequently in liquid-state NMR applications, we find a surprising result that the commutator term vanishes due to certain symmetries of the Hamiltonian, leaving the circuit structure and depth identical to the case of mean Hamiltonians.

{\bf Numerical results.} Numerical simulations of NMR spectra offer insights into molecular structures, interactions, and dynamics. Their significance is highlighted by the considerable attention they have received in the scientific community, as evidenced by the proliferation of dedicated software packages tailored for this purpose. Notable examples include SPINACH \cite{spinach}, EasySpin \cite{easyspin}, SpinEvolution \cite{spinevolution}, and SIMPSON \cite{simpson}.  Motivated by the importance of NMR simulations, we present concrete Trotterized circuits for our fourth-order Magnus-based approach and demonstrate its efficacy in simulating spin dynamics in NMR.

In particular, the numerical evidence shows a significant advantage over the second-order approximation and an ability to take time-steps larger than the wavelength of the external pulse. The latter ability stems from our approach to decouple the resolution of the external pulses from the time-step of the numerical scheme, along the lines of \cite{iserles2019compact,iserles2019solving,singh2019sixth}: Unlike traditional applications of the Magnus expansion, which discretize the integrals using a fixed number of Gauss--Legendre quadrature nodes, and therefore tie the resolution of the external pulse to the time-step, in these approaches the one and two-dimensional integrals in the fourth-order Magnus expansion are computed to high precision on a classical computer. In the context of many-body spin systems, the investigations by \cite[Section~4.2]{goodacrethesis}, provided initial evidence of this phenomenon under weak fields.

The proposed approach is compared to first-order methods, second-order methods, and other fourth-order methods that are free of commutators, including a commutator-free quasi-Magnus approach and an autonomization approach described in \Cref{sec: methods without commutators}. The proposed approach is found superior in all cases -- producing shorter circuits for any given accuracy, with the circuit compression ratios higher for more oscillatory pulses.

Notable differences are found from the simulation of time-independent versions of the Hamiltonians, where the second-order method produces the shortest circuits for accuracies of practical interest ($\sim 10^{-2}$ -- $10^{-5}$), with the fourth-order method only becoming more efficient for higher accuracies, which are not expected to be achievable in the near term. In contrast, for highly-oscillatory pulses that appear in control applications, we find that our fourth-order method produces the shortest circuits even when very low levels of accuracies ($\sim 10^{-1}$) are required. Nevertheless,  the circuit depths required for the time-dependent case are found to be two orders of magnitude ($\sim 110$) higher, which suggests that quantum simulation of time-dependent Hamiltonians that appear in the context of practical applications is a significantly harder task, and may remain a challenge even when `{\em quantum advantage}' has been demonstrated on carefully chosen time-independent Hamiltonians, especially when taking into account the sparse couplings typically considered in recent attempts \cite{smith2019simulating,IBM23nisq}.

\para{Outline. }
In \Cref{sec: theory}, we establish a theoretical framework for describing the dynamics of many-body two-level quantum systems using spin-$\nicefrac{1}{2}$ systems as a primary example. This section introduces the Magnus expansion technique and highlights challenges in simulating time-dependent Hamiltonians.
In \Cref{sec: matrix exponential}, we discuss methods for approximating matrix exponentials using quantum computers, with a focus on a novel approach for eliminating commutators in the fourth-order Magnus expansion.
\Cref{sec: splitting and circuit} details the construction of Trotterized quantum circuits for the fourth-order Magnus expansion, and provides gate and circuit complexities.
In \Cref{sec: numerical results}, we present numerical results comparing our Hamiltonian simulation method against other approaches. This section emphasizes the effectiveness of the proposed approach across different Hamiltonian types and pulse influences, and benchmarks the behaviour under a range of oscillatory pulses and coupling strengths.
Finally, \Cref{sec: conclusions} concludes the paper with a brief discussion.

%% file: 2_theory.tex
\section{Theory} \label{sec: theory}

\subsection{The Hamiltonian structure}

We consider the evolution of an $M$-body quantum system, where each subsystem is a two-level system.
In particular, we consider spin-$\nicefrac{1}{2}$ systems,
which serve as prominent examples of a two-level system.
However, we emphasize that our technique
extends to any two-level system,
of which spin-$\nicefrac{1}{2}$ systems are just one illustrative instance.

The state of an $M$-body spin system at time $t$ is described by a density matrix $\varrho(t) \in \mathbb{C}^{2^M \times 2^M}$.
In the absence of dissipative terms ($\ie$, decoherence), the dynamics of the density matrix are governed by the Liouville-von Neumann equation,
\begin{equation}
    \frac{\partial \varrho(t)}{\partial t} = -\ii [\ha(t),\varrho(t)], \qquad \varrho(0) = \varrho_0, \qquad t \in \mathbb{R},
\end{equation}
where $\ha(t)$ is a time-dependent Hamiltonian. Since the density matrix $\varrho(t)$ is an ensemble of pure states $\psi_j(t)$,
\begin{equation}
    \varrho(t) = \sum_j p_j\ket{\psi_j(t)}\bra{\psi_j(t)}, \qquad \sum_j p_j = 1,
\end{equation}
its evolution in the non-dissipative case can also be understood by studying the evolution of $\ket{\psi_j(t)}$, whose dynamics are described by the Schrödinger equation
\begin{equation}
    \ii \frac{\partial}{\partial t} \ket{\psi(t)} = \ha(t) \ket{\psi(t)}, \qquad \ket{\psi(0)} = \ket{\psi_0}\in \mathbb{C}^{2^M}, \qquad t \in \mathbb{R}.
    \tag{\ref{eq: Schrodinger}}
\end{equation}

We consider the Hamiltonian $\ha(t)$ to consist of a single-spin ($\ie$, non-interacting) sub-Hamiltonian $\hs(t)$ and a coupling or interaction sub-Hamiltonian $\hi$,
\begin{equation}
    \label{eq: Hamiltonian}
    \ha(t) = \hs(t) + \hi.
\end{equation}

For an $M$-body spin system, the Hamiltonian for non-interacting spins can be written as
\begin{equation}
    \hs(t) = \ve(t)^\top\vcs \quad = \quad \sum_{k=1}^M  \ \sum_{\alpha \in \{X,Y,Z\}} \mathrm{e}^{\alpha}_k(t)\, \alpha_k, \label{eq: single Hamiltonian}
\end{equation}
where
\begin{align*}
    \ve(t) = \begin{pmatrix} \vex(t) \\ \vey(t) \\ \vez(t) \end{pmatrix}  \in \mathrm{C}^\infty(\mathbb{R}; \mathbb{R}^{3M}),
\end{align*}
are the {\em pulses} or {\em controls} and
\begin{align*}
    \vcs = \begin{pmatrix} {\vcs^X} \\ {\vcs^Y} \\ {\vcs^Z} \end{pmatrix} \in \mathbb{C}^{3M \times (2^M \times 2^M)}, \qquad
    \vcs^{\alpha} = \begin{pmatrix} \alpha_1 \\ \alpha_2 \\ \vdots \\ \alpha_M \end{pmatrix}  \in \mathbb{C}^{M \times (2^M \times 2^M)}, \quad \alpha \in \{X,Y,Z\},
\end{align*}
are vectors of tensorized Pauli matrices. Specifically, the matrices $\alpha \in \{X,Y,Z\}$ are the
Pauli spin--$\frac12$ matrices,
\begin{equation}
X = \begin{pmatrix}
0 & 1 \\
 1 & 0
\end{pmatrix},\quad
Y = \begin{pmatrix}
0 & -\ii \\
\ii & 0
\end{pmatrix},\quad
Z = \begin{pmatrix}
1 & 0 \\
0 & -1
\end{pmatrix},
\label{eq: Pauli matrices}
\end{equation}
and for a matrix $\alpha\in \mathbb{C}^{2\times 2}$ we use $\alpha_k$ to denote the tensor
\begin{equation}
    \alpha_k = \underbrace{\oneb\tp \cdots \tp \oneb}_{M-k\ \text{times}} \tp \underbrace{\alpha}_{k^\text{th}}\tp  \underbrace{\oneb\tp \cdots \tp \oneb}_{k-1\ \text{times}} \quad \in \CC^{2^M\times2^M}, \label{eq: tensor}
\end{equation}
whose action is to apply $\alpha$ to the $k$th spin and leave other spins unaffected --
$\ie$, the subscript $k$ denotes the spin on which the matrix $\alpha$ acts. Here $\oneb$ is the $2\times 2$ identity matrix.

\begin{Remark}
In this paper, bold letters such as $\ve$, $\vo$, and $\vone$ denote column vectors.  $\px$, $\py$, $\pz$, $A$ and $I$ denote matrices. Matrices with a single subscript such as $\px_k$ and $A_k$ denote a tensor product of the matrix and identity matrices, as defined in \cref{eq: tensor}. $\vcs$, $\vca$, and $\vcb$ denote vectors of tensor products.
\end{Remark}

\begin{Remark}
For $\va \in \mathbb{C}^{3M}$, we often write $\va = \lb {\va^X}^\top, {\va^Y}^\top, {\va^Z}^\top\rb^\top$ with the subcomponents $\va^{\alpha} \in \mathbb{C}^M, \alpha \in \{X,Y,Z\}$.
\end{Remark}

\begin{Remark}
Note that in NMR literature it is common to use $\customI{k}{x}$, $\customI{k}{y}$, and $\customI{k}{z}$ in place of $X_k$, $Y_k$ and $Z_k$, respectively, and to normalize them by a factor of $\frac{1}{2}$.
\end{Remark}

\begin{Remark}
\label{rmk: smooth pulses}
Throughout this work, we assume that the controls $\ve \in \mathrm{C}^\infty(\mathbb{R}; \mathbb{R}^{3M})$ are smooth functions of time. However, by splitting the temporal window of integration appropriately, our approach can be extended in a straightforward way to piecewise smooth controls.
\end{Remark}

Given an interaction tensor, %
\begin{equation}
    \jg = \begin{pmatrix}
    C^{X,X} & C^{X,Y} & C^{X,Z}\\
    C^{Y,X} & C^{Y,Y} & C^{Y,Z}\\
    C^{Z,X} & C^{Z,Y} & C^{Z,Z} \end{pmatrix}, \qquad C^{\alpha,\beta} \in \mathbb{R}^{M \times M}, \quad \alpha,\beta \in \{X,Y,Z\}, \label{eq: most general C}
\end{equation}
the interaction sub-Hamiltonian,
\begin{equation}
    \hi \quad = \quad \frac12 \vcs^\top \jg \vcs \quad = \quad \frac12  \sum_{j=1}^M \ \sum_{k=1}^M \sum_{\alpha \in \{X,Y,Z\}} \sum_{\beta \in \{X,Y,Z\}} C^{\alpha,\beta}_{j,k}\,  \alpha_j \, \beta_k, \label{eq: interaction Hamiltonian}
\end{equation}
encodes all possible pairwise interactions between the $M$ spins. We assume throughout that $C^{\alpha,\beta}_{j,j}=0$ for all $\alpha,\beta \in \{X,Y,Z\}$ and $j=1,\ldots,M$, $\ie$, the interaction terms are strictly pairwise. The $C^{\alpha,\beta}_{j,j}\neq 0$ case leads to scaled copies of the identity tensor in the Hamiltonian, which results in a global phase factor that is trivial to handle.

\begin{Remark}
    The typical practice in NMR is to express the interaction tensor as an $M\times M$ matrix of $3 \times 3$ submatrices.
    Instead, in \cref{eq: most general C}, we organize the interaction tensor as a $3 \times 3$ matrix $\jg$ of $M \times M$ submatrices $C^{\alpha,\beta}$.
\end{Remark}

The overall Hamiltonian considered in this work is
\begin{equation}
    \ha(t) \quad = \quad \underbrace{\ve(t)^\top\vcs}_{\hs} \quad + \quad \underbrace{\frac12 \vcs^\top \jg \vcs}_{\hi}. \label{eq: full Hamiltonian}
\end{equation}
Hamiltonians of this form appear in many areas including nuclear magnetic resonance (NMR) \cite{Smith1992NMRHamiltoniansPI, Smith1992NMRHamiltoniansPII, Smith1993NMRHamiltoniansPIII}, spin chains \cite{heisenberg1985theorie,hubbard1964electron}, Ising models \cite{ising1925beitrag, brush1967history,cipra1987introduction}, and honeycomb Kitaev model \cite{kitaev2006anyons,trebst2017kitaev}.
The complexity of the Hamiltonian may be substantially reduced in specific applications. For instance, couplings are typically symmetric, $\ie$, $C^{\alpha,\alpha} = {C^{\alpha,\alpha}}^\top$ and interactions can often be very sparse -- $\eg$, $C^{\alpha,\beta}_{j,k}=0$ in spin chains unless $|j-k|=1$.

\begin{Remark}
\label{rmk: mixed couplings}
Our most general algorithm, developed in \Cref{sec: commutator elimination}, involves no assumptions on $\jg$. The Trotterization procedure described in \Cref{sec: matrix exponential,sec: splitting and circuit} requires the absence of mixed coupling, $\ie$, $C^{\alpha,\beta} = 0$ for $\alpha \neq \beta$.
In principle, mixed couplings can be eliminated by performing a local change of basis \cite{Klassen2019twolocalqubit,consani2020effective}, and this is not a severely restrictive assumption. In \Cref{sec: homonuclear} we consider the special case of isotropic couplings, where $C^{X,X}=C^{Y,Y}=C^{Z,Z}=C=C^\top$, and mixed couplings are absent.
This case often appears in the context of liquid-state NMR.
\end{Remark}

A crucial quantity in the accuracy (and difficulty) of the Hamiltonian simulation problem is the maximum spectral radius of the Hamiltonian over the interval of simulation $[0,T]$,
\begin{equation}
    \label{eq: hamiltonian spectral radius}
    \rho = \max_{t \in [0,T]} \| \ha(t)\|, %
\end{equation}
where $\| \cdot \|$ is the $\ell^2$ norm.
The norm $\rho$ can be bounded from above as
\begin{equation}
    \label{eq: Hamiltonian norm time dependent}
    \rho \leq  3M \max_{t \in [0, T]} \| \ve(t) \|_{\infty} + \frac12 |\jg|\, \|\jg\|_{\max},
\end{equation}
\ie, $\rho = \order \lb \| \ve \|_{\infty} M + |\jg|\, \|\jg\|_{\max} \rb$,
where $|\jg| \leq (9/2) M (M-1)$ is the number of non-zero entries in the interaction or coupling tensor, and the matrix max norm, $\|A\|_{\max} := \|\mathrm{vec}(A)\|_\infty$, is the maximum over the absolute values of all entries in the matrix $A$.

\subsection{The Magnus expansion}
\label{sec: Magnus expansion}
The Schrödinger equation, \cref{eq: Schrodinger}, can be written in the form
\begin{equation}
    \partial_t \ket{\psi(t)} = \A(t)\, \ket{\psi(t)}, \qquad \ket{\psi(0)} = \ket{\psi_0}, \label{eq: time-dep linear ODE}
\end{equation}
where $\A(t) = -\ii\ha(t)$. Since $\mathcal{A}(t) \in \mathfrak{su}(2^M)$ is a skew-Hermitian matrix for all $t\in \mathbb{R}$, the solution of \cref{eq: time-dep linear ODE} is given by
\[ \ket{\psi(T)} = U(T, 0)\ \ket{\psi(0)}, \]
where $U(T,0) \in \mathrm{U}(2^M)$ is a unitary matrix for all $T \in \mathbb{R}$.

The problem of Hamiltonian simulation of \cref{eq: time-dep linear ODE} effectively reduces to the approximation of $U(T,0)$, and the development of quantum algorithms for its approximation is currently an area of intense activity.
Among the various algorithms for solving the Schrödinger equation on classical computers, some of the most effective methods utilize the Magnus expansion \cite{magnus1954exponential}.

The Magnus expansion is a Lie group method \cite{iserles2000lie}, which seeks an approximation of $U(T, 0)\in \mathrm{U}(2^M)$ in the Lie algebra $\mathfrak{su}(2^M)$. In practice we seek a matrix $\Theta(T, 0)$ such that
\begin{equation}
    \label{eq: Magnus Lie}
    U(T, 0) = \exp(\Theta(T, 0)), \qquad \Theta(T, 0) \in \mathfrak{su}(2^M).
\end{equation}
Since the exponential map `$\exp$' maps the Lie algebra to its Lie group, $\exp(\Theta(T, 0))$ is in $\mathrm{U}(2^M)$. Moreover, any approximation of $\Theta(T, 0)$ in the Lie algebra $\mathfrak{su}(2^M)$ retains this unitarity property.

However, a matrix with the above properties is not always guaranteed to exist. The existence (or convergence) of the Magnus expansion for ordinary differential equations (ODEs) such as \cref{eq: Schrodinger} has been widely studied \cite{strichartz1987campbell, moan2008convergence},  and very tight bounds for the convergence are known. Namely, if
\begin{equation}
\label{eq: Magnus convergence}
\int_{0}^{T} \|\A(t)\| \, \mathrm{d}t \leq \pi,
\end{equation}
where $\| \cdot \|$ is the $\ell^2$ norm,
then there is always a $\Theta(T, 0)$ which satisfies the property \eqref{eq: Magnus Lie}, $\ie$, the solution operator $U(T,0)$ can be expressed as the exponential of a matrix in $\mathfrak{su}(2^M)$. Moreover, this is a tight bound, $\ie$, there are examples of $\A(t)$ that violate the bound in \cref{eq: Magnus convergence} and for which there is no $\Theta(T, 0)$ that satisfies \cref{eq: Magnus Lie}.

As a consequence, the Magnus expansion cannot provide global solutions in general ($\ie$, over a long interval $[0,T]$ with a large $T$) and has to be used as a time-stepping scheme with a small time-step $h$ over a time grid $t_n = n h$, $n \in \mathbb{N}_0$,
\begin{equation}
    \ket{\psi(t_{n+1})} = U_n\ \ket{\psi(t_n)}, \quad n \in \mathbb{N}_0, \quad \ket{\psi(t_0)} = \ket{\psi_0}, \qquad U_n := U(t_{n+1}, t_n) = \ee^{\Theta(t_{n+1},t_n)}
\end{equation}
Equivalently, the propagator $U(T, 0)$ is computed as
\begin{equation}
    \label{eq: propagator form}
    U(T, 0) = \prod_{n=0}^{N-1} U_n = \prod_{n=0}^{N-1} \ee^{\Theta(t_{n+1},t_n)}, \qquad t_N = T,
\end{equation}
where the product $\prod_{n=0}^{N-1} U_n\ :=\ U_{N-1}U_{N-2} \ldots U_{1} U_{0}$ needs to be considered as ordered.

For practical purposes we consider the Magnus expansion written as an infinite sum in the Lie algebra $\mathfrak{su}(2^M)$,
\begin{equation}
    \label{eq: Magnus series}
    \Theta(t_{n+1},t_n) = \lim_{k \rightarrow \infty}\Theta_k(t_{n+1},t_n), \quad  \Theta_k(t_{n+1},t_n) = \sum^k_{j=1}\Theta^{[j]}(t_{n+1},t_n) ,
\end{equation}
where the  terms
$\quad \Theta^{[j]}(t_{n+1},t_n) \in \mathfrak{su}(2^M)$
are progressively smaller in size (in terms of their spectral radii) and the finite truncations $\Theta_k$ provide progressively accurate approximations to $\Theta$. A practical algorithm for approximating $U(T,0)$ in \cref{eq: propagator form} involves computing the exponential of a finite truncation of the Magnus expansion,
\[ \exp\left(\Theta_k(t_{n+1},t_n)\right) \quad \approx \quad U(t_{n+1}, t_n) = U_n. \]
Since $\Theta_k(t_{n+1},t_n) \in \mathfrak{su}(2^M)$, its exponential also provides a unitary approximation to $U_n$.

The first two terms of the Magnus expansion which we will utilize in this work are
\begin{align}
    \Theta^{[1]}(t_{n+1},t_n) =& \int^h_0\mathcal{A}(t_n+\zeta)\mathrm{d}\zeta \quad = \order\left(h\right),\label{eq: first term in Magnus}\\
    \Theta^{[2]}(t_{n+1},t_n) =& - \frac{1}{2}\int^h_0\int^{\zeta}_0[ \mathcal{A}(t_n+\xi), \mathcal{A}(t_n+\zeta) ] \mathrm{d}\xi \mathrm{d}\zeta \quad = \order\left(h^3\right). \label{eq: second term in Magnus}
\end{align}
Here the notation $[\ \cdot \ , \ \cdot \ ]$ represents the commutator of two matrices, defined as
\begin{equation}
    \label{eq: commutator}
    [P,Q] = PQ - QP.
\end{equation}
Note that the occurrence of two integrals in $\Theta^{[2]}(t_{n+1},t_n)$ suggests that this term should scale as $\order\left(h^2\right)$ in the time-step $h$. However, due to time-symmetry and due to the cancellation of certain terms in the Taylor series, terms in the Magnus expansion tend to be smaller than expected when $\A$ is smooth in time \cite{iserles01tsa}. In particular, it has been shown \cite{iserles2000lie} that
\begin{align*}
\Theta(t_{n+1},t_n) & = \Theta_1(t_{n+1},t_n) + \order\left(h^3\right),\\
\Theta(t_{n+1},t_n) & = \Theta_2(t_{n+1},t_n) + \order\left(h^5\right).
\end{align*}
\begin{Remark}
For ease of notation, we will often write  $\Theta_k(t_n)$ or simply $\Theta_k$ for $\Theta_k(t_{n+1}, t_n)$ in the rest of the manuscript, taking either the time-step $h$ or both $h$ and $t_n$ as implicitly implied. The same holds in general for any function $\va(t_n)$ which should be understood to be a function of both $t_n$ and $h$.
\end{Remark}

Time-stepping with the truncated Magnus expansions $\Theta_k, k=1,2$,
\begin{equation}
    \label{eq: timestep}
    \ket{\widetilde{\psi}^k_{n+1}} \ = \ \ee^{\Theta_k(t_n)}\ \ket{\widetilde{\psi}^k_n}, \qquad n \in \mathbb{N}_0, \qquad \ket{\widetilde{\psi}^k_0} = \ket{\psi_0},
\end{equation}
results in a second-order method for $k=1$ and a fourth-order method for $k=2$. Specifically, for $k=1,2$, there exist constants $c_{k}$ independent of $h$ and $n$ such that
\begin{equation}
    \left \| \ket{\widetilde{\psi}^k_n} - \ket{\psi(t_n)} \right\|_2 \ \leq \ c_{k} T \gamma \rho^{k+1} h^{2k}, \qquad h \rightarrow 0, \quad nh \leq T,
\end{equation}
where $\rho$, as defined in \cref{eq: hamiltonian spectral radius,eq: Hamiltonian norm time dependent},
is the maximum spectral radius of the Hamiltonian on the time interval of simulation $[0,T]$, and
\begin{equation}
    \label{eq: control frequency max}
    \gamma = \max_{t \in [0,T]} \|\ve'(t)\|_\infty
\end{equation}
depends on the maximum frequency of the controls in this interval.
Equivalently, the second and fourth-order approximations to $U(T, 0)$ have the {\em propagator errors}
\begin{equation}\label{eq: propagator approximation}
  \left\| U(T, 0) - \prod_{n=0}^{N-1} \ee^{\Theta_k(t_n)} \right\| \ \leq \ c_{k} T \gamma \rho^{k+1} h^{2k}, \qquad h \rightarrow 0, \quad Nh = T.
\end{equation}
For the derivation of these estimates, we refer the reader to \cite{iserles2000lie}.

In the following subsections, we present the concrete forms of the second and fourth-order Magnus expansions, $\Theta_1$ and $\Theta_2$, for the Schrödinger equation \eqref{eq: Schrodinger} with the Hamiltonian \eqref{eq: full Hamiltonian}. These are developed for an arbitrary interaction tensor $\jg$, \cref{eq: most general C}, and under the assumption of smooth controls, $\ve \in \mathrm{C}^\infty(\mathbb{R}; \mathbb{R}^{3M})$ (see \cref{rmk: smooth pulses}).

\subsection{Second-order Magnus expansion}
The second-order Magnus expansion is
\begin{equation}
    \label{eq: Theta_1}
    \Theta_1(t_n) = \Theta^{[1]}(t_n) = \int^h_0\mathcal{A}(t_n+\zeta)\mathrm{d}\zeta .
\end{equation}
We obtain the expansion for the Schrödinger equation \eqref{eq: Schrodinger} with the Hamiltonian \eqref{eq: full Hamiltonian} by substituting $\A(t_n+\zeta) = -\ii\ha(t_n + \zeta) = -\ii\ve(t)^\top\vcs - \ii \frac12 \vcs^\top \jg \vcs$,
\begin{equation}
    \Theta_1(t_n) = \int^h_0\mathcal{A}(t_n+\zeta)\mathrm{d}\zeta = -\ii \vmu(t_n)^\top\vcs  -\ii \frac{h}{2} \vcs^\top \jg \vcs  , \label{eq: theta_1 final}
\end{equation}
where
\begin{equation}
\vmu(t_n) = \int_0^h \ve(t_n+\zeta) \dd \zeta.
\label{eq: second-order Magnus parameters without new notation}
\end{equation}

Since $\ve \in \mathrm{C}^\infty(\mathbb{R}; \mathbb{R}^{3M})$, $\A(t)$ is smooth in time and using $\Theta_1$ in the time-stepping scheme \eqref{eq: timestep} or the propagator approximation in \cref{eq: propagator approximation} leads to a second-order method.

\subsection{Fourth-order Magnus expansion}\label{sec: fourth-order Magnus}
Since $\ve \in \mathrm{C}^\infty(\mathbb{R}; \mathbb{R}^{3M})$ and $\A(t)$ is smooth in time, the fourth-order Magnus expansion results from truncating the Magnus series \eqref{eq: Magnus series} to two terms, $\Theta^{[1]}$ and $\Theta^{[2]}$,
\begin{equation}
    \Theta_2(t_n) = \int^h_0\mathcal{A}(t_n+\zeta)\mathrm{d}\zeta - \frac{1}{2}\int^h_0\int^{\zeta}_0[ \mathcal{A}(t_n+\xi), \mathcal{A}(t_n+\zeta) ]\, \mathrm{d}\xi \mathrm{d}\zeta \label{eq: Theta_2}.
\end{equation}

The additional term $-\frac{1}{2}\int^h_0\int^{\zeta}_0[ \mathcal{A}(t_n+\xi), \mathcal{A}(t_n+\zeta) ]\, \mathrm{d}\xi \mathrm{d}\zeta$ in the fourth-order Magnus expansion is a source of numerous difficulties, arising primarily from the presence of the commutator term. When it comes to the $M$-body Hamiltonian \eqref{eq: full Hamiltonian}, $\mathcal{A}(t) = \ii \ha(t)$ has $3M$ single spin terms and $|\jg|$ pairwise coupling terms, which is reducible to $\frac12|\jg|$ in case of symmetry, ${C^{\alpha,\beta}}^\top = C^{\alpha,\beta}$ for $\alpha, \beta \in \{X,Y,Z\}$. For dense interactions, $|\jg| = \order\left(M^2\right)$, so that in general $\mathcal{A}(t)$ has $\order\left(M^2\right)$ pairwise coupling terms.

To estimate the complexity of the additional commutator ${[\mathcal{A}(t_n+\xi), \mathcal{A}(t_n+\zeta)]}$, a very na\"{i}ve estimate can be obtained by considering the commutator definition of $[P,Q]$, \cref{eq: commutator}, and individually analysing the subcomponents $PQ$ or $QP$. The term ${\mathcal{A}(t_n+\xi)\mathcal{A}(t_n+\zeta)}$, for instance, involves $\order\left(M^4\right)$ terms with four-way couplings of the form $\alpha_j \beta_k \gamma_\ell \delta_m$, where $\alpha,\beta,\gamma,\delta \in \{X,Y,Z\}$, in contrast to the pairwise couplings seen in the Hamiltonian \eqref{eq: full Hamiltonian}. This would make any application of the fourth-order Magnus expansion infeasible in the near term. In \Cref{thm: general Theta_2}, we see that this estimate is highly pessimistic and all four-way coupling terms in the commutator cancel due to the interactions in the Hamiltonian \eqref{eq: full Hamiltonian} being time-independent.

\begin{Definition}\label{def: cross product}
\normalfont
We define the wedge product between $\va(t) = \lb\va^X(t)^\top, \va^Y(t)^\top, \va^Z(t)^\top\rb^\top$ and $\vb(t) = \lb\vb^X(t)^\top, \vb^Y(t)^\top, \vb^Z(t)^\top\rb^\top$ as
\begin{equation}
\vc(t,s) = (\va \wedge \vb) (t,s) = \lb\vc^X(t,s)^\top, \vc^Y(t,s)^\top, \vc^Z(t,s)^\top\rb^\top,
\label{eq: wedge product definition}
\end{equation}
where
\begin{align*}
\vc^X(t,s) & =  \va^Y(t) \odot \vb^Z(s) - \vb^Z(t) \odot \va^Y(s),\\
\vc^Y(t,s) & = \va^Z(t) \odot \vb^X(s) - \vb^X(t) \odot \va^Z(s),\\
\vc^Z(t,s) & = \va^X(t) \odot \vb^Y(s) - \vb^Y(t) \odot \va^X(s),
\end{align*}
where $\odot$ is the Hadamard product.
\end{Definition}

\begin{Theorem}\label{thm: general Theta_2}
\normalfont
The fourth-order Magnus expansion for the Schrödinger equation \eqref{eq: Schrodinger} with the Hamiltonian \eqref{eq: full Hamiltonian} can be expressed as
\begin{equation}
\Theta_2(t_n) = -\ii \vr(t_n)^\top\vcs  -\ii \frac{h}{2}
\vcs^\top \jg \vcs
+ \lbb  \vu(t_n)^\top\vcs\, ,\, \vcs^\top \jg \vcs \rbb \label{eq: Theta_2 general with commutator},
\end{equation}
where
\begin{equation}
\vu(t_n) = -  \frac12 \int_0^h \left(\zeta - \frac{h}{2}\right) \ve(t_n+\zeta) \mathrm{d} \zeta
    , \label{eq: extra commutator rotation gates weight}
\end{equation}
\begin{equation}
    \label{eq: magnus fourth order r}
    \vr(t_n) = \vmu(t_n) - \int_0^h \int_0^\zeta (\ve \wedge \ve)(t_n+\xi, t_n+\zeta)\ \mathrm{d} \xi \mathrm{d} \zeta.
\end{equation}
Overall, the fourth-order Magnus expansion involves $3M$ single-spin terms of the form $\alpha_j$, $|\jg|$ pairwise coupling terms of the form $\alpha_j \beta_k$, and $3 M |\jg|$ three-way coupling terms of the form $\alpha_j \beta_k \gamma_\ell$.
\end{Theorem}
\begin{proof}[Proof of \Cref{thm: general Theta_2}]
See \Cref{sec: proof of 4th Magnus}.
\end{proof}
\Cref{thm: general Theta_2} is a minor extension of \cite[Theorem 3.5]{goodacrethesis},  differing only in terms of the more compact form of the integrals derived here and the treatment of $\ve^Z(t)$ as time-dependent, which requires the $\wedge$ product notation.

The reduced complexity of the fourth-order Magnus expansion is due to the most expensive part of the commutator $[\mathcal{A}(t_n+\xi), \mathcal{A}(t_n+\zeta)]$ vanishing in \Cref{thm:eliminating}. The remaining part of the commutator,
\begin{equation}
\label{eq: extra commutator}
\lbb  \vu(t_n)^\top\vcs\, ,\, \vcs^\top \jg \vcs \rbb,
\end{equation}
is somewhat simpler but nevertheless involves an additional $\order \lb |\jg| M \rb$ three-way coupling terms, which keeps the fourth-order Magnus expansion prohibitively more complex than the second-order Magnus expansion \eqref{eq: theta_1 final} and the time-independent Hamiltonian. In \Cref{sec: commutator elimination} we will develop techniques to eliminate this remaining commutator term. In \Cref{sec: homonuclear} we consider a special but important case where this commutator vanishes on its own.

\subsection{Approximating the integrals}
\label{sec:quadrature}

It is a well-established convention in the application of Magnus expansions \cite{iserles2000lie} that the integrals appearing in the Magnus expansion should be approximated with a quadrature method of a matching order of accuracy. Quadrature methods approximate an integral as
a weighted sum of function values at specifically chosen points. For integrating over the interval $[0, h]$ as in \cref{eq: first term in Magnus}, the rule takes the form
\begin{equation}
    \int _{0}^{h}\mathcal{A}(\xi)\,\mathrm{d} \xi\approx \sum _{i=1}^{k}w_{i}\mathcal{A}(x_{i}),
\end{equation}
where $k$ is the number of sample points used, $w_i$ are the quadrature weights, and $x_i$ are the quadrature {\em knots}. The  quadrature weights can be computed by integrating the Lagrange interpolation functions $\ell_i(\xi)$,
\begin{equation}
    w_i = \int_{0}^{h} \ell_i(\xi) \mathrm{d}\xi,\quad  \mathrm{where}\ \ell_i(x) = \prod^n_{\substack{j=1 \\ j\neq i}}\frac{x - x_j}{x_i - x_j}.
\end{equation}
The choice of quadrature knots $x_i$ as the roots of Legendre polynomials leads to the Gauss--Legendre quadrature, which has the highest asymptotic accuracy among all quadrature methods with a given number of nodes. In the context of the Magnus expansion,
Gauss–Legendre quadrature methods achieve $\order\lb h^{2n+1}\rb$ accuracy for $n$ knots, which is one order higher than usually expected out of them in general \cite{iserles2000lie}.
The fourth-order Magnus expansion also features a term~\eqref{eq: second term in Magnus} with a double integral, a corresponding two-dimensional quadrature rule for which is given by
\begin{equation}
    \int_0^h \int_0^\zeta \lbb\mathcal{A}(\xi),\mathcal{A}(\zeta)\rbb  \mathrm{d} \xi \mathrm{d} \zeta = \sum_{i=1}^{k}\sum_{j=1}^{k} w_{i,j} \lbb\mathcal{A}(x_i),\mathcal{A}(x_j)\rbb,
    \label{eq: 2d GL knots}
\end{equation}
where the weights $w_{i,j}$ are computed via double integrals of the  Lagrange interpolation functions,
\begin{equation}
    \int_0^h \int_0^\zeta  \lb \ell_i(\xi) \ell_j(\zeta) - \ell_j(\xi) \ell_i(\zeta) \rb \mathrm{d} \xi \mathrm{d} \zeta.
\end{equation}

In the context of the second-order Magnus expansion, $\Theta_1(t_n)$, \cref{eq: Theta_1}, the prescribed quadrature is the Gauss--Legendre quadrature with a single knot, specifically the midpoint, $\frac{t_n + t_{n+1}}{2}$. For the
fourth-order Magnus expansion, $\Theta_2(t_n)$, \cref{eq: Theta_2}, the prescribed quadrature is the Gauss--Legendre quadrature with two knots in each interval, $[t_n, t_{n+1}]$.
However, in the presence of highly-oscillatory driving pulses, \cite{iserles2018magnus,iserles2019compact,iserles2019solving,singh2019sixth} demonstrate the advantage of approximating the integrals in the Magnus expansion with quadratures with much higher accuracies than conventionally prescribed.
In this manuscript, to facilitate the use of more quadrature knots, we use GL$k$ to denote the use of a Gauss--Legendre quadrature with $k$ knots for computing the integrals appearing in the Magnus expansion.

%% file: 3_approx_matrix_exp.tex
\section{Approximating the matrix exponential}
\label{sec: matrix exponential}
Time-stepping with the Magnus expansion \eqref{eq: timestep}, or equivalently, approximating the propagator $U(T,0)$ in
\cref{eq: propagator approximation}, requires efficient procedures for approximating the matrix exponentials of the second and fourth-order Magnus expansions, $\Theta_1$ and $\Theta_2$, respectively.

\subsection{Trotterization}
\label{sec: Trotterization}
We start by considering the Trotterization technique for approximating the exponential of matrices of the form,
\begin{equation}
    \label{eq: common structure}
       \Omega =  -\ii \va^\top\vcs  - \ii \frac{h}{2} \vcs^\top \jg \vcs.
\end{equation}
In the context of Trotterization techniques, we assume that mixed couplings are not present in the interaction tensor, $C^{\alpha,\beta}=0$ for $\alpha \neq \beta$, and matrices of the form \eqref{eq: common structure} can be separated into $X$, $Y$ and $Z$ components, 
\[\Omega = \Omega^X + \Omega^Y + \Omega^Z,\]
where
\begin{equation}
    \Omega^\alpha = -\ii {\va^\alpha}^\top\vcs^\alpha - \frac{h}{2}{\vcs^\alpha}^\top C^{\alpha,\alpha}\ \vcs^\alpha,\qquad \alpha\in \{X,Y,Z\}. \label{eq: one x or y or z part of Hamiltonian}
\end{equation}
As a consequence of the following lemma, any two terms in a given component $\Omega^\alpha$ commute with each other.
\begin{Lemma} 
\begin{equation}
    \lbb A_j , A_k \rbb = 0,\quad  \lbb A_i , A_jA_k \rbb = 0,\quad \lbb A_iA_\ell , A_jA_k \rbb = 0, \qquad i,j,k,\ell\in \{1,\cdots,M\}.
\end{equation}\label{lemma: AA commutators}
\end{Lemma}
\begin{proof}
These follow immediately from the definition of $A_k$, \cref{eq: tensor}, and the fact that a matrix commutes with itself and with the identity matrix.
\end{proof}
As a consequence, the exponential of each component $\Omega^\alpha$ scaled by an arbitrary real scalar $s$ can be computed exactly as
\begin{equation}
    \label{eq: single component of splitting general}
    \ee^{s \Omega^\alpha} = \prod_{\ell=1}^M \ee^{-\ii s \va^\alpha_\ell \alpha_\ell} \prod_{j=1}^M\prod_{\substack{k=1\\k\neq j}}^M \ee^{-\ii s \frac{h}{2} C^{\alpha,\alpha}_{j,k}  \alpha_{j}\alpha_{k}}, \qquad s \in \RR,
\end{equation}
which can be implemented using $M$ rotation gates (\cref{fig: rotation gate}) and $|C^{\alpha,\alpha}|$ {\em Ising} coupling gates (\cref{fig: coupling gate}) with different gate parameters. Here $|C^{\alpha,\alpha}| \leq M(M-1)$ is the number of non-zero entries of $C^{\alpha,\alpha}$.
When ${C^{\alpha,\alpha}}^\top = C^{\alpha,\alpha}$, we can combine half the exponents and compute
\begin{equation}
    \label{eq: single component of splitting}
    \ee^{s \Omega^\alpha} = \prod_{\ell=1}^M \ee^{-\ii s \va^\alpha_\ell \alpha_\ell} \prod_{j=1}^M\prod_{k=j+1}^M \ee^{-\ii s h C^{\alpha,\alpha}_{j,k}  \alpha_{j}\alpha_{k}}, \qquad s \in \RR,
\end{equation}
so that the number of Ising coupling gates required are $\frac12 |C^{\alpha,\alpha}|\leq \frac12 M(M-1)$. 

As noted in \cref{rmk: mixed couplings}, the absence of mixed couplings is not a serious restriction. In \Cref{sec: splitting and circuit}, each component $\ee^{\Omega^\alpha}$ is implemented as described in \cref{eq: single component of splitting} and depicted by a block of gates in the circuit presented in \cref{fig: trotter circuit} (c).

However, this assumption is crucial in the context of Trotterization since it allows us to split $\Omega$ in components $\Omega^X, \Omega^Y$ and $\Omega^Z$, each of whose exponential can be computed exactly, as outlined above. 
A first-order time-stepping method can be obtained by approximating the exponential of $\Omega$ using the {\em Trotter} {\em splitting},
\begin{equation}
\label{eq:trotter omega}
\ee^{h(A+B+C)} \ \ = \ \  \ee^{h A} \ee^{h B} \ee^{h C} + \order \lb h^2 \rb.
\end{equation}
Higher-order time-stepping methods can be derived by using higher-order {\em splittings} \cite{mclachlan2002splitting,blanes2008splitting}, which are collectively also called {\em Trotterizations}. A second-order method is given by the {\em Strang} splitting \cite{strang1968construction},
\begin{equation}
\label{eq:strang omega}
\ee^{h(A+B+C)} \ \ = \ \  \ee^{\frac{h}{2} A} \ee^{\frac{h}{2} B} \ee^{h C} \ee^{\frac{h}{2} B} \ee^{\frac{h}{2} A} + \order \lb h^3 \rb,
\end{equation}
and a fourth-order method by the {\em Yoshida} splitting \cite{yoshida1990construction},
\begin{equation}
\begin{aligned}
\ee^{h(A+B+C)} =& \ee^{\frac{x h}{2}A}\ee^{\frac{x h}{2}B}\ee^{x hC}\ee^{\frac{x h}{2}B}\ee^{\frac{(x+y)h}{2}A}\ee^{\frac{yh}{2}B}\ee^{yh  C}\ee^{\frac{yh}{2}B}\ee^{\frac{(x+y)h}{2}A}\ee^{\frac{x h}{2}B}\ee^{x hC}\ee^{\frac{x h}{2}B}\ee^{\frac{x h}{2}A}\\
&+ \order \lb h^5 \rb,
\end{aligned}
\label{eq: 4th splitting}
\end{equation}
where
\begin{equation}
    x =  \frac{1}{2 - 2^{1/3}}, \quad y = 1-2 x.
\end{equation}
The error in a single time-step of a $k$th order splitting $\mathcal{S}_k$ ($\eg$, the product of the exponentials in the right hand sides of eqs. \eqref{eq:trotter omega}, \eqref{eq:strang omega} or \eqref{eq: 4th splitting}) is bounded from above by 
\begin{equation} 
    \label{eq:splitting error}
    \left\|\ee^{h(A+B+C)} - \mathcal{S}_k\right\| \leq \tilde{c}_k \left(\max\{\|A\|, \|B\|, \|C\|\}\right)^k h^{k+1},
\end{equation}
for some constant $\tilde{c}_k$.

When repeated for multiple steps, the first and last occurrences of $\ee^{c A}$ can be combined together in the case of the Strang splitting and the fourth-order Yoshida splitting. On average, Strang splitting requires $4$ exponentials per time-step and the fourth-order Yoshida splitting required $12$ exponentials. In priciple, for the exponentiation of $\Omega$, we may assign $\Omega^X, \Omega^Y, \Omega^Z$ to $A,B,C$ in any order. However, different choices can lead to different circuit depths and gate counts. In the fourth-order Yoshida splitting, on average $3$ exponentials each of $A$ and $C$ are required per time-step and $6$ of $B$. Ideally the component assigned to $B$ should be the one involving the simplest circuit -- this is expected to vary with the architecture as well as the coupling tensor $\jg$.

Trotterization techniques can be used for computing the matrix exponential of a time-independent Hamiltonian, 
\[
\ha_0 = \ve_0 \, + \, \frac12 \vcs^\top \jg \vcs,
\] 
by time-stepping,
\begin{equation}
    \label{eq: constant}
    U(T,0) = \exp\left(-\ii T \ha_0\right) \quad  \approx \quad  \prod_{n=0}^{N-1}  \exp\left(-\ii h \ha_0\right), \qquad N h = T,
\end{equation}
where each $\exp\left(-\ii h \ha_0\right)$ is approximated by an appropriate splitting, \cref{eq:trotter omega,eq:strang omega,eq: 4th splitting}. 
In the case of a time-dependent Hamiltonian, \cref{eq: full Hamiltonian}, time-stepping with a piecewise-constant approximation of the Hamiltonian $\ha(t_n)$,
\begin{equation}
    \label{eq: piecewise constant}
    U(T,0) \quad  \approx \quad  \prod_{n=0}^{N-1}  \exp\left(-\ii h \ha(t_n)\right), \qquad N h = T,
\end{equation}
where each $\exp\left(-\ii h \ha(t_n)\right)$ is approximated by the Trotter splitting \eqref{eq:trotter omega}, leads to a first-order method for \cref{eq: Schrodinger}.
We note that in both cases in eqs \eqref{eq: constant} and \eqref{eq: piecewise constant}, the matrices to be exponentiated have the form \eqref{eq: common structure}, with
 $\va = h \ve_0$ and $\va=h \ve(t_n)$, respectively. 
 
The second-order Magnus expansion, $\Theta_1(t_n)$ in \cref{eq: theta_1 final}, also has the same structure \eqref{eq: common structure}, with $\va=\vmu(t_n)$, and we can approximate its exponential by utilizing the appropriate splitting -- in this case the Strang splitting \eqref{eq:strang omega}, which is a second-order method. Due to the common structure \eqref{eq: common structure}, each step of the second-order Magnus-based time-stepping method \eqref{eq: propagator approximation} is exactly as expensive as a time-step of Strang splitting method with a time-independent Hamiltonian \eqref{eq: constant}, and only $33\%$ more expensive than the first-order method \eqref{eq: piecewise constant} for time-dependent Hamiltonians.

Moreover, the circuit used for the Strang splitting of $\exp\left(-\ii h \ha_0\right)$ can be re-used, with only the values of the coefficients `$\va$' requiring a change -- this involves changing some tuneable parameters of the quantum gates. Note that the integrals in $\vmu(t_n)$, \cref{eq: second-order Magnus parameters without new notation}, are one-dimensional and can be easily computed a priori to a high accuracy on a classical computer. 

The integrals, \cref{eq: extra commutator rotation gates weight,eq: magnus fourth order r},
appearing in the fourth-order Magnus expansion $\Theta_2(t_n)$ in \cref{eq: Theta_2 general with commutator} look more complicated, but are only two-dimensional and can also be computed easily to high accuracy on a classical computer. 
A much more significant problem for the utilization of the fourth-order Magnus expansion is posed by the presence of the commutator term
\begin{equation}
    \lbb  \vu(t_n)^\top\vcs\, ,\, \vcs^\top \jg \vcs \rbb \tag{\ref{eq: extra commutator}}
\end{equation}
in \cref{eq: Theta_2 general with commutator}, which means that $\Theta_2(t_n)$ does not have the structure \eqref{eq: common structure}. 
Thus, unlike the case of the second-order Magnus expansion, we cannot utilize the splittings in \crefrange{eq:trotter omega}{eq: 4th splitting} directly, and in particular, we cannot re-use the circuit for the time-independent Hamiltonian by only changing the tunable gate parameters.

Moreover, by \Cref{thm: general Theta_2}, the commutator term \eqref{eq: extra commutator} has $\order \lb |\jg|M \rb \leq \order \lb M^3 \rb$ three-way coupling terms  of the form $\alpha_j \beta_k \gamma_\ell$. A Trotterized approach for the Magnus expansion in this form is highly problematic for multiple reasons: (i) the circuit depth and gate count must increase by a factor of $\order \lb M \rb$, (ii) the additional terms require three-qubit gates, and (iii) involve mixed three-way couplings. Due to the mixed couplings, there is no straightforward way of splitting $\Theta_2$ into a sum of a small number of subcomponents where any two subterms commute.

In \Cref{sec: commutator elimination}, we present a technique for eliminating this commutator, and in \Cref{sec: homonuclear} we consider a special case where the commutator simplifies. 
This overcomes all the aforementioned challenges and allows the application of a Trotterized approach to the fourth-order Magnus expansion. A detailed Trotterized quantum circuit for the fourth-order Magnus expansion is presented in \Cref{sec: splitting and circuit}.

The Trotterization approach presented in this section for exponentiating \eqref{eq: common structure} involves time-stepping. Higher accuracy requires smaller time-steps, which corresponds to larger circuit depths. In practice, Trotterization aims to achieve a reasonable balance -- producing simple and relatively shallow circuits for reasonable accuracy in short-to-medium time simulation problems. This balance makes Trotterized algorithms among the few feasible candidates for Hamiltonian simulation on near-term quantum devices, and first-order Trotterized methods have been utilized in early attempts to explore utility of quantum computing in presence of noise \cite{IBM23nisq}. Since the main emphasis in this work is shorter circuit depths for reasonable accuracy, Trotterisation is our method of choice. Nevertheless, we note that other alternatives do exist and comment upon some of them in \Cref{sec: qubitization} briefly.

\subsection{Other approaches for matrix exponential}
\label{sec: qubitization}

A significant contrast to the time-stepping approach in Trotterization is the recently developed qubitization procedure  \cite{low2019hamiltonian} which is a global approximation technique that does not require time-stepping. Qubitization achieves `{\em additive}' query complexity for Hamiltonian simulation of time-independent Hamiltonians, 
\begin{equation}
    \label{eq:additive complexity}
\order\left(\rho T+ \frac{\log (1/\varepsilon)}{\log(\ee + \log (1/\varepsilon)/\rho T)} \right),
\end{equation}
where $[0,T]$ is the time window of simulation, $\varepsilon$ is the allowable error, $\rho$ is the spectral radius \eqref{eq: hamiltonian spectral radius} of the Hamiltonian, and the Hamiltonian to be exponentiated is given by a linear combination of unitaries as 
\begin{equation}
    \label{eq: LCU norm}
    \widetilde{\ha} = \sum^{d}_{j = 1} u_j\mathcal{U}_j,  \qquad u_j \in \RR, \qquad \mathcal{U}_j^*\mathcal{U}_j = \mathcal{U}_j\mathcal{U}_j^* = I, \qquad \text{and} \quad \rho =  \|\widetilde{\ha}\| \leq \sum^{d}_{j = 1} |u_j|. 
\end{equation} While qubitization achieves the optimal complexity for Hamiltonian simulation and should be the method of choice when high accuracy is required over a long temporal window of simulation, it requires fully fault-tolerant quantum computers and involves larger overheads and circuit depths, potentially making it unsuitable for near term applications. The time-ordering approach of \cite{watkins2022timedependent} and the Floquet-based approach of \cite{mizuta2023optimal} extend the optimal (additive) complexity of qubitization to periodic-in-time Hamiltonians. However, no approach currently exists for combining qubitization with non-periodic time-dependent Hamiltonian and achieving additive complexity.

While qubitization can be combined with time-stepping methods \eqref{eq: propagator approximation} in a straightforward way at a less optimal query complexity, 
\begin{equation}
    \label{eq:non-additive complexity} 
    \order\left(\frac{\rho T^2}{h} +  \frac{T\log (1/\varepsilon)}{h\log(\ee + \log (1/\varepsilon)/\rho T)} \right),
\end{equation}
where $h$ is the time-step size,
the simpler approach of BCCKS (Berry-Childs-Cleve-Kothari-Somma) \cite{BCCKS_berry2015simulating} works equally well for small time-steps. 

The overall gate complexity, with the exception of some overheads, can be estimated as a product of the query complexity and the gate complexity of the {\em oracle}. In this section, we briefly consider the gate complexity of the oracle required for time-stepping with \cref{eq: common structure} and the fourth-order Magnus expansion \eqref{eq: Theta_2 general with commutator}.

Key to the applicability of qubitization and BCCKS to the exponentiation of \cref{eq: common structure}, is the fact that the {\em effective} Hamiltonian $\widetilde{\ha} := \frac{\ii}{h} \Omega$ can also be express as a linear combination of unitaries,
\begin{equation}
    \widetilde{\ha} = \sum^{d}_{j = 1} u_j\mathcal{U}_j \quad = \quad  \sum_{k=1}^M  \ \sum_{\alpha \in \{X,Y,Z\}} \frac{\va^{\alpha}_k}{h}\, \alpha_k \ + \ \frac12  \sum_{j=1}^M \ \sum_{k=1}^M \sum_{\alpha \in \{X,Y,Z\}} \sum_{\beta \in \{X,Y,Z\}} C^{\alpha,\beta}_{j,k}\,  \alpha_j \, \beta_k, \label{eq: LCU}
\end{equation}
since $\alpha_k$ are unitary, and products of the unitaries $\alpha_j \, \beta_k$ is also unitary. 
In particular,  $\widetilde{\ha}$ is a linear combination of $d=3M + |\jg|$ unitaries, where $|\jg| \leq 9 M(M-1)$ is the number of non-zero elements of the interaction tensor. If we assume symmetry of couplings, ${C^{\alpha,\beta}}^\top = C^{\alpha,\beta}$, we can express $\widetilde{\ha}$ as a linear combination of $d=3M + \frac12 |\jg|$ unitaries. Further reductions are possible if mixed-couplings are eliminated by a suitable change of basis \cite{Klassen2019twolocalqubit}.

Let $C_1$ be the cost of implementing the unitary operator $\alpha_k$ in terms of the number of single-qubit gates (typically $C_1=1$), and let $C_{2,1}$ and $C_{2,2}$ be the cost of implementing the unitary operator  $\alpha_j \, \beta_k$ in terms of the number of single-qubit and two-qubit gates, respectively, then a single query of the Hamiltonian oracle in the BCCKS or qubitization procedure requires $3M C_1 + |\jg| C_{2,1}$ single-qubit gates and $|\jg| C_{2,2}$ two-qubit gates (with $|\jg|$ being replaced by $\frac12 |\jg|$ if couplings are symmetric).

The exponential of the time-independent Hamiltonian $\ha_0$ can be computed directly ($\ie$, without time-stepping) using the qubitization procedure at the optimal query complexity \eqref{eq:additive complexity}. However, the second-order Magnus expansion $\Theta_1(t_n)$ and the piecewise-constant Hamiltonian $\ha(t_n)$ both require time-stepping with a time-step $h$ that is expected to be small. Both BCCKS and qubitization can be used in the time-stepping procedures \eqref{eq: propagator approximation} and \eqref{eq: piecewise constant}, at the suboptimal query complexity \eqref{eq:non-additive complexity}. The gate complexity of the oracle in all three cases (time-independent, piecewise constant, and second-order Magnus expansion) is identical since they all conform to the structure \eqref{eq: common structure}. We note that the error in the Magnus expansions increases with the spectral radius, \cref{eq: propagator approximation}, and as the number of spins, the connectivity, strength of controls, or coupling strengths increase, we need smaller time-steps due to an increase in spectral radius $\rho$, \cref{eq: Hamiltonian norm time dependent}.
As mentioned previously, in the small time-step regime BCCKS is as effective as qubitization, and arguably simpler.

The direct application of BCCKS or qubitization to computing the exponential of the fourth-order Magnus expansion \eqref{eq: Theta_2 general with commutator} in the context of the time-stepping procedure \eqref{eq: propagator approximation} becomes substantially more expensive due to the commutator term,
\begin{equation}
    \lbb  \vu(t_n)^\top\vcs\, ,\, \vcs^\top \jg \vcs \rbb. \tag{\ref{eq: extra commutator}}
\end{equation}
By \Cref{thm: general Theta_2}, the commutator term involves $\order \left(|\jg| M \right) \leq \order \lb M^3 \rb$ three-way unitary terms of the form $\alpha_j \beta_k \gamma_\ell$. The direct implementation of the effective Hamiltonian, $\widetilde{\ha} =\frac{\ii}{h} \Theta_2(t_n)$, in this case requires an additional $\order \left(|\jg| M \right)$ three-qubit gates. This makes the gate complexity of the oracle extremely high and a direct application of BCCKS and qubitization to a fourth-Magnus expansion becomes prohibitively expensive in gate count and circuit depth. However, similar to the Trotterization case discussed in \Cref{sec: Trotterization}, this prohibitive cost is possible to avoid by eliminating the commutator \eqref{eq: extra commutator} using the technique presented in \Cref{sec: commutator elimination}, or when considering the special case described in \Cref{sec: homonuclear}.

\subsection{Eliminating commutators}
\label{sec: commutator elimination}
As we have seen in \Cref{sec: Trotterization,sec: qubitization}, the commutator \eqref{eq: extra commutator} is highly undesirable due to the prohibitive increase in circuit depths and gate counts, in both Trotterization and qubitization procedures. It seems reasonable to assume that once a commutator appears in a Magnus expansion, it must also appear in its splitting, and the additional cost due to this commutator is unavoidable in fourth-order Magnus-based methods. However, by composition with carefully chosen exponentials it is possible to eliminate commutators even in sixth-order Magnus expansions \cite{singh2019sixth}. We show that an asymmetric composition can eliminate the undesirable commutator term in our fourth-order Magnus expansion \eqref{eq: Theta_2 general with commutator}.

\begin{Theorem}
\normalfont
    \label{thm:eliminating}
     The exponential of the fourth-order Magnus expansion $\Theta_2(t_n)$, \cref{eq: Theta_2 general with commutator}, can be approximated by the composition
    \begin{equation}
        \ee^{-\eli(t_n)} \ee^{W(t_n)}\ee^{\eli(t_n)} = \ee^{\Theta_2(t_n)} + \order \lb h^5 \rb, \label{eq: Theta_2 general with eliminator}
    \end{equation}
    where
    \begin{equation}
    \label{eq: eli}
        \eli(t_n) = -\ii \frac{2}{h} \vu(t_n)^\top\vcs
    \end{equation}
    involves only single-spin terms, and
    \begin{equation}
        \label{eq: Theta_2 hetero W}
        W(t_n) = -\ii \tilde{\vr}(t_n)^\top\vcs  -\ii \frac{h}{2}  \vcs^\top \jg \vcs
    \end{equation}
    is a term which conforms to the structure \cref{eq: common structure}, with
    \begin{equation}
        \label{eq:r tilde}
        \tilde{\vr}(t_n) = \vr(t_n) + \frac{4}{h}(\vu(t_n) \times \vr(t_n)) \quad \in \CC^{3M}.
    \end{equation}
    Here the cross product
    $\vc = (\va \times \vb)$ is such that $\vc_k  = \va_k \times \vb_k $ for $k=1,\ldots,3M$, is the usual cross product on $\RR^3$, $\ie$ $\times$ is the cross product with the Hadamard product $\odot$ being underlying product.

    Moreover, there exist constants $\hat{c}_1$ and $\hat{c}_2$ such that
    \begin{equation}
        \left\| \ee^{-\eli(t_n)} \ee^{W(t_n)}\ee^{\eli(t_n)} - \ee^{\Theta_2(t_n)} \right\| \leq \left(\hat{c}_1 M^2 \gamma_n^2 \rho_n + \hat{c}_2 M \gamma_n \rho_n^3 \right)h^5, \label{eq: Theta_2 general with eliminator estimate}
    \end{equation}    
    where $\gamma_n = \max_{t \in [t_n,t_{n+1}]} \|\ve'(t)\|_\infty$ depends on the maximum frequency of the controls in the interval $[t_n,t_{n+1}]$, 
    and $\rho_n = \max_{t \in [t_n,t_{n+1}]} \|\ha(t)\|$ is the maximum spectral radius of the time-dependent Hamiltonian over this interval, which is the single time-step version of \cref{eq: hamiltonian spectral radius}.
\end{Theorem}

\begin{proof}
The right hand side of \cref{eq: Theta_2 general with eliminator} can be expressed in terms of the {\em Baker--Campbell--Hausdorff} (BCH) expansion \cite{oteo1991baker,hall2013lie},
\begin{equation}
    \label{eq:eliminating}
    \ee^{-\eli(t_n)} \ee^{W(t_n)} \ee^{\eli(t_n)} = \ee^{\widetilde{\Theta}_2(t_n)},
\end{equation}
where
\begin{equation}
    \label{eq:BCH}
    \widetilde{\Theta}_2(t_n) := \bch(\bch(-\eli(t_n), W(t_n)),\eli(t_n)).  
\end{equation}
The asymmetric splitting in \cref{eq:eliminating,eq:BCH} is typically associated with a first-order approximation. However, for the specific choice of $E(t_n)$ and $W(t_n)$, we demonstrate that $\widetilde{\Theta}_2(t_n)$ provides a fourth-order approximation of the fourth-order Magnus expansion $\Theta_2(t_n)$. We note in passing that the composition in \cref{eq:eliminating} is also a similarity transformation, although we do not explicitly exploit this fact here.

Since we seek a fourth-order approximation, we can ignore $\order \lb h^5 \rb$ and smaller terms in the BCH expansions. In particular, commutators with four or more letters can be ignored since these must feature at least one occurrence of $E(t_n)=\order \lb h^2 \rb$, and since $W(t_n)=\order \lb h \rb$.
Expanding both BCH series in \cref{eq:BCH}, we find
\begin{equation}
\begin{aligned}
\widetilde{\Theta}_2(t_n) &=
W+ \underbrace{\eli -\eli}_{=0}+{\frac {1}{2}}[-\eli,W]+{\frac {1}{2}}[W,\eli] \underbrace{-{\frac {1}{12}}[W,[-\eli,W]]
+{\frac {1}{12}}[W,[W,\eli]]}_{=0}
\\
&
\quad \underbrace{+{\frac {1}{2}}[-\eli,\eli]
+{\frac {1}{12}}[-\eli,[-\eli,\eli]]
+{\frac {1}{12}}[W,[-\eli,\eli]]
-{\frac {1}{12}}[\eli,[-\eli,\eli]]}_{=0}
\\
&
\quad \underbrace{
+{\frac {1}{12}}[-\eli,[-\eli,W]]
+{\frac {1}{4}}[[-\eli,W],\eli]
+{\frac {1}{12}}[-\eli,[W,\eli]]
-{\frac {1}{12}}[\eli,[W,\eli]]}_{=\order \lb h^5 \rb}
+\order \lb h^5 \rb\\
&=W(t_n)-[\eli(t_n),W(t_n)] + \order \lb h^5 \rb,
\end{aligned}
\label{eq: Magnus 4th BCH expansion}
\end{equation}
where we have written $E$ and $W$ as short form for $E(t_n)$ and $W(t_n)$, respectively. In the commutators involving three letters, instances with two occurrences of $E$ are also $\order \lb h^5 \rb$, and have been discarded.

Substituting \cref{eq: Theta_2 hetero W,eq: eli,eq:r tilde} into \cref{eq: Magnus 4th BCH expansion},
\begin{align*}
\widetilde{\Theta}_2(t_n)
&= W(t_n) 
+\frac{2}{h} [ \vu(t_n)^\top\vcs,\vr(t_n)^\top\vcs]
+\frac{8}{h^2} [ \vu(t_n)^\top\vcs,(\vu(t_n) \times \vr(t_n))^\top\vcs]\\
&\quad +[\vu(t_n)^\top\vcs,\vcs^\top \jg \vcs]
+ \order \lb h^5 \rb \\
&= W(t_n) 
+\frac{2}{h} [ \vu(t_n)^\top\vcs,\vr(t_n)^\top\vcs]
+[\vu(t_n)^\top\vcs,\vcs^\top \jg \vcs]
+ \order \lb h^5 \rb \\
&= \underbrace{-\ii \vr(t_n)^\top\vcs -\ii \frac{h}{2}
\vcs^\top \jg \vcs +[\vu(t_n)^\top\vcs,\vcs^\top \jg \vcs]}_{\Theta_2(t_n)}\\
& \quad 
-\ii \frac{4}{h}(\vu(t_n) \times \vr(t_n))^\top \vcs +\frac{2}{h} [ \vu(t_n)^\top\vcs,\vr(t_n)^\top\vcs]
+ \order \lb h^5 \rb \\
\end{align*}
where $\frac{8}{h^2} [ \vu(t_n)^\top\vcs,(\vu(t_n) \times \vr(t_n))^\top\vcs]=\order \lb h^5 \rb$ is discarded since $\vu(t_n) = \order \lb h^3 \rb$ and $\vr(t_n)=\order \lb h \rb$.

In \Cref{sec: u and r commutator} we show that
\begin{equation}
\lbb\vu(t_n)^\top\vcs,\vr(t_n)^\top\vcs\rbb = 2 \ii (\vu(t_n) \times \vr(t_n))^\top \vcs.
\label{eq: u and r commutator} 
\end{equation}

Thus,
\[ \widetilde{\Theta}_2(t_n) = \Theta_2(t_n) + \order \lb h^5 \rb,\]
and
\[ \ee^{-\eli(t_n)} \ee^{W(t_n)}\ee^{\eli(t_n)} = \ee^{\widetilde{\Theta}_2(t_n)} = \ee^{\Theta_2(t_n)} + \order \lb h^5 \rb. \]

Overall, in the BCH formula, the leading terms that are discarded either have two occurrences of $\eli(t_n)$ and one of $W(t_n)$, or one occurrence of $\eli(t_n)$ and three of $W(t_n)$. The only other term discarded is 
$\frac{8}{h^2} [ \vu(t_n)^\top\vcs,(\vu(t_n) \times \vr(t_n))^\top\vcs]$ which is certainly smaller than $\|\eli(t_n)\| \|W(t_n)\|$. Thus the leading error term is proportional to $\|\eli(t_n)\|^2 \|W(t_n)\|$ and $\|\eli(t_n)\| \|W(t_n)\|^3$. 

By Taylor expanding $\ve(t+\zeta)$ at $t$, $\ve(t+\zeta) = \ve(t) + \zeta \ve'(t+\xi)$ for some $\xi \in [0,\zeta] \subseteq [0,h]$, and substituting in \cref{eq: extra commutator rotation gates weight}, we can get the bound
\[
\|E(t_n)\| \leq 
\frac{3M}{h} \gamma_n
\int^h_0 \left(\zeta - \frac{h}{2}\right) \zeta \mathrm{d} \zeta =
\frac{M\gamma_n}{4}h^2,
\]
where $\gamma_n = \max_{t \in [t_n,t_{n+1}]} \|\ve'(t)\|_\infty$.
By a similar application of Taylor series to the approximation of the integral in $\widetilde{\vr}$ \eqref{eq:r tilde} and the definition of the wedge product \eqref{eq: wedge product definition} we can show that there exists $\hat{c}_3$ such that 
\begin{equation*}
\|W(t_n)\| \leq \| \vmu(t_n)^\top\vcs + \frac{h}{2}  \vcs^\top \jg \vcs \|  + \| \left(\tilde{\vr}(t_n)-\vmu(t_n)\right)^\top \vcs \| \leq \rho_n h +  \hat{c}_3 M \nu_n  \gamma_n h^3,
\end{equation*}
where $\nu_n= \max_{t \in [t_n,t_{n+1}]} \|\ve(t)\|_\infty$ is the maximum intensity of the controls in this interval. $\ie$, up to $\order(h^3)$, the norm of $W(t_n)$ is approximated by the norm of $h \ha(t_n)$. 
\Cref{eq: Theta_2 general with eliminator estimate} follows from substituting these bounds in the leading error terms, $\|\eli(t_n)\|^2 \|W(t_n)\|$ and $\|\eli(t_n)\| \|W(t_n)\|^3$.

\end{proof}

The crucial achievement in \Cref{thm:eliminating} is that $W(t_n)$ has the structure \eqref{eq: common structure}. In particular, unlike the fourth-order Magnus expansion $\Theta_2$, \cref{eq: Theta_2 general with commutator}, it does not have any commutators.
When utilizing qubitization in a time-stepping procedure, the cost of exponentiating $W(t_n)$ is no more expensive than a single step when time-stepping with a piecewise-constant Hamiltonian or the second-order Magnus expansion $\Theta_1$. In a Trotterized approach, the exponential of $W(t_n)$ needs to be computed to fourth-order accuracy using the Yoshida splitting \cref{eq: 4th splitting}, the cost of which is identical to a single step of a fourth-order Trotterized time-stepping procedure for a time-independent Hamiltonian.

By approximating the exponential of the fourth-order Magnus expansion using \cref{eq:eliminating}, only a small structural modification ($\ie$, up to changes in gate parameters) to a fourth-order Trotterized circuit for a time-independent Hamiltonian is required: Namely, the additional layers to account for the $\ee^{\eli(t_n)}$ and $\ee^{-\eli(t_n)}$ terms. Similarly, these additional layers either side of a qubitized time-stepping circuit for a piecewise constant Hamiltonians result in a qubitized approach for the fourth-order Magnus expansion (up to changes in gate parameters). 

\newcommand{\vlam}{\boldsymbol{\lambda}}

\begin{Theorem}
\label{thm:single spin}
\normalfont
The exponential of a term $-\ii \va^T \vcs$, which only contains single-spin terms, can be implemented exactly using three layers with $M$ single-spin gates each,
\begin{equation} 
    \label{eq:exact single spin}
    \ee^{-\ii \va^T \vcs} = \ee^{-\ii {\vlam^X}^T \vcs^X}\ee^{-\ii {\vlam^Y}^T \vcs^Y}\ee^{-\ii {\vlam^Z}^T \vcs^Z} , \qquad \ee^{-\ii {\vlam^\alpha}^T \vcs^\alpha} = \bigotimes _{m=1}^M   \ee^{-\ii  \lambda_m^\alpha \alpha},
\end{equation}  
where the coefficients $\lambda_m^\alpha$ satisfy the equations,
\begin{equation}
\begin{aligned}
\va_m^X &= q(\vlam_m) \lb \sin(\lambda_m^Z) \sin(\lambda_m^Y) \cos(\lambda_m^X) + \sin(\lambda_m^X)\cos(\lambda_m^Z)\cos(\lambda_m^Y) \rb, \\
\va_m^Y &= q(\vlam_m) \lb\sin(\lambda_m^Y)\cos(\lambda_m^Z)\cos(\lambda_m^X) - \sin(\lambda_m^Z)\cos(\lambda_m^Y)\sin(\lambda_m^X) \rb,\\
\va_m^Z &= q(\vlam_m) \lb\sin(\lambda_m^Z)\cos(\lambda_m^Y)\cos(\lambda_m^X) + \sin(\lambda_m^Y)\cos(\lambda_m^Z)\sin(\lambda_m^X) \rb,
\end{aligned}
\label{eq: Trotter type E parameters}
\end{equation}
and
\[ q(\vlam_m)  = p\lb\prod_{\alpha \in \{X,Y,Z\}}\cos(\lambda_m^\alpha)  \ \ - \prod_{\alpha \in \{X,Y,Z\}}\sin(\lambda_m^\alpha)\rb, \qquad  p(x) = \frac{\arccos(x)}{\sin\lb\arccos(x)\rb}.\]
\end{Theorem}
\begin{proof}[Proof of \Cref{thm:single spin}]
See \Cref{sec: Trotter type E parameters derivation}.
\end{proof}

It follows from \Cref{thm:single spin} that $\ee^{\eli(t_n)}$ and $\ee^{-\eli(t_n)}$ can be computed using $3M$ single-qubit gates each, which represents a very marginal increase in gate count and circuit depth for the fourth-order Magnus expansion. The non-linear system of equations, \cref{eq: Trotter type E parameters}, can be solved on a classical computer a priori, once the integral $\vu(t_n)$ has been computed. In the context of its application in the fourth-order Magnus-based method developed here, the solution of $\lambda_m^\alpha$ only needs to be computed to an accuracy of $\order \lb h^5 \rb$, but it can be computed to much higher accuracies cheaply, if needed.

\subsection{Isotropic Hamiltonians with identical controls}
\label{sec: homonuclear}

In this section, we demonstrate that for a large class of Hamiltonians appearing in NMR applications, the additional commutator \eqref{eq: extra commutator} vanishes, and does not require the elimination procedure outlined in \Cref{sec: commutator elimination}. Specifically, we consider Hamiltonians with {\em isotropic} couplings and identical controls. Hamiltonians of this form appear frequently in the context of liquid-state NMR when there is only one {\em specie} of NMR active nuclei. In such cases, the time-dependent control pulses that accompany $X$ and $Y$ are identical across all spins and the component accompanying $Z$ is a spin-dependent but time-independent scalar quantity $\Omega_k$ which is called a {\em spin offset} or a {\em chemical shift},
\[ \ve(t) = (p_x(t)\vone^\top, p_y(t) \vone^\top, \vo^\top)^\top, \qquad \vo = (\Omega_1,\cdots,\Omega_M)^\top\in \RR^M,\]
where $\vone = (1,\ldots,1)^\top \in \RR^M$. The {\em isotropic} nature of the coupling strengths corresponds to $C^{X,X} = C^{Y,Y} = C^{Z,Z} = C = C^\top\in \RR^{M\times M}$ and $C^{\alpha,\beta}=0$ for $\alpha \neq \beta$, $\ie$ coupling strengths are symmetric and equal in all directions, while mixed couplings are absent. Overall, the interaction tensor $\jg$ becomes 
$$ \jhomo = \begin{pmatrix}
    C & 0 & 0\\
    0 & C & 0\\
    0 & 0 & C \end{pmatrix}, \qquad C^\top=C, \qquad C \in \mathbb{R}^{M \times M}. $$
This is in contrast to the strongly {\em anisotropic} case commonly encountered in solid-state NMR. In liquid-state NMR, anisotropic interactions are typically averaged out due to molecular motion, especially for isotropic solutions, which simplifies the treatment to isotropic couplings. The same phenomena is responsible for the absence of mixed couplings.

We start by stating an important observation in \Cref{lem: homonuclear reduction} that proves helpful in simplifying the fourth-order Magnus expansion in the isotropic case.
\begin{Lemma}
\begin{equation}
    \lbb  \vone^\top \vcs^{\alpha} , \sum_{\beta \in\{ X,Y,Z \}} {\vcs^{\beta}}^\top C \vcs^{\beta} \rbb = 0 , \qquad \alpha \in\{ X,Y,Z \}.
\end{equation}\label{lem: homonuclear reduction}
\end{Lemma}

\begin{proof}[Proof of \Cref{lem: homonuclear reduction}]
See \Cref{sec: proof of homonuclear reduction theorem}.
\end{proof}

\begin{Theorem}
    \label{thm: homonuclear reduction}
    \normalfont
    For isotropic Hamiltonians with identical controls, the fourth-order Magnus expansion in \cref{eq: Theta_2 general with commutator} reduces to the form 
    \begin{equation}
        \Theta_2(t_n) = -\ii \vr(t_n)^\top\vcs  -\ii \frac{h}{2}\vcs^\top \jhomo \vcs \ , \label{eq: Theta_2 homo}
    \end{equation}
    where 
    \begin{align}
        \nonumber \vr^X(t_n) & = \left(\int^h_0  p_x(t_n+\zeta)  \mathrm{d}\zeta\right) \vone + \lb \int^h_0 (2\zeta - h)p_y(t_n+\zeta)\mathrm{d}\zeta\rb  \vo, \\
        \nonumber \vr^Y(t_n) & = \left(\int^h_0  p_y(t_n+\zeta)  \mathrm{d}\zeta\right) \vone - \lb \int^h_0 (2\zeta - h)p_x(t_n+\zeta)\mathrm{d}\zeta\rb  \vo, \\
        \vr^Z(t_n) & = \lb- \int^h_0\int^{\zeta}_0\lb p_x(t_n+\xi)p_y(t_n+\zeta)-p_x(t_n+\zeta)p_y(t_n+\xi)\rb\mathrm{d}\xi \mathrm{d}\zeta \rb \vone + h \vo.
        \label{eq: r reduced homo}
    \end{align}
\end{Theorem}

\begin{proof}[Proof of \Cref{thm: homonuclear reduction}]
    The integral $\vr$ is as defined in \cref{eq: magnus fourth order r} but assume a simpler form due to the identical controls in the Hamiltonian -- in particular, we only need four scalar integrals over an interval and one scalar integral over a triangular region. Similarly, $\vu(t_n)$, \cref{eq: extra commutator rotation gates weight}, simplifies to
    \begin{align} 
        \nonumber \vu^X(t_n) &= \lb \frac{1}{2} \int^h_0 (2\zeta - h)p_x(t_n+\zeta)\mathrm{d}\zeta \rb \vone, \\
        \nonumber  \vu^Y(t_n) &= \lb \frac{1}{2} \int^h_0 (2\zeta - h)p_y(t_n+\zeta)\mathrm{d}\zeta \rb  \vone, \\
        \vu^Z(t_n) &= \frac{1}{2} \int^h_0 (2\zeta - h)\vo\mathrm{d}\zeta  = 0.
        \label{eq: u reduced homo}
    \end{align}

    Due to $\vu^Z(t_n)=0$, the commutator \eqref{eq: extra commutator} becomes
    \begin{align}
        \nonumber &\frac{1}{2}\lb \int^h_0 (2\zeta - h)p_x(t_n+\zeta)\mathrm{d}\zeta \rb \lbb \vone^\top\vcs^X ,  \vcs^\top \jhomo \vcs \rbb \\
        & \qquad \qquad + \frac{1}{2} \lb \int^h_0 (2\zeta - h)p_y(t_n+\zeta)\mathrm{d}\zeta \rb \lbb  \vone^\top\vcs^Y , \vcs^\top \jhomo \vcs \rbb. \label{eq: cancelling out 1}
    \end{align}
    However, by \Cref{lem: homonuclear reduction} both commutators in \cref{eq: cancelling out 1} vanish.
\end{proof}

As a consequence of \Cref{thm: homonuclear reduction}, the fourth-order Magnus expansion for isotropic Hamiltonians, 
\cref{eq: Theta_2 homo}, has the structure \eqref{eq: common structure}. Consequently, time-stepping with the fourth-order Magnus expansion \eqref{eq: Theta_2 homo} combined with the fourth-order Yoshida splitting \eqref{eq: 4th splitting} is precisely as costly as time-stepping with a time-independent Hamiltonian using the fourth-order Yoshida splitting. When combined with a qubitization procedure, it is as expensive time-stepping with piecewise-constant Hamiltonian, while providing a fourth-order accuracy in contrast to the first-order accuracy of the latter. Moreover, we can utilize circuits with an identical structure.

\begin{Remark}
While the structure of the fourth-order Magnus expansion $\Theta_2(t_n)$, \cref{eq: Theta_2 homo}, is the same as the second-order Magnus expansion $\Theta_1(t_n)$, \cref{eq: theta_1 final}, it should be noted that the values of the singles-spin parameters are different -- while these are given by the integral $\vmu(t_n)$, \cref{eq: second-order Magnus parameters without new notation}, for the second-order Magnus expansion, the integrals $\vr(t_n)$, \cref{eq: magnus fourth order r} and \cref{eq: r reduced homo}, for the fourth-order Magnus expansion involve  additional terms. These additional terms are essential for the fourth-order accuracy.
\end{Remark}

\subsection{Other methods without commutators}
\label{sec: methods without commutators}

In \Cref{sec: commutator elimination}, we developed a novel technique for eliminating commutators appearing in the fourth-order Magnus method, while in \Cref{sec: homonuclear} we find that for Hamiltonians of certain types these commutators vanish by themselves. 
The appearance of commutators is a problem that occurs specifically in high-order Magnus-based methods for time-dependent Hamiltonians, which has been a hurdle in their adoption for Hamiltonian simulation using quantum circuits. 
However, these are not the only methods for time-dependent Hamiltonians that are able to avoid commutators. In this subsection we mention two notable alternatives that can handle arbitrary time-dependent (not just time-periodic) Hamiltonians without using commutators, and which lead to comparable circuits.

{\bf Commutator-free or quasi-Magnus methods.}
Commutator-free methods\cite{alvermann2011high}, also called quasi-Magnus methods \cite{blanes2017high}, are specifically designed to sidestep the need for computing commutators in the Magnus expansion. Specifically, they approximate the exponential of the Magnus expansion by a product of two or more exponentials, each of which is free of commutators but involves linear combinations of the Hamiltonian evaluated at different time knots or of its integrals. In particular, we will consider the CF42 integrator \cite{alvermann2011high}, 
\begin{equation}
    \label{eq: CF42 form}
    U(T, 0) \approx \prod_{n=0}^{N-1} U_n = \prod_{n=0}^{N-1} \exp\lb\frac12 \mathcal{A}_1(t_n) + \frac13 \mathcal{A}_2(t_n) \rb \exp\lb\frac12 \mathcal{A}_1(t_n) - \frac13 \mathcal{A}_2(t_n) \rb,
\end{equation}
where 
\begin{equation*}
\mathcal{A}_1(t_n) = \int^h_0 \mathcal{A}\lb t_n + \zeta\rb \mathrm{d} \zeta \, , \qquad \mathcal{A}_2(t_n) = 3 \int^h_0 \lb\frac{2\zeta}{h} - 1\rb\mathcal{A}\lb t_n + \zeta\rb  \mathrm{d} \zeta \, .
\end{equation*}
The prefix `CF4' in CF42 indicates its status as a fourth-order integrator, whereas the suffix `2' highlights an optimization in the integrator's design. 
It is notable that in this approach, despite the vanishing of nested commutator, the number of exponential stages doubles, which is undesirable in terms of the depth of the quantum circuit.

{\bf Autonomization. } A standard technique for converting a non-autonomous system of ordinary differential equations (ODEs), 
\[ x' = f(x, t), \qquad x(0)=x_0 \in \mathbb{R}^d,\]
to an autonomous system is to consider a coupled system of ODEs,
\begin{equation}
\label{eq:autonomous}
\begin{tabular}{r l r l}
    $x'$ &$= f(x, \tau),$  & $x(0)$&$=x_0 \in \mathbb{R}^d,$\\
    $\tau'$ &$= 1,$   & $\tau(0)$ &$= 0 \in \mathbb{R},$
\end{tabular}     
\end{equation}
where the derivative in $x'$ and $\tau'$ are with respect to $t$. The technique of introducing an `artificial' time variable $\tau$ to transform nonautonomous problems into autonomous problems is a well-established concept \cite{Wiggins1990,guckenheimer2013nonlinear}, which has been applied widely in classical algorithms for Hamiltonian simulation \cite{sanz1996classical}.
When considering 
\[f(x,t) = f_1(x,t) + f_2(x) + f_3(x, t), \]
$\ie$, where one of the components (here $f_2$) is not explicitly time-dependent, we can approximate the solution of \cref{eq:autonomous} by creating an augmented state $(x, \tau)$, and considering the Trotter splitting in terms of {\em flows} \cite{hairer2006geometric,mclachlan2002splitting,blanes2008splitting},
\begin{equation} 
\label{eq:split flow Trotter}
(x_{n+1}, \tau_{n+1}) \approx \Phi_h^{[3]} \circ \Phi_h^{[2]} \circ \Phi_h^{[1]}\  (x_n, \tau_n), 
\end{equation}
where $\Phi_h^{[j]}$ is the flow under $f_j$, for $j=1,2,3$. In particular, the computation of 
\[(x^{[1]}_{n+1}, \tau^{[1]}_{n+1}): = \Phi_h^{[1]} (x_n, \tau_n)\]
corresponds to freezing $\tau=t_n$ and solving 
\begin{equation} 
\label{eq:flow1} y' = f_1(y, \tau), \qquad y(0) = x_n, 
\end{equation}
exactly for time $h$ with initial condition  $x_n$, $\ie$, $x^{[1]}_{n+1}:= y(h)$ and $\tau^{[1]}_{n+1}=t_n$. The second step of \cref{eq:split flow Trotter} involves computing $(x^{[2]}_{n+1}, \tau^{[2]}_{n+1}) = \Phi_h^{[2]}(x^{[1]}_{n+1}, \tau^{[1]}_{n+1})$ by solving the system of equations,
\begin{equation}
\label{eq:flow2}
\begin{tabular}{r l r l }
    $y'$ &$= f_2(y),$  & $y(0)$ & $= x^{[1]}_{n+1},$ \\
    $s'$ &$= 1,$ & $s(0)$ &$= \tau^{[1]}_{n+1} = t_n,$ \\
\end{tabular}     
\end{equation}
for time $h$ with initial conditions  $(x^{[1]}_{n+1},t_n)$, $\ie$, $x^{[2]}_{n+1}:= y(h)$ and $\tau_{n+1}:=s(h)=t_{n+1}$. Lastly, we solve 
\begin{equation} 
\label{eq:flow3} y' = f_2(y, \tau), \qquad y(0) = x^{[2]}_{n+1}, 
\end{equation}
for time $h$ with initial condition $(x^{[2]}_{n+1}, t_{n+1})$ and with $\tau=t_{n+1}$ frozen, $\ie$, $x_{n+1} := y(h)$, completing one step of \cref{eq:split flow Trotter}. Note that \cref{eq:flow1,eq:flow2,eq:flow3} are all autonomous. 

When considering linear non-autonomous equations, with 
\[f_1(\ket{\psi},\tau) = -\ii \mathcal{H}^x(\tau) \ket{\psi},\qquad f_2(\ket{\psi}) = -\ii \mathcal{H}^z \ket{\psi}, \qquad f_3(\ket{\psi},\tau) = -\ii \mathcal{H}^y(\tau) \ket{\psi}, \]
the corresponding flows are
\begin{align*}
    &\Phi_{\alpha h}^{[1]}\, (\ket{\psi}, \tau) = \left(\ee^{-\ii \alpha h \mathcal{H}^x(\tau)} \ket{\psi} , \tau \right), \\
    &\Phi_{\alpha h}^{[2]}\, (\ket{\psi}, \tau) = \left(\ee^{-\ii \alpha h \mathcal{H}^z} \ket{\psi} , \tau + \alpha h\right), \\
    &\Phi_{\alpha h}^{[3]}\, (\ket{\psi}, \tau) = \left(\ee^{-\ii \alpha h \mathcal{H}^y(\tau)}  \ket{\psi} , \tau \right),
\end{align*}
$\ie$, the {\em internal} time variable $\tau$ only advances with the flow of $\mathcal{H}^z$, which is the autonomous component, and does so proportional to $\alpha h$ where $\alpha \in \mathbb{R}$ is a splitting coefficient.

Effectively, the autonomization approach involves applying a series of exponential operators in a time-sequential manner, each corresponding to the Hamiltonian sampled at specific times. For instance, the autonomization approach for the Yoshida 4th-order method in \cref{eq: 4th splitting} yields the splitting,
\begin{equation}
\label{eq: time-orderng}
\textrm{e}^{-\mathrm{i} \alpha_{13} \mathcal{H}^x(\tau_6)} \textrm{e}^{-\mathrm{i} \alpha_{12} \mathcal{H}^z} \textrm{e}^{-\mathrm{i} \alpha_{11} \mathcal{H}^y(\tau_5)} \cdots \textrm{e}^{-\mathrm{i} \alpha_3 \mathcal{H}^y(\tau_1)} \textrm{e}^{-\mathrm{i} \alpha_2 \mathcal{H}^z} \textrm{e}^{-\mathrm{i} \alpha_1 \mathcal{H}^x(\tau_0)},  
\end{equation}
where the internal time progression is defined by
\[
\tau_k = \tau_{k-1} + h \alpha_{2k}, \quad \tau_0 = t_n, \qquad k \in \{1,\cdots,6\},
\]
where $\alpha_k$ are the coefficients of the splitting in \cref{eq: 4th splitting}.

%% file: 4_trotterized_circuit.tex
\section{Trotterized quantum circuit construction}
\label{sec: splitting and circuit}

As we have seen in \Cref{sec: homonuclear}, in the case of isotropic Hamiltonians with identical controls, the fourth-order Magnus expansion does not have any commutators, and has the form \eqref{eq: common structure}. For general Hamiltonians, the commutator in the fourth-order Magnus expansion can be eliminated due to \Cref{thm:eliminating}, and each time-step involves the computation of $\ee^{-\eli(t_n)} \ee^{W(t_n)}\ee^{\eli(t_n)}$, where $\ee^{-\eli(t_n)}$ and $\ee^{\eli(t_n)}$ are single-spin terms and $W(t_n)$ has the form \eqref{eq: common structure}.
Thus, central exponential $\ee^{W(t_n)}$ in \cref{eq: Theta_2 general with eliminator} or the exponential of the fourth-order Magnus expansion for isotropic Hamiltonians  with identical controls, \cref{eq: Theta_2 homo}, can be implemented on the QPU by Trotterization, BCCKS \cite{BCCKS_berry2015simulating} or qubitization \cite{low2019hamiltonian}, as outlined in \Cref{sec: Trotterization,sec: qubitization}.

In this section, we discuss how to design a Trotterized quantum circuit to implement the time-stepping procedure in \cref{eq: timestep} or \cref{eq: propagator approximation} for the fourth-order Magnus expansions, \cref{eq: Theta_2 general with eliminator,eq: Theta_2 homo}. Along the lines of \Cref{sec: Trotterization}, in this section we assume the absence of mixed couplings,  $C^{\alpha,\beta}=0$ for $\alpha \neq \beta$.

We start by presenting a fourth-order Trotterized circuit for computing the exponential of a matrix of the form \cref{eq: common structure}. In the absence of mixed couplings, $\Omega = \Omega^X + \Omega^Y + \Omega^Z$,
scaled exponentials of each component, $\ee^{s \Omega^\alpha}$, can be computed to fourth-order accuracy using \cref{eq: single component of splitting,eq: 4th splitting}.
In \cref{eq: single component of splitting} we have assumed symmetric couplings, ${C^{\alpha,\alpha}}^\top = C^{\alpha,\alpha}$, which has allowed us to halve the exponents in the coupling term. A generalization of the circuit presented in this section to non-symmetric couplings \eqref{eq: single component of splitting general} is straightforward.

\begin{figure}[htbp]
\iftoggle{rotationgate}{
\centerline{
\Qcircuit @C=.5em @R=0.1em @!R {
 & & \cdots & \\
\lstick{k^{\mathrm{th}}  \mathrm{qubit}} & \qw& \gate{R_\alpha(\theta)} &\qw  & \qw& \\
 & & \cdots &}
}
}{}
\caption{$\alpha$ rotation gates implement the single-spin terms $\ee^{-\ii c \alpha_k}$, with $\theta=2c$. }
\label{fig: rotation gate}
\end{figure}
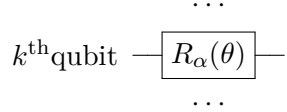

The exponential of a single component, $\ee^{s \Omega^\alpha}$ in \cref{eq: single component of splitting}, contains terms of the form $\ee^{-\ii c \alpha_k}$ and $\ee^{-\ii c \alpha_j\alpha_k}$, where $c$ are some real-valued scalars. The circuit implementation of these terms is hardware dependent. The single-spin term $\ee^{-\ii c \alpha_k}$ can be implemented in a quantum circuit as a  rotation or a Pauli-$\alpha$ gate, $\alpha\in\{X,Y,Z\}$, acting on the $k$-th qubit as shown in \Cref{fig: rotation gate}. Rotation gates are available on many quantum computing architectures as elementary single-qubit gates, but can always be created with the composition of a few single-qubit gates on all architectures. A term $\ee^{-\ii c \alpha_k}$ with scalar parameter $c$ is implemented by the rotation gate $R_\alpha(\theta)$ with $\theta = 2c$.

\begin{figure}[htbp]
\iftoggle{couplinggates}{
\centerline{
\Qcircuit @C=.5em @R=0.1em @!R {
\lstick{j^{\mathrm{th}}  \mathrm{qubit}} & \qw& \cgateRxx{2} &\qw &\qw
& &
\lstick{j^{\mathrm{th}}\ \mathrm{qubit}}&  \ctrl{2} & \gate{R_x(\theta)} & \ctrl{2} & \qw
\\
&\cdots & &\cdots &
&\push{\rule{.3em}{0em}=\rule{5em}{0em}} &
& &\cdots & &
\\
\lstick{k^{\mathrm{th}}  \mathrm{qubit}} & \qw& \gate{R_{xx}(\theta)} & \qw &\qw
& &
\lstick{k^{\mathrm{th}}\ \mathrm{qubit}}& \targ & \qw & \targ & \qw
}
}
\vspace{3em}
\centerline{\hspace{6.7em}
\Qcircuit @C=.5em @R=0.1em @!R {
\lstick{j^{\mathrm{th}}  \mathrm{qubit}} & \qw& \cgateRyy{2} &\qw &\qw
& &
\lstick{j^{\mathrm{th}}\ \mathrm{qubit}}& \qw& \ctrl{2} & \gate{R_x(\theta)} & \ctrl{2} & \qw & \qw
\\
&\cdots & &\cdots &
&\push{\rule{.3em}{0em}=\rule{5em}{0em}} &
&& &\cdots & & &
\\
\lstick{k^{\mathrm{th}}  \mathrm{qubit}} & \qw& \gate{R_{yy}(\theta)} & \qw &\qw
& &
\lstick{k^{\mathrm{th}}\ \mathrm{qubit}}&\gate{P(\frac{\pi}{2})}& \targ & \qw & \targ &\gate{P(\frac{\pi}{2})}& \qw
}
}
\vspace{3em}
    \centerline{
\Qcircuit @C=.5em @R=0.1em @!R {
\lstick{j^{\mathrm{th}}  \mathrm{qubit}} & \qw& \cgateRzz{2} &\qw &\qw
& &
\lstick{j^{\mathrm{th}}\ \mathrm{qubit}}&  \ctrl{2} & \qw & \ctrl{2} & \qw
\\
&\cdots & &\cdots &
&\push{\rule{.3em}{0em}=\rule{5em}{0em}} &
& &\cdots & &
\\
\lstick{k^{\mathrm{th}}  \mathrm{qubit}} & \qw& \gate{R_{zz}(\theta)} & \qw &\qw
& &
\lstick{k^{\mathrm{th}}\ \mathrm{qubit}}& \targ & \gate{R_z(\theta)} & \targ & \qw
}
}
}{}
\caption{$\alpha \alpha$ Ising coupling gates implement the exponential of two-spin couplings, $\ee^{-\ii c \alpha_j\alpha_k}$, with $\theta=2c$. }
\label{fig: coupling gate}
\end{figure}
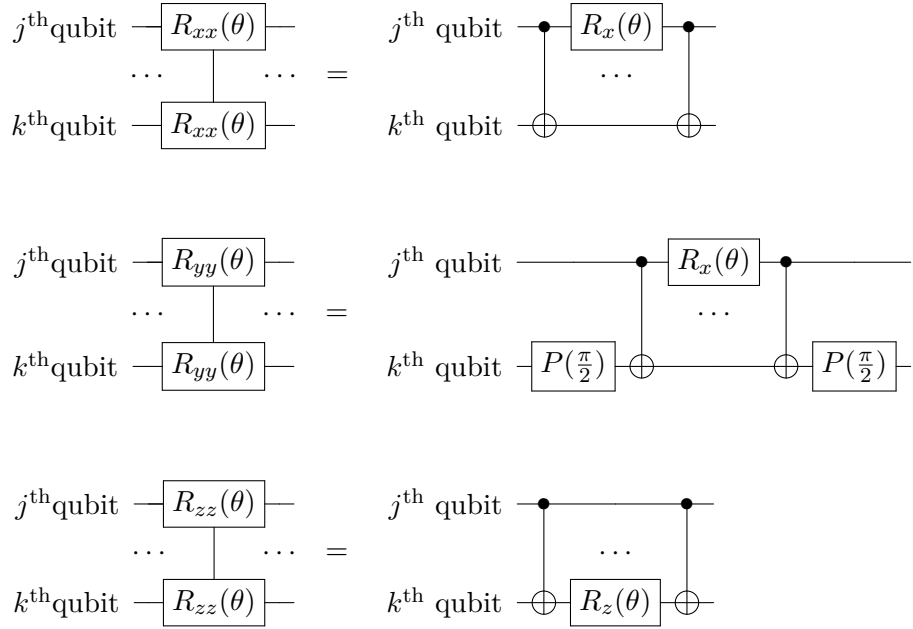

The pairwise coupling term $\ee^{-\ii c\alpha_j\alpha_k}$ can be implemented in a quantum circuit as an Ising coupling gate acting on two qubits, the $j$-th qubit and the $k$-th qubit, as shown on the left of \Cref{fig: coupling gate}. A term $\ee^{-\ii c \alpha_j\alpha_k}$ with parameter $c$ can be implemented by the Ising coupling gate $R_{\alpha\alpha}(\theta)$ with $\theta = 2c$. Ising coupling gates are less commonly available natively, $\ie$, as elementary gates. On such architectures, they can be implemented as a combination of elementary gates, for example as shown on the right of \Cref{fig: coupling gate}. Here we have assumed an architecture with full connectivity. In practise, many swap gates are required to overcome the limited connectivity which is more typical of many current quantum architectures. In the case that coupling interactions are not dense, $\ie$, each spin is coupled only to a few other spins, the optimal assignment of spins to physical qubits (in terms of swap gate counts) can depend heavily on the underlying connectivity as well as the sparsity pattern of the interaction tensor $\jg$. %

\begin{figure}[htbp]
\centering
\iftoggle{circuit}{
\begin{minipage}{0.9\textwidth}
\begin{subcolumns}[0.13\textwidth]
  \subfloat[]{\includegraphics[width=0.6\subcolumnwidth]{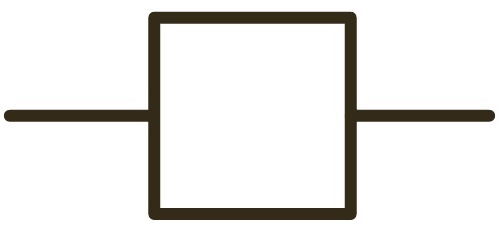}}\\
  \subfloat[]{\includegraphics[width=0.6\subcolumnwidth]{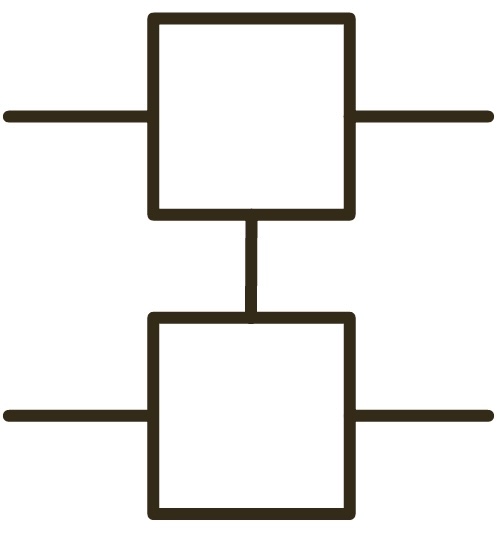}}
\nextsubcolumn[0.85\textwidth]
  \subfloat[]{\includegraphics[width=\subcolumnwidth]{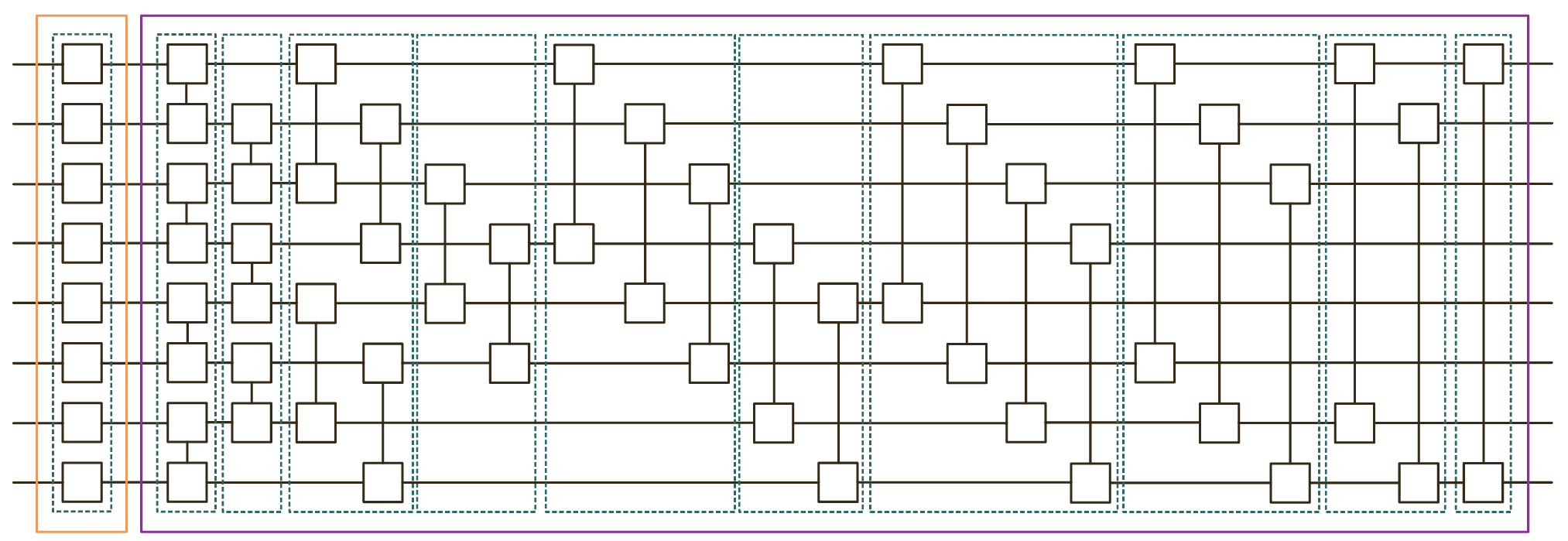}}
\end{subcolumns}
\end{minipage}
}{}
\caption{Block example for implementing \cref{eq: single component of splitting} for 8 qubits. (a) and (b) represent single rotation gates and coupling gates respectively. Gates circled in the orange box in (c) are the single-qubit rotation gates, and gates circled in the purple box in (c) are the two-qubit coupling gates. In each green box, the gates could be conducted parallelly, so each green box adds depth 1 to the circuit.}
\label{fig: trotter circuit}
\end{figure}

Overall, each component $\ee^{s \Omega^\alpha}$ in \cref{eq: single component of splitting} can be implemented by a block of single-qubit (\cref{fig: rotation gate}) and two-qubit (\cref{fig: coupling gate}) gates as depicted in the circuit in \cref{fig: trotter circuit} (c).

A fourth-order approximation of the exponential of \cref{eq: common structure} requires the fourth-order Yoshida splitting, \cref{eq: 4th splitting}.
As noted in \Cref{sec: matrix exponential}, time-stepping with \cref{eq: 4th splitting} involves (on average) half the number of $\ee^{hC}$ and $\ee^{hA}$ terms compared to $\ee^{hB}$ ($3$ vs $6$). Thus it is desirable to keep the splitting component $B$ the simplest in terms of gate and circuit complexity. The optimal choices varies with the qubit connectivity and elementary gates available in the quantum architecture, as well as the sparsity of spin couplings in the interaction tensor $\jg$. Given fully dense interactions and a fully connected architecture where the coupling gates are not available and need to be implemented as the compositions shown on the right side of \cref{fig: coupling gate}, for instance, the $YY$ Ising gate $R_{yy}$ is the most expensive in terms of elementary gates, while $R_{xx}$ and $R_{zz}$ are less expensive. In such a case a reasonable choice of components in the splitting \eqref{eq: 4th splitting} is
\begin{equation}
    hA = \Omega^X, \quad hB = \Omega^Z, \quad hC = \Omega^Y,
\end{equation}
resulting in the following Trotterization,
\begin{equation}
    \label{eq: yoshida for Omega}
    \begin{aligned}
    \Upsilon &:= \ee^{\frac{x}{2}\Omega^X}\, \Gamma \ \ee^{\frac{x}{2}\Omega^X} \quad \approx \ee^{\Omega},\\
    \Gamma & = \ee^{\frac{x}{2}\Omega^Z}\ee^{x\Omega^Y}\ee^{\frac{x}{2}\Omega^Z}\ee^{\frac{x+y}{2}\Omega^X}\ee^{\frac{y}{2}\Omega^Z}\ee^{y\Omega^Y}\ee^{\frac{y}{2}\Omega^Z}\ee^{\frac{x+y}{2}\Omega^X}\ee^{\frac{x}{2}\Omega^Z}\ee^{x\Omega^Y}\ee^{\frac{x}{2}\Omega^Z}.
    \end{aligned}
\end{equation}

For the isotropic Hamiltonians considered in \Cref{sec: homonuclear}, the Trotterization \eqref{eq: yoshida for Omega} suffices for exponentiating the fourth-order Magnus expansion, \cref{eq: Theta_2 homo} by setting $\Omega := \Theta_2(t_n) = -\ii \vr(t_n)^\top\vcs  -\ii \frac{h}{2}\vcs^\top \jhomo \vcs$. As mentioned previously, the circuit structure and gate count is identical to a fourth-order Trotterized circuit for a time-independent Hamiltonian.

For general Hamiltonians, where a single time-step involves computing $\ee^{-\eli(t_n)} \ee^{W(t_n)}\ee^{\eli(t_n)}$, we set $\Omega := W(t_n)= -\ii \tilde{\vr}(t_n)^\top\vcs  -\ii \frac{h}{2}  \vcs^\top \jg \vcs$, \cref{eq: Theta_2 hetero W}, for the central exponent.
Due to \Cref{thm:single spin}, the single-spin term $\ee^{\eli(t_n)}$ can be implemented as
\[\ee^{\eli(t_n)} = \ee^{\widetilde{\eli}(t_n)^x} \ee^{\widetilde{\eli}(t_n)^y} \ee^{\widetilde{\eli}(t_n)^z}, \]
using three layers of $M$ single-qubit rotation gates each, where the parameters of the rotation gates in \cref{eq:exact single spin} are computed by solving the non-linear system of equations \cref{eq: Trotter type E parameters}. $\ee^{-\eli(t_n)}$ is obtained from the expression for $\ee^{\eli(t_n)}$ by negating the sign and reversing the order of the exponentials.
Overall, each time-step of the fourth-order Magnus expansion for general Hamiltonians has the form
\begin{equation}
    \widetilde{\Upsilon} := \ee^{-\widetilde{\eli}^Z} \ee^{-\widetilde{\eli}^Y} \ee^{-\widetilde{\eli}^X + \frac{x}{2}\Omega^X}\, \Gamma\ \ee^{\frac{x}{2}\Omega^X + \widetilde{\eli}^X}\ee^{\widetilde{\eli}^Y} \ee^{\widetilde{\eli}^Z} \quad \approx \ee^{\Theta_2}.
\label{eq: Trotter type E}
\end{equation}
where $\Gamma$ is as defined in \cref{eq: yoshida for Omega}, and the $-\widetilde{\eli}^X$ and $\widetilde{\eli}^X$ terms are subsumed alongwith $\frac{x}{2}\Omega^X$, with which they commute due to \Cref{lemma: AA commutators}.

\subsection{Error estimates}

\begin{Theorem}
    Let $\widetilde{\Upsilon}(t_n)$ be the Magnus-based method described in \cref{eq: Trotter type E} for time-stepping over the interval $[t_n, t_{n+1}]$. There exists a constant $c$ such that
    \begin{equation}
        \label{eq: main error bound}
        \left\| U(T,0) - \prod_{n=1}^{N-1} \widetilde{\Upsilon}(t_n) \right\| \leq c T \left[ \rho^4 +  M \gamma \rho \left(M \gamma +  \rho^2 \right)  + \gamma \rho^3 \right] h^4 , \qquad h \rightarrow 0, \quad Nh = T,
    \end{equation}
    where $\gamma= \max_{t\in [0,T]} \|\ve'(t)\|_\infty$
    is related to the maximum frequency of the controls in the interval $[0,T]$,  and $\rho$ is the maximum spectral radius of the Hamiltonian over $[0,T]$, \cref{eq: hamiltonian spectral radius}.

\end{Theorem}
\begin{proof}
The proof follows in a straightforward way from the local error estimates of the splitting error \eqref{eq:splitting error}, the local error estimates of the commutator elimination error \eqref{eq: Theta_2 general with eliminator estimate}, and the global error estimates of the Magnus expansion \eqref{eq: propagator approximation}.

In particular, for the splitting error, \cref{eq:splitting error}, we use the fact that $\|\ha^\alpha(t) \| \leq \|\ha(t)\|$ for $\alpha \in \{X,Y,Z\}$ so that $\max \{\|A\|,\|B\|,\|C\|\} \leq \rho$, and its contribution to the global error is $CT\rho^4 h^4$. Note that since the equation \cref{eq: Schrodinger} is a linear equation, and the time-evolution operators and their approximations are unitary, the methods described in this paper are unconditionally stable. In fact, due to unitarity, the local errors accumulate linearly \cite{hochbruck2003magnus}.
\end{proof}

From \cref{eq: main error bound} we note that unless $\gamma$ is very large ($\ie$, the frequency of the controls increases excessively), we can assume that the largest term is the $\rho^4 h^4$ error due to Yoshida splitting, giving an upper bound
$c T (\rho h)^4$. Thus for accuracy $c T (\rho h)^4 \leq \varepsilon$, we need a time-step size
\[ h = \order\left(\frac{1}{\rho}\left( \frac{\varepsilon}{T} \right)^{1/4}\right)\]
and a total of
\begin{equation}
    \label{eq: time-step complexity of Trotterization}
    N = \order\left(\rho \left( \frac{T}{\varepsilon} \right)^{1/4}\right)
\end{equation}
time-steps of \cref{eq: Trotter type E}. The complexity of the number of time steps obtained here is multiplicative in terms of the accuracy and spectral radius, unlike the additive complexity of the qubitization procedure, \cref{eq:non-additive complexity}.

.

\begin{Remark}
    As before, in a time-stepping procedure, the first and last exponents can be combined. It should be noted that in \cref{eq: Trotter type E}, $\eli^\alpha$ stands for $\eli^\alpha(t_n)$, $\ie$, its value changes across time-steps. Thus, while $\ee^{-\frac12 \eli^Z(t_n)}$ can be combined with $\ee^{-\frac12 \eli^Z(t_{n+1})}$ from the next time-step, they do not cancel out.
\end{Remark}

\begin{Remark}
    Compared to the circuit for fourth-order Trotterized time-stepping with a time-independent Hamiltonian, the additional cost of each time-step for time-stepping with \cref{eq: Trotter type E}, on average, is $3M$ single-qubit gates due to the $3$ terms of the form $\ee^{s \eli^\alpha}$, and $M$ single-qubit gates and $\frac12 |C^{X,X}|$ two-qubit gates due to the $\Omega^X$ that can no longer be combined across consecutive time-steps. %
\end{Remark}

\subsection{Circuit complexity per time-step}

The overall time-stepping procedure \eqref{eq: propagator approximation} involves $N$ steps of the fourth-order Trotterization -- \cref{eq: Trotter type E} for general Hamiltonians and \cref{eq: yoshida for Omega} for isotropic Hamiltonians. We pre-compute the vectors of integrals --  $\vr(t_n), \vu(t_n) \in \CC^{3M}$, \cref{eq: magnus fourth order r,eq: extra commutator rotation gates weight}, for the general case and $\vr(t_n)\in \CC^{3M}$, \cref{eq: r reduced homo}, for the isotropic case -- for each time-step, $n = 1,\cdots,N$, on a classical computer, and input these into the quantum computer as the parameters of the various rotation and coupling gates, as outlined earlier in this subsection.

\newcommand{\jin}{C_{\mathrm{in}}}

\renewcommand{\arraystretch}{1.5}
\begin{table}[h]
\small
    \centering
    \begin{tabular}{|c|c|c|c|}
        \hline
        {\bf Method} & {\bf Order} & {\bf 1-qubit gates} & {\bf 2-qubit gates} \\
        \toprule
        \hline
        \multicolumn{4}{c}{{\bf (a) Time-stepping with $\Omega$, \cref{eq: common structure}. }} \\
        \multicolumn{4}{c}{{\em (i)} {\em Coupling gates available, $\Omega^\alpha$ assigned to $hA, hB, hC$ arbitrarily.}} \\
        \hline
        Trotter, \cref{eq:trotter omega} & 1 & $3 M N$ & $3 N |C_\mathrm{in}|$\\
        Strang, \cref{eq:strang omega} & 2 & $4 M N + M$ & $4 N |C_\mathrm{in}| + |C_\mathrm{in}|$\\
        Yoshida, \cref{eq: 4th splitting} & 4 & $12 M N + M$ & $12 N |C_\mathrm{in}| + |C_\mathrm{in}|$\\
        \hline
        \multicolumn{4}{c}{}\\
        \multicolumn{4}{c}{{\em (ii)} {\em  Coupling gates implemented as shown in \cref{fig: coupling gate}, $hA = \Omega^X, hB = \Omega^Z, hC = \Omega^Y$.}} \\
        \hline
        Trotter, \cref{eq:trotter omega} & 1 & $ \left(3 M + 5 |C_\mathrm{in}|\right)N$ & $6 |C_\mathrm{in}| N$\\
        Strang, \cref{eq:strang omega} & 2 & $\left(4 M + 6 |C_\mathrm{in}|\right) N + M +  |C_\mathrm{in}|$ & $8 |C_\mathrm{in}| N + 2 |C_\mathrm{in}|$\\
        Yoshida, \cref{eq: 4th splitting} & 4 & $\left(12 M + 18 |C_\mathrm{in}|\right) N + M + |C_\mathrm{in}|$ & $24 |C_\mathrm{in}| N + 2 |C_\mathrm{in}|$ \\
        \hline
        \multicolumn{4}{c}{}\\
        \multicolumn{4}{c}{{\bf (b) Time-stepping with {\normalfont $\ee^{\eli}\ee^{W}\ee^{-\eli}$}, \cref{eq: Theta_2 general with eliminator}. }}\\
        \multicolumn{4}{c}{{\em (i)} {\em Coupling gates available, $\Omega^\alpha$ assigned to $hA, hB, hC$ arbitrarily.} } \\
        \hline
        Modified Yoshida, \cref{eq: Trotter type E} & 4& $16 M N + M$ & $13 |C_\mathrm{in}| N$ \\
        \hline
        \multicolumn{4}{c}{}\\
        \multicolumn{4}{c}{{\em (ii)} {\em  Coupling gates implemented as shown in \cref{fig: coupling gate},  $hA = \Omega^X, hB = \Omega^Z, hC = \Omega^Y$. }} \\
        \hline
        Modified Yoshida, \cref{eq: Trotter type E}& 4 & $\left(16 M + 19 |C_\mathrm{in}|\right) N + M$ & $26 |C_\mathrm{in}| N$ \\
        \hline
    \end{tabular}\\
    \vspace{10pt}
    \caption{Gate counts for $N$ time-steps of Trotterized time-stepping for $M$ spins for: $\Omega$, \cref{eq: common structure}  {\em (top two sub-tables, (a) (i) \& (a) (ii))} and  {\normalfont $\ee^{\eli}\ee^{W}\ee^{-\eli}$}, \cref{eq: Theta_2 general with eliminator} {\em (bottom two sub-tables, (b) (i) \& (b) (ii))}.
    }
    \label{tbl: gate count}
\end{table}
\def\arraystretch{1}

\renewcommand{\arraystretch}{1.5}
\begin{table}[h]
\small
    \centering
    \begin{tabular}{|c|c|c|c|}
        \hline
        {\bf Method} & {\bf Order} & {\bf Circuit Depth, M odd} & {\bf Circuit Depth, M even} \\
        \toprule
        \hline
        \multicolumn{4}{c}{{\bf (a) Time-stepping with $\Omega$, \cref{eq: common structure}. }} \\
        \multicolumn{4}{c}{{\em (i)} {\em Coupling gates available, $\Omega^\alpha$ assigned to $hA, hB, hC$ arbitrarily.}} \\
        \hline
        Trotter, \cref{eq:trotter omega} & 1 & $\left(\frac{9 M}{2} - \frac{3}{2}\right) N$ & $\left(\frac{9 M}{2} - 3\right) N$\\
        Strang, \cref{eq:strang omega} & 2  & $\left(6 M - 2\right) N + \frac{3 M}{2}  - \frac{1}{2}$ & $\left(6 M - 4\right) N + \frac{3 M}{2} - 1$\\
        Yoshida, \cref{eq: 4th splitting} & 4 & $\left(18 M - 6\right) N + \frac{3 M}{2} - \frac{1}{2}$ & $\left(18 M - 12\right) N + \frac{3 M}{2} - 1$\\
        \hline
        \multicolumn{4}{c}{}\\
        \multicolumn{4}{c}{{\em (ii)} {\em  Coupling gates implemented as shown in \cref{fig: coupling gate}, $hA = \Omega^X, hB = \Omega^Z, hC = \Omega^Y$.}} \\
        \hline
        Trotter, \cref{eq:trotter omega} & 1  & $\left(\frac{33 M}{2} - \frac{27}{2}\right) N$ & $\left(\frac{33 M}{2} - 19\right) N$\\
        Strang, \cref{eq:strang omega} & 2 & $\left(21 M - 17\right) N + \frac{9 M}{2} - \frac{7}{2}$ & $\left(21 M - 24\right) N + \frac{9 M}{2}- 5$\\
        Yoshida, \cref{eq: 4th splitting} & 4  & $\left(63 M - 51\right) N + \frac{9 M}{2} - \frac{7}{2}$ & $\left(63 M - 72\right) N + \frac{9 M}{2}- 5$\\
        \hline
        \multicolumn{4}{c}{}\\
        \multicolumn{4}{c}{{\bf (b) Time-stepping with {\normalfont $\ee^{\eli}\ee^{W}\ee^{-\eli}$}, \cref{eq: Theta_2 general with eliminator}. }}\\
        \multicolumn{4}{c}{{\em (i)} {\em Coupling gates available, $\Omega^\alpha$ assigned to $hA, hB, hC$ arbitrarily.} } \\
        \hline
        Modified Yoshida, \cref{eq: Trotter type E} & 4 & $\left(\frac{39 M}{2} - \frac{7}{2}\right) N + 1$ & $\left(\frac{39 M}{2} - 10\right) N + 1$\\
        \hline
        \multicolumn{4}{c}{}\\
        \multicolumn{4}{c}{{\em (ii)} {\em  Coupling gates implemented as shown in \cref{fig: coupling gate},  $hA = \Omega^X, hB = \Omega^Z, hC = \Omega^Y$. }} \\
        \hline
        Modified Yoshida, \cref{eq: Trotter type E}& 4  & $\left(\frac{135 M}{2} - \frac{103}{2}\right) N + 1$ & $\left(\frac{135 M}{2} - 74\right) N + 1$\\
        \hline
    \end{tabular}\\
    \vspace{10pt}
    \caption{Circuit depths for $N$ time-steps of Trotterized time-stepping for $M$ spins for: $\Omega$, \cref{eq: common structure}  {\em (top two sub-tables, (a) (i) \& (a) (ii))} and  {\normalfont $\ee^{\eli}\ee^{W}\ee^{-\eli}$}, \cref{eq: Theta_2 general with eliminator} {\em (bottom two sub-tables, (b) (i) \& (b) (ii))}. Here we have assumed that $|\jg|=9M(M-1)$, $\ie$, the spins are fully connected, $C^{\alpha,\beta} = 0, \alpha \neq \beta$, and ${C^{\alpha,\alpha}}^\top = C^{\alpha,\alpha}$. For sparse couplings, $\eg$, $|\jg|=\order \lb M \rb$,
    the circuit depth can depend strongly on the connectivity. For spin chains, for example, circuit depth is independent of $M$, while for a star topology, where one {\em central} spin is connected to all other spins, circuit depth remains $\order \lb M \rb$.
    }
    \label{tbl: circuit depth}
\end{table}
\def\arraystretch{1}

The overall gate count and circuit depth required for $N$ time steps of different Trotterized time-stepping methods are enumerated in \Cref{tbl: gate count,tbl: circuit depth}, respectively. For a time-independent Hamiltonian, $\ha_0$, a Trotterized time-stepping procedure can involve a first, second, or fourth-order method, depending on the accuracy required. For time-dependent Hamiltonians, a piecewise-constant Hamiltonian, $\ha(t_n)$, constitutes a first-order approximation, and should be combined with the first-order method -- the Trotter splitting, \cref{eq:trotter omega}. Similarly, the second-order Magnus expansion, $\Theta_2(t_n)$ in \cref{eq: theta_1 final}, constitutes a second-order approximation, and should be combined with the second-order method -- the Strang splitting, \cref{eq:strang omega}. Since the exponent in all of these cases has the common form $\Omega$, \cref{eq: common structure}, the relevant subtables for these cases are (a)(i) and (a)(ii), depending on whether coupling gates are natively available or not, respectively.

For fourth-order Magnus expansion $\Theta_2(t_n)$, as we have seen in \Cref{sec: commutator elimination,sec: homonuclear}, the situation differs depending on whether the Hamiltonian is isotropic or not. For isotropic Hamiltonians, $\Theta_2(t_n)$ in \cref{eq: Theta_2 homo} has the common form $\Omega$, \cref{eq: common structure}, and the relevant subtables are again (a)(i) and (a)(ii). For more general cases, to eliminate the extra commutator, we need the asymmetric splitting $\ee^{\eli(t_n)}\ee^{W(t_n)}\ee^{-\eli(t_n)}$, \cref{eq: Theta_2 general with eliminator}, which is implemented as the modified Yoshida splitting \cref{eq: Trotter type E}, with $\Omega = W(t_n)$. The relevant subtables in this case are (b)(i) and (b)(ii).

\subsection{Overall circuit complexity.}
\label{sec: overall complexity}
For simulating Hamiltonians \eqref{eq: full Hamiltonian} with fully-dense interactions, $\ie$ where each spin interacts with every other spin, we can see from \cref{eq: Hamiltonian norm time dependent} that the Hamiltonian spectral radius grows quadratically with $M$, $\ie$, $\rho = \order(M^2)$. Consequently, using the time-step complexity estimate in \cref{eq: time-step complexity of Trotterization}, the number of time-steps required for achieving a prescribed accuracy also grow quadratically as $N = \order(M^2)$. Referring to the circuit depth analysis given in \Cref{tbl: circuit depth}, the Yoshida method yields a circuit depth of $\order(MN)$. Overall, for achieving a prescribed accuracy for fully-dense interactions in an $M$ body two-level system, the
circuit depth needs to increase as $\order(M^3)$ and quantum volume as $\order(M^4)$.

For sparse Hamiltonians where each spin interacts with a fixed number of spins ($\eg$, spin chains, Ising models, and Kitaev models), the Hamiltonian spectral radius only grows linearly, $\rho=\order(M)$, so that the circuit depth grows as $\order(M)$ and quantum volume as $\order(M^2)$. This makes simulation of sparse Hamiltonians a significantly easier problem, especially for large $M$, where any quantum advantage is expected to be demonstrated.

%% file: 5_results.tex
\section{Numerical results}
\label{sec: numerical results}

\subsection{Chirped pulse}\label{sec: Chirped pulse}
Chirped pulses belong to a broader group of swept-frequency pulses and are parametric pulses in which the frequency of irradiation varies over time, typically in a linear fashion. They are frequently used in the control of spin systems, superconducting qubits \cite{Setiawan2023PhysRevApplied}, NMR \cite{BOHLEN1990183,FOROOZANDEH2020106768}, and atomic as well as molecular processes \cite{AmstrupChirped}.
The broadband motion of the pulse allows efficient control of spins with a wide range of frequencies. However, the highly oscillatory nature of these pulses makes them especially challenging to handle numerically.

A general expression for a chirped pulse is
\begin{equation}
p(t)=p_{x}(t)+\mathrm{i} p_{y}(t) = A(t) \exp (\mathrm{i} \phi(t)),
\label{eq: pxpy start}
\end{equation}
where the amplitude envelope $A(t)$ and the phase $\phi(t)$ are arbitrary real-valued functions of time. For a chirped pulse with peak amplitude $A_{\max}$, bandwidth $\Delta F$, duration $\tau_{p}$, and overall phase $\phi_{0}$, the amplitude envelope is given by
\begin{equation}
\label{eq:eq3}
A(t)=A_{\max} \exp \left[-4\left(\frac{2 t}{\tau_p}-1\right)^\eta\right]  , \qquad \eta \in 2\mathbb{N}_0,\qquad t\in [0,\tau_p],
\end{equation}
which is a {\em super-Gaussian} envelope. If the frequency of the pulse is swept linearly
\begin{equation}
\omega(t)=\pi \Delta F\left(\frac{2 t}{\tau_p}-1\right),
\label{eq: frequency}
\end{equation}
the phase is
\begin{equation}
\phi(t)=\int \omega(t) d t=\phi_0+\pi \Delta F t\left(\frac{t}{\tau_p}-1\right).
\end{equation}
The maximum amplitude of a chirped pulse can be calculated using the three parameters $\Delta F$, $\tau_p$, and $\mathcal{Q}_0$ as
\begin{equation}
\label{eq:rfqchirp}
A_{\max}=\sqrt{\frac{2 \pi \Delta F \mathcal{Q}_0 }{ \tau_p}} \ ,
\end{equation}
where $\mathcal{Q}_0$ is the {\em adiabaticity} factor at time $\nicefrac{\tau_\text{p}}{2}$, where $t \in [0, \tau_p]$,
\begin{equation}
\mathcal{Q}_0=\frac{2}{\pi}\ln{\left(\frac{2}{\cos{(\theta)}+1}\right)},
\label{eq: pxpy end}
\end{equation}
for a desired {\em pulse flip angle} $\theta$. Since the effective flip angle of a chirped pulse approaches $180^{\circ}$ asymptotically as $A_{\max}$ increases, for most practical purposes a value of $\mathcal{Q}_0$ is chosen (normally 5) to satisfy the adiabatic condition while avoiding excessive pulse amplitudes.

\subsection{Spin systems}

\begin{Example}[isotropic coupling with identical pulses]
    \normalfont
    \label{ex:isotropic homonuclear}

    We consider a three-spin-$\nicefrac{1}{2}$ isotropic system appearing in liquid-state NMR applications, where the pulse is identical for each spin,

    \begin{align}
     \nonumber \ha(t) &=\frac{1}{2} p_x(t) \lb\XYZ{X}{1} + \XYZ{X}{2} + \XYZ{X}{3}\rb +  \frac{1}{2}p_y(t) \lb\XYZ{Y}{1} + \XYZ{Y}{2} + \XYZ{Y}{3}\rb + \pi \Omega_1 \XYZ{Z}{1} + \pi\Omega_2 \XYZ{Z}{2} + \pi\Omega_3 \XYZ{Z}{3}    \\
     \label{eq: isotropic homonuclear example} &
     + \frac{\pi}{2}  \sum_{\alpha \in \{X,Y,Z\}} J_{1,2}\,  \alpha_1 \, \alpha_2 + \frac{\pi}{2}  \sum_{\alpha \in \{X,Y,Z\}} J_{2,3}\,  \alpha_2 \, \alpha_3,+ \frac{\pi}{2}  \sum_{\alpha \in \{X,Y,Z\}} J_{1,3}\,  \alpha_1 \, \alpha_3
     ,
    \end{align}
    where $\Omega_{1}$, $\Omega_{2}$, and $\Omega_{3}$ are the {\em spin offsets}
    .
    This Hamiltonian can be described in the form \eqref{eq: full Hamiltonian}, with
    \[\ve(t) = \frac{1}{2}\left(p_x(t)\vone^\top , p_y(t) \vone^\top , 2\pi\vo ^\top \right)^\top, \quad \vo = (\Omega_1, \Omega_2, \Omega_3)^\top, \]
    and
    \[C^{\alpha,\beta}=0\ \ \text{for}\ \ \alpha \neq \beta, \qquad C^{X,X} = C^{Y,Y} = C^{Z,Z} = C = \frac{\pi}{2} \begin{bmatrix}
        0 & J_{1,2} & J_{1,3} \\
        J_{1,2} & 0 & J_{2,3} \\
        J_{1,3} & J_{2,3} & 0 \\
    \end{bmatrix}.\]

    The three-spin system we consider in this example has the offsets $\Omega_{1}=2$ kHz, $\Omega_{2}=1.5$ kHz, and $\Omega_{3}=1.6$ kHz, and the coupling constants are defined by $J_{1,2}=7$ Hz, $J_{1,3}=12 $ Hz, and $J_{2,3}=20 $ Hz.
    We consider three different choices for the set of parameters defining the pulses $ p_x(t)$ and $ p_y(t)$:
    \begin{enumerate}[label=(\roman*)]
        \item\label{case: indep} {\bf Time-independent Hamiltonian.}
        $ p_x(t)= p_y(t)=0$, which corresponds to $ A_{\text{max}}=0$.

        \item\label{case: Gaussian} {\bf Gaussian pulse.}
        A general expression of the Gaussian pulse is given by
        \begin{equation}
        \label{eq: Gaussian pulse}
            p(t)=p_{x}(t)+\mathrm{i} p_{y}(t) =
            \theta
            \frac{\exp \left[-4\left(\frac{2 t}{\tau_p}-1\right)^\eta\right]}
            {\stretchint{4ex}_{\!\!\! 0}^{\tau_p} \exp \left[-4\left(\frac{2 t}{\tau_p}-1\right)^\eta\right]  \dd t}
            \exp \left[\mathrm{i} \left(\phi_0+ 2 \omega t\right)\right].
        \end{equation}
        Gaussian pulses can be selective in nature -- by tuning $\omega = \Omega$, the pulse in \cref{eq: Gaussian pulse} can control spins with offset $\Omega$, without affecting other spins. This can also be a limitation when we seek to control multiple spins with varying offsets, where the pulse may only influence one specific spin.
        In this section we consider a Gaussian pulse with $\theta = \pi$, $\eta = 2$, $\omega = \Omega_1 = 2000$ Hz, $\phi_0=0$ rad, and $\tau_p = 10$ ms. The real and imaginary components of the pulse, $ p_x(t)$ and $ p_y(t)$ are shown in \cref{fig: Gaussian pulse}.

        \begin{figure}[htbp]
        \centering
        \iftoggle{Gaussian_pulse}{
            \resizebox{\textwidth}{!}
            {\input{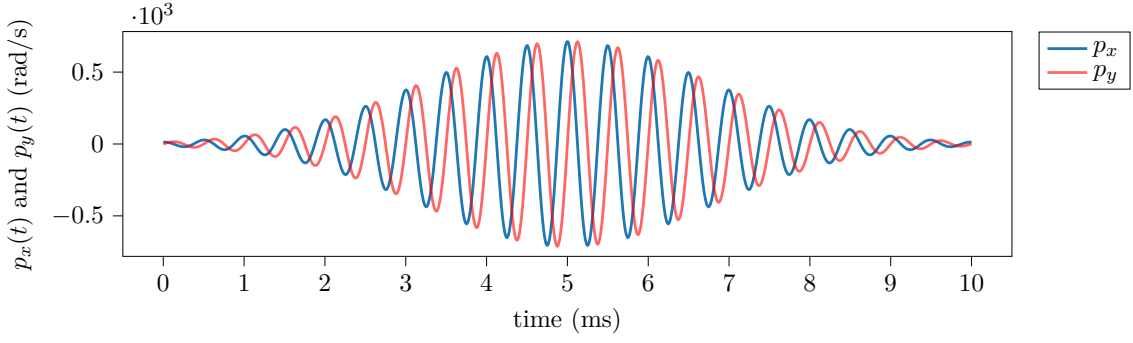}}
            \caption{An exemplar Gaussian pulse, \cref{eq: Gaussian pulse}, with $\tau_p=10$ ms, $\phi_0=0$ rad, and $\eta=2$, $\omega = \Omega_1$.
            }
            }{}
            \label{fig: Gaussian pulse}
        \end{figure}

        \item\label{case: chirped} {\bf Chirped pulse.}
        $ A_{\text{max}}= 2\pi \times 1545$ rad/s, $\Delta F=30$ kHz, $\tau_p=10$ ms, and $\eta=40$. The real and imaginary components of the pulse, $ p_x(t)$ and $ p_y(t)$, are shown in \cref{fig: chirped pulse}.

        \begin{figure}[htbp]
        \centering
        \iftoggle{chirped_pulse}{
            \resizebox{\textwidth}{!}
            {\input{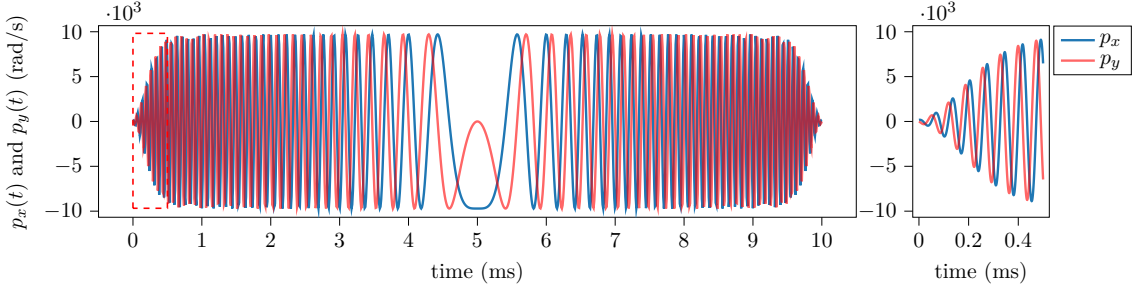}}
            \caption{An exemplar chirp pulse, \crefrange{eq: pxpy start}{eq: pxpy end}, with $\Delta F=30$ kHz, $\tau_p=10$ ms, $\phi_0=0$ rad, and $\eta=40$. On the right, we show a zoomed view of the highly oscillatory part of the pulse, from $0$ ms to $0.5$ ms.
            }
            }{}
            \label{fig: chirped pulse}
        \end{figure}
\end{enumerate}

\end{Example}

\begin{figure}[htbp]
    \centering
    \iftoggle{observables}{
        \resizebox{\textwidth}{!}{\input{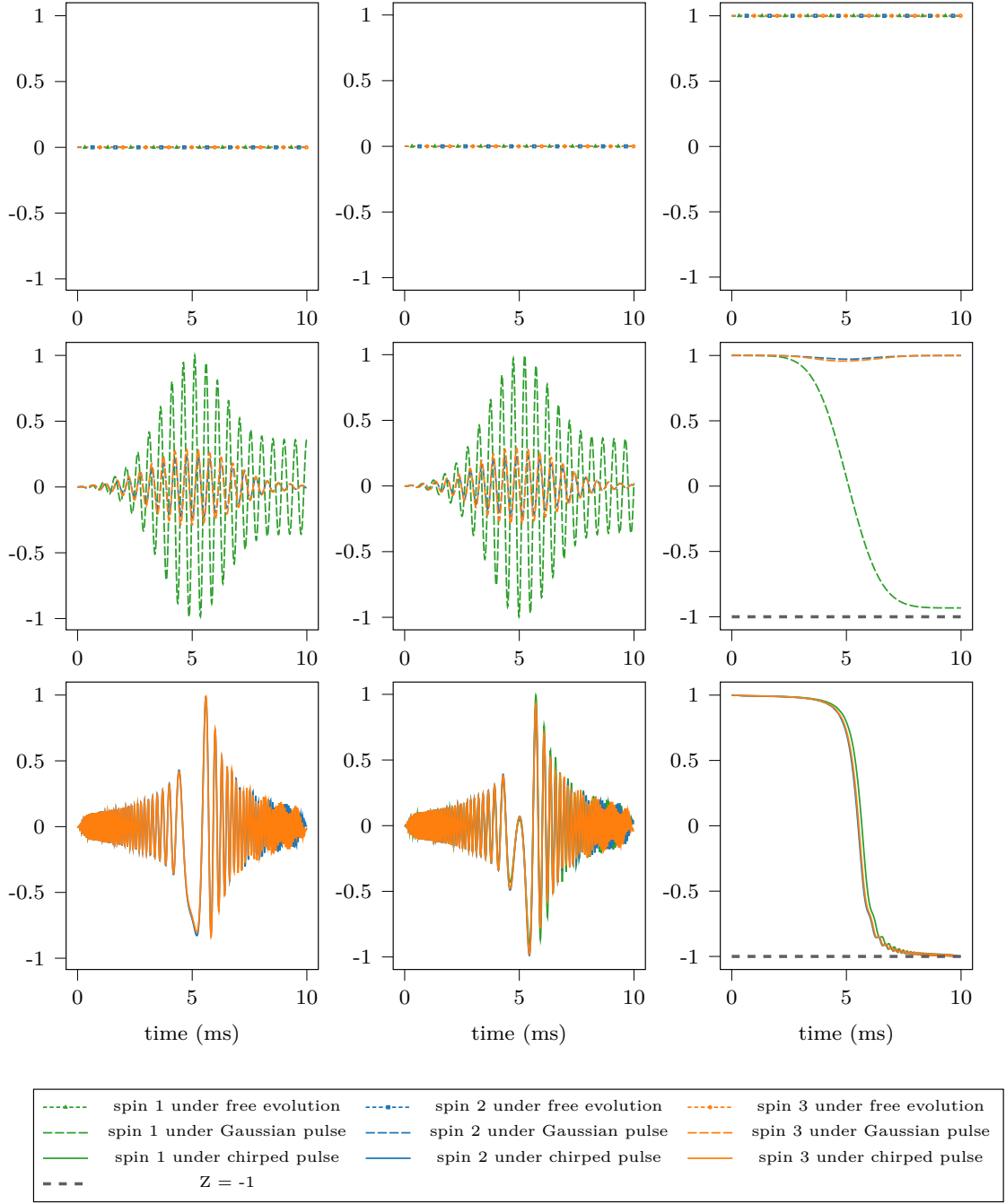}}
        }{}
\caption{The time evolution of the X, Y and Z component observables, respectively, for a system of 3 interacting spins described in \Cref{ex:isotropic homonuclear}, each initially in the Z-up direction. Three choices of pulses are considered:  Free evolution -- $\ie$, under \Cref{ex:isotropic homonuclear}, pulse \ref{case: indep}; a Gaussian pulse -- \Cref{ex:isotropic homonuclear}, pulse \ref{case: Gaussian}; a chirped pulse -- \Cref{ex:isotropic homonuclear}, pulse \ref{case: chirped}.}
\label{fig: observables}
\end{figure}

In \cref{fig: observables}, we observe three interacting spins under the three different pulses defined in \Cref{ex:isotropic homonuclear}. All three spins start from the Z-up position. The figure shows that the behaviour of the spins under a time-independent pulse is trivial. The Gaussian pulse can only flip one spin, specifically the one it is tuned for. However, it fails to completely drive the spin to the Z-down position -- this failure is due to a non-trivial interaction between the spins. The other two spins are minimally impacted and return to their original position eventually. In contrast, a chirped pulse is capable of flipping all three spins simultaneously -- doing so despite the interactions that are not explicitly taken into account in its design. While a sequence of three Gaussian pulses could theoretically achieve a similar result, it would require a specifically tuned pulse for each spin, significantly increasing the required time as the number of spins with different offsets grows. Moreover, the efficacy is expected to be limited in the presence of coupling terms.

On the other hand, compared to the Gaussian pulse shown in \cref{fig: Gaussian pulse}, which has a longer wavelength, the chirped pulse shown in \cref{fig: chirped pulse} is much more oscillatory, making it more challenging to resolve numerically. A similar phenomenon is observed in the dynamics of spins, which is much more oscillatory under a chirped pulse compared to the case of a Gaussian pulse. This is evident, for instance, in \cref{fig: observables}, where the $X$ and $Y$ observables are particularly oscillatory under the chirped pulse.

\begin{Example}[General Hamiltonian]
    \normalfont
    \label{ex:general}
    We consider a more general three-spin system
    with individual control pulses for each spin and non-isotropic couplings.
    The parameters of the Hamiltonian in \eqref{eq: full Hamiltonian} are given by
    \[\ve(t) =\frac{1}{2}\left(( p_{x,1}(t),  p_{x,2}(t),  p_{x,3}(t))^\top , ( p_{y,1}(t),  p_{y,2}(t),  p_{y,3}(t))^\top , 2 \pi \vo ^\top \right)^\top, \]
    where
    $p_{x,k},p_{y,k}$ are $p_{x},p_{y}$ defined in \crefrange{eq: pxpy start}{eq: pxpy end} with parameters $\{{\phi_{0}}, \Delta F\}$ replaced by
    $\{{\phi_{0,k}}, \Delta F_k\}$,
    where ${\phi_{0,1}} = \pi$, $\Delta F_1 = 30$ kHz, ${\phi_{0,2}} = \pi/6$, $\Delta F_2 = 15$ kHz, and ${\phi_{0,3}} = -\pi/6$, $\Delta F_3 = 45$ kHz. $\vo$ remains the same as in \Cref{ex:isotropic homonuclear}. We consider two cases for the couplings:
    \begin{enumerate}[label=(\roman*)]
        \item\label{case: no mixed}{\bf Without mixed couplings.}
        \[
        C^{\alpha,\beta}=0\ \ \text{for}\ \ \alpha \neq \beta, \qquad
        C^{\alpha,\alpha} = \frac{\pi}{2} \begin{bmatrix}
        0 & J^{\alpha,\alpha}_{1,2} & J^{\alpha,\alpha}_{1,3} \\
        J^{\alpha,\alpha}_{1,2} & 0 & J^{\alpha,\alpha}_{2,3} \\
        J^{\alpha,\alpha}_{1,3} & J^{\alpha,\alpha}_{2,3} & 0 \\
    \end{bmatrix},\quad \alpha\in\{X,Y,Z\},
        \]
        where the values of $J^{\alpha,\beta}_{j,k}$ (in Hz) for $\alpha,\beta \in \{X,Y,Z\}$ and $j,k \in \{1,2,3\}$ can be found in \Cref{table: no mixed coupling} in row labeled $j,k$ and column labeled $\alpha,\beta$.
\begin{table}[H]
\centering
\caption{Coupling constants in Hz, \Cref{ex:general}, case \ref{case: no mixed}}
\begin{tabular}{|c|c|c|c|}
\hline
\diagbox[width=5em,height=3em]{$j,k$}{$\alpha,\beta$}
 & $X,X$ & $Y,Y$ & $Z,Z$ \\
\hline
$1,2$ & $10$ & $5$ & $12$ \\
\hline
$1,3$ & $2$ & $8$ & $4$ \\
\hline
$2,3$ & $4$ & $9$ & $11$ \\
\hline
\end{tabular}
\label{table: no mixed coupling}
\end{table}

        \item\label{case: mixed}{\bf With mixed coupling.}
        \[
        C^{\alpha,\beta} = \frac{\pi}{2} \begin{bmatrix}
        0 & J^{\alpha,\beta}_{1,2} & J^{\alpha,\beta}_{1,3} \\
        J^{\alpha,\beta}_{1,2} & 0 & J^{\alpha,\beta}_{2,3} \\
        J^{\alpha,\beta}_{1,3} & J^{\alpha,\beta}_{2,3} & 0 \\
    \end{bmatrix},\quad \alpha,\beta \in\{X,Y,Z\},
        \]
        where the values of $J^{\alpha,\beta}_{j,k}$ (in Hz) for $\alpha,\beta\in \{X,Y,Z\}$ and $j,k \in \{1,2,3\}$ can be found in \Cref{table: mixed coupling} in row labeled $j,k$ and column labeled $\alpha,\beta$.

\begin{table}[H]
\centering
\caption{Coupling constants in Hz, \Cref{ex:general}, case \ref{case: mixed}}
\begin{tabular}{|c|c|c|c|c|c|c|c|c|c|}
\hline
\diagbox[width=5em,height=3em]{$j,k$}{$\alpha,\beta$} & $X,X$ & $Y,Y$ & $Z,Z$ & $X,Y$ & $X,Z$ & $Y,Z$ & $Y,X$ & $Z,X$ & $Z,Y$ \\
\hline
$1,2$ & $10$ & $5$ & $12$ & $13$ & $18$ & $4$ & $15$ & $15$ & $6$ \\
\hline
$1,3$ & $2$ & $8$ & $4$ & $19$ & $10$ & $14$ & $9$ & $13$ & $9$ \\
\hline
$2,3$ & $4$ & $9$ & $11$ & $2$ & $11$ & $10$ & $6$ & $4$ & $5$ \\
\hline
\end{tabular}
\label{table: mixed coupling}
\end{table}

    \end{enumerate}
\end{Example}

\begin{Remark}
In all cases we restrict out attention to
\[
\lb J^{\alpha, \beta} \rb ^\top = J^{\alpha, \beta}, \qquad \ie, \qquad \lb C^{\alpha, \beta} \rb ^\top = C^{\alpha, \beta},
\]
since this is typically the case in practical applications. We note, however, that our derivations do not require these assumptions, and our algorithm also works for cases with $\lb C^{\alpha, \beta} \rb ^\top \neq C^{\alpha, \beta}$.
\end{Remark}

\subsection{Results}

{\bf Naming and other conventions}

{\em Splittings.} In this section, for the case of time-dependent Hamiltonians, we will be pairing  Magnus expansions with various quadrature approximations for the integrals and appropriate Trotterization methods. For clarity, unless otherwise specified, first-order methods will be paired with Trotter splitting, \cref{eq:trotter omega}, second-order methods with Strang splitting, \cref{eq:strang omega}, and fourth-order methods with Yoshida splitting, \cref{eq: yoshida for Omega}.

{\em Integrals.} Our standard approach involves using \texttt{scipy} to approximate integrals in \cref{eq: extra commutator rotation gates weight,eq: magnus fourth order r} in the Magnus expansion as well as the commutator-free method CF42, \cref{eq: CF42 form}, to machine precision. However, for $\Theta_1$, \cref{eq: theta_1 final}, which is normally a second-order Magnus expansion, we will also explore using the starting point, $\ie$, $\Theta_1 \approx -\ii h \ha(t_n)$, which leads to a first-order method, as well as the midpoint $\Theta_1 \approx -\ii h \ha(t_n + h/2)$, which leads to a second-order method. For $\Theta_2$, \cref{eq: Theta_2 homo}, we will also examine the effects of using Gauss--Legendre (GL) quadratures, which are described in \Cref{sec:quadrature}. In particular, GL$k$ will refer to the Gauss--Legendre quadrature with $k$ knots.

{\em Naming conventions.} All Magnus expasion-based methods are named in the format
\[\Theta_k(t_n),\text{ eq. (n), {\em quadrature-method}}\]
for $k\in \{1,2\}$, where `eq. (n)' is the equation number for the Magnus expansion (and occassionally for the concrete form, when combined with an appropriate splitting),  and `{\em quadrature-method}' refers to the approach for approximating the integrals in the Magnus expansion. As described earlier in this subsection, the splitting to be used is governed by the order. The commutator-free method CF42 \eqref{eq: CF42 form} follows a similar convention, with $\Theta_k(t_n)$ replaced by CF42, while the autonomized Yoshida splitting~\eqref{eq: time-orderng} is written without denoting a quadrature method since it does not feature integrals.

{\em Error.} Throughout this manuscript,
the numerical simulations of the quantum circuits are carried out on a classical computer and
the error depicted is the $\ell^2$ error in the propagator, as defined, for example in \cref{eq: propagator approximation}.

{\bf Time-independent pulse, isotropic Hamiltonian} (\Cref{ex:isotropic homonuclear}, pulse \ref{case: indep})

    \begin{figure}[htbp]
    \centering
    \iftoggle{indep_result}{
        \resizebox{\textwidth}{!}{\input{fig_indep_result}}
        }{}
        \caption{{\em (Time-independent isotropic  Hamiltonian; splitting methods)} The accuracy of Hamiltonian simulation for isotropic Hamiltonian \eqref{eq: isotropic homonuclear example} under free evolution, \ie, \Cref{ex:isotropic homonuclear} with pulse \ref{case: indep}.}
        \label{fig: accuracy time independent}
    \end{figure}
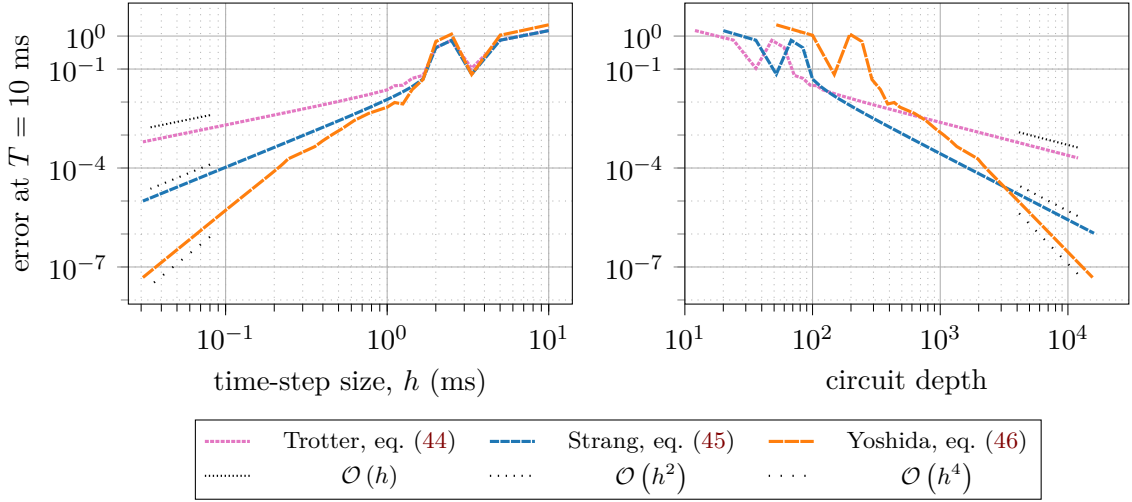

    \Cref{fig: accuracy time independent} shows the time-step size and circuit depth required for achieving a prescribed accuracy for the case of the time-independent Hamiltonian described in \Cref{ex:isotropic homonuclear}, $\ie$, for pulse \ref{case: indep}: $ p_x(t)= p_y(t)=0$. Higher accuracies require either smaller time-steps, which correspond to longer circuits, or the use of higher-order splittings.

    Even for a low accuracy of $10^{-2}$, the second-order Strang splitting, \cref{eq:strang omega}, requires a shallower circuit, with circuit depth $180$ and quantum volume $540$, than the first-order Trotter splitting, whose circuit depth is around $252$ ($\ie,$ a $40\%$ longer circuit) and quantum volume around $756$.
    Strang splitting stays competitive till an accuracy of $10^{-5}$, at which point Yoshida splitting, \cref{eq: 4th splitting}, becomes a more efficient approach.
    For five digits of accuracy, Yoshida splitting requires a circuit depth of $4,036$ and quantum volume of $12,108$, as compared to the circuit depth of $5,188$ required for the Strang splitting. These are roughly $7$ and $28$ times longer than those required for an accuracy of $10^{-2}$ using Yoshida (with circuit depth requirement of $532$) and Strang, respectively.

    {\em Relevance to practical circuits.} These observations suggest that for simulation of time-independent Hamiltonians, second-order methods may lead to shorter circuits compared to first-order methods, even when relatively low accuracies are sought, for example on near-term quantum devices. Higher-order methods are not expected to be useful for the time-independent Hamiltonians considered in this paper until accuracies at the level of $10^{-5}$ are desired and feasible in terms of circuit depths as well as circuit noise levels, which may require fault-tolerant quantum computing. For $M\gg 3$ spin systems, where any quantum advantage is expected to be demonstrated, the error \eqref{eq: time-step complexity of Trotterization} is expected to grow linearly with the number of spins for sparse Hamiltonians and quadratically for Hamiltonians with dense couplings \cref{eq: Hamiltonian norm time dependent}. However, since these error bounds are not tight, precise error levels for large $M$ are hard to estimate.

{\bf Chirped pulse, isotropic Hamiltonian } (\Cref{ex:isotropic homonuclear}, pulse \ref{case: chirped})

    \begin{figure}[htbp]
    \centering
    \iftoggle{chirped_result}{
        \resizebox{\textwidth}{!}{\input{fig_chirped_result}}
        }{}
        \caption{{\em (Isotropic Hamiltonian under chirped pulse; Magnus-based methods)}
        The accuracy of Hamiltonian simulation  for isotropic Hamiltonian \eqref{eq: isotropic homonuclear example} under a chirped pulse, \ie, \Cref{ex:isotropic homonuclear} with pulse \ref{case: chirped}, which is highly oscillatory as shown in \cref{fig: chirped pulse}. The methods compared are the first-order method using piecewise constant Hamiltonian $\ha(t_n)$, $\ie$, at the start pointing, combined with Trotter splitting \eqref{eq:trotter omega}, the second-order method using piecewise constant Hamiltonian $\ha(t_n+h/2)$, $\ie,$ at the mid-point, combined with Strang splitting \eqref{eq:strang omega}, the second-order Magnus expansion \eqref{eq: theta_1 final} with integrals computed using \texttt{scipy} and combined with the Strang splitting \eqref{eq:strang omega}, and the fourth-order Magnus expansion \eqref{eq: Theta_2 homo} combined with Yoshida splitting \eqref{eq: 4th splitting}. The integrals in the fourth-order Magnus expansions are either computed to machine precision using the Python package  \texttt{scipy} or approximated with two or three {\em Gauss--Legendre} quadrature nodes. The corresponding plots are labeled \texttt{scipy}, GL2, and GL3 respectively.}
        \label{fig: accuracy chirped isotropic homonuclear}
    \end{figure}
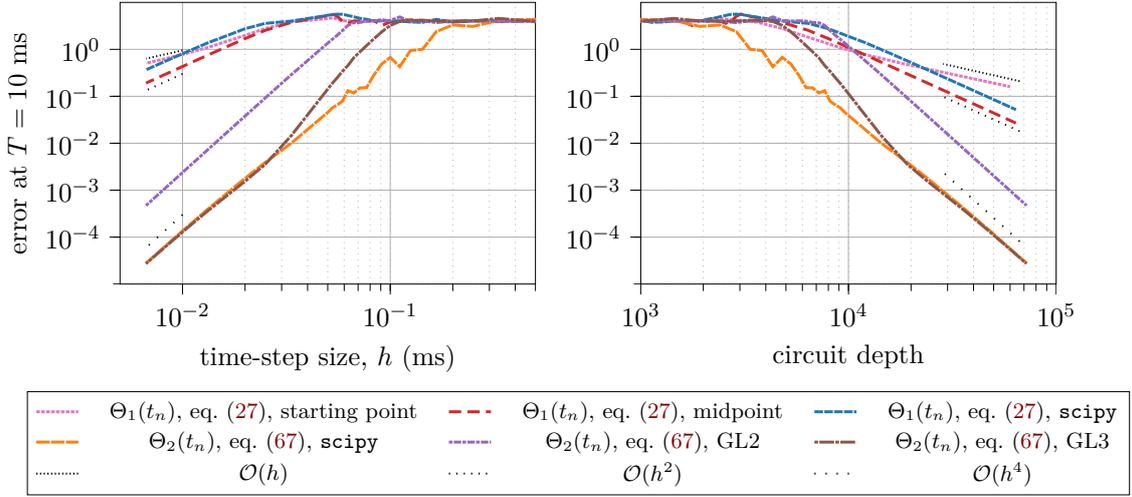

    \Cref{fig: accuracy chirped isotropic homonuclear} shows the time-step size and circuit depth required for achieving a prescribed accuracy for the case of the highly-oscillatory chirped pulse, $\ie$, pulse \ref{case: chirped} in \Cref{ex:isotropic homonuclear}. As evident from this figure, even achieving a $10^{-1}$ accuracy, where the numerical results are barely reliable, becomes extremely challenging in the case of the chirped pulse. In case of the time-independent Hamiltonian -- \Cref{ex:isotropic homonuclear}, pulse \ref{case: indep} -- the accuracy of which is depicted in \Cref{fig: accuracy time independent}, the first-order method requires a circuit depth of $72$ for an accuracy of $10^{-1}$, while the second-order method requires a circuit depth of $52$. In contrast, in \Cref{fig: accuracy chirped isotropic homonuclear} which shows the accuracy under the chirped pulse, the first-order method ($\Theta_1(t_n)$ at starting point combined with Trotter splitting) requires a circuit depth of $98,028$ and quantum volume of $294,084$, while the second-order method ($\Theta_1(t_n)$ at midpoint combined with Strang splitting) requires a circuit depth of $33,108$ and quantum volume of $99,324$ for the same accuracy -- $\ie$,  circuit depths that are $1,361$ and $636$ times longer, respectively. This highlights the significantly increased difficulty of Hamiltonian simulation under time-dependent pulses compared to time-independent Hamiltonians, particularly when the pulses are as oscillatory as chirped pulses.

    The fourth-order method $\Theta_2(t_n)$ with integrals computed using \texttt{scipy} and combined with Yoshida splitting achieves an accuracy of $10^{-1}$ with a circuit depth of roughly $7,876$, which is $76.2\%$ shorter than that required by the best second-order method and $92\%$ shorter than that required by the first-order method. Nevertheless, this is roughly $3.48$ times the circuit depth required for $10^{-4}$ accuracy in the time-independent case shown in \Cref{fig: accuracy time independent}. An advantage of a higher-order method is that achieving $10^{-2}$ accuracy requires a relatively smaller increase in circuit depth -- in particular, the method ``$\Theta_2(t_n)$, \cref{eq: Theta_2 homo}, \texttt{scipy}'' requires a circuit depth of $14,596$ and quantum volume of $43,788$ for $10^{-2}$ accuracy, compared to a circuit depth of $104,820$ expected for ``$\Theta_1(t_n)$, \cref{eq: theta_1 final}, midpoint'', \ie, $86.1\%$ shorter than the best second-order method.

    {\em Relevance to practical circuits.} In contrast to the case of time-independent Hamiltonians, Hamiltonian simulation for time-dependent pulses such as chirped pulses seems to benefit from higher-order methods even when extremely low accuracies are sought, for example when using near-term quantum devices, which are characterised by high circuit noise. On the other hand, the circuit depths required for Hamiltonian simulation applications with time-dependent Hamiltonians are very large even for the $3$ spin systems considered here. Moreover, the circuit depths are expected to grow linearly in $M$ for sparse Hamiltonians and cubically for densely connected Hamiltonians (cf. \cref{sec: overall complexity}), making the Hamiltonian simulation problem for time-dependent Hamiltonians significantly more challenging for large $M$.

    {\em Large time-steps. } Consistent with the observations of \cite{iserles2019compact,iserles2019solving,singh2019sixth} for the cases of electronic and nuclear Schr\"odinger equations under highly-oscillatory external pulses, and the preliminary confirmation of the same phenomenon in the context of many-body spin systems under weak fields \cite[Section~4.2]{goodacrethesis},
    we find that the Magnus expansion combined with a highly-accurate approximation of the integrals allows time-steps that can easily span a whole wavelength of the driving pulse. In  \Cref{fig: accuracy chirped isotropic homonuclear}, time-stepping with a time-step of $0.061$ ms leads to an accuracy of  $10^{-1}$, when using $\Theta_2(t_n)$ with \texttt{scipy} integrals. In \Cref{fig: chirped pulse} (b), which covers an interval of $0.5$ ms, there are roughly $7.5$ wavelengths. A time-step of $0.061$ ms corresponds to nearly one whole wavelength. At twice the bandwidth, $\Delta F=60$ kHz, while the number of wavelengths double, the time-step required for an accuracy of $10^{-1}$ decreases only marginally to $0.051$, highlighting the resilience of the proposed method to highly oscillatory pulses.

\begin{figure}[h]
    \centering
    \iftoggle{bandwidthTheta2}{
        \resizebox{\textwidth}{!}{\input{fig_bandwidth_size_depth_best_method}}
        }{}
        \caption{Plots comparing time-step sizes and circuit depths required to reach accuracies of $10^{-1}$, $10^{-2}$, and $10^{-3}$ against various bandwidths, using $\Theta_2(t_n)$, \cref{eq: Theta_2 homo}, \texttt{scipy}.}
        \label{fig: size depth Theta2 different bandwidth}
    \end{figure}
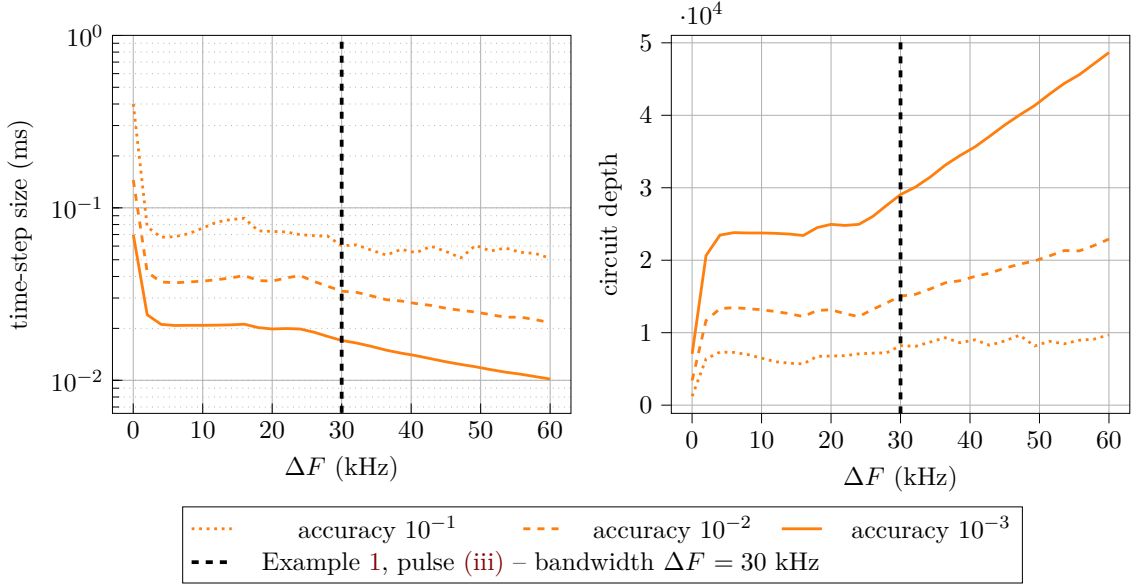

\begin{figure}[h]
    \centering
    \iftoggle{bandwidth_integrals}{
        \resizebox{\textwidth}{!}{\input{fig_bandwidth_depth_speedup_different_integrals}}
        }{}
        \caption{{\em (Circuit compression ratios for oscillatory pulses; effect of quadrature on fourth-order Magnus)} Comparison of the fourth-order Magnus expansion $\Theta_2$, \cref{eq: Theta_2 homo}, when combined with GL2 and GL3 integration methods for approximating the integrals in \cref{eq: extra commutator rotation gates weight,eq: magnus fourth order r}, against the baseline provided by $\texttt{scipy}$, which allows us to compute the integrals to machine precision. The y-axis represents the circuit compression ratio, highlighting the relative efficiency of the GL2 and GL3 integrations in comparison to the standard $\texttt{scipy}$ implementation. Black dashed vertical lines represent \Cref{ex:isotropic homonuclear}, pulse \ref{case: chirped}, where the bandwidth is $\Delta F= 30$ kHz. The horizontal line at $y = 1$ marks the point where the efficiency of the methods being compared is equivalent to that of the proposed method. Higher values signify a less efficient method.}
        \label{fig: comparison across bandwidth different integrals}
\end{figure}
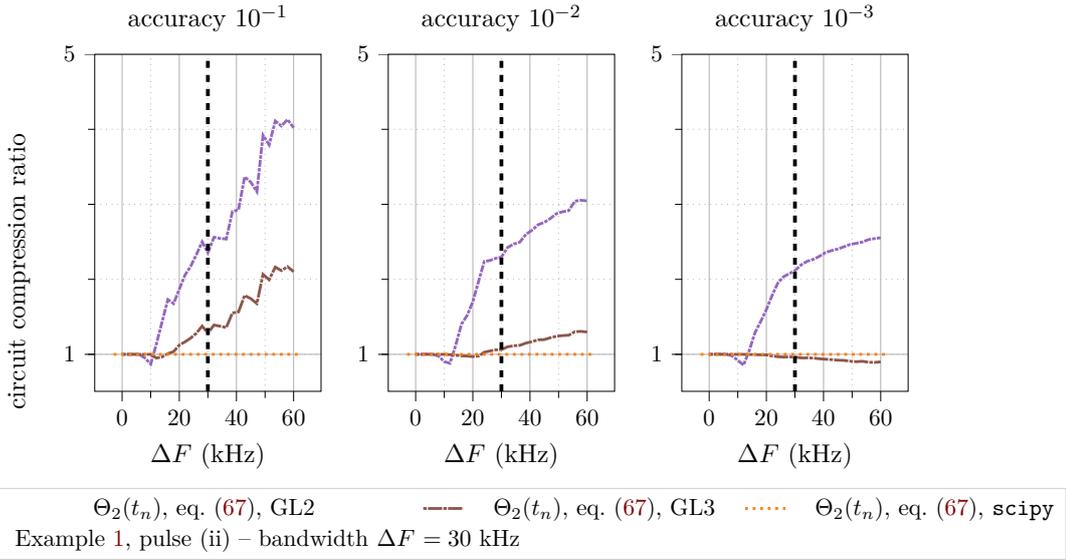

{\em The effect of approximating integrals.}
By computing the integrals in \cref{eq: extra commutator rotation gates weight,eq: magnus fourth order r} to machine precision using integration routines in the Python package \texttt{scipy}, we find that we achieve higher accuracy when compared to GL2, or, equivalently, can take a substantially larger time-step for a prescribed accuracy. For instance, in \cref{fig: accuracy chirped isotropic homonuclear} we see that when utilizing GL2, we require a time-step of $0.0255$ ms for accuracy of $10^{-1}$, which is roughly $58.1\%$ smaller than $0.0609$ ms that is required with \texttt{scipy} integrals, resulting in a $138\%$ longer circuit (circuit depth of $7,876$ with \texttt{scipy} and $18,772$ with GL2). For low accuracies, the GL3 method has a similar behavior, requiring longer circuit than \texttt{scipy}. However, for accuracies higher than $10^{-2}$, it starts to become as effective as the \texttt{scipy} approach for computing integrals. The specifics of this transition are expected to be heavily dependent on the oscillatory nature of the pulse, with more oscillatory pulses requiring higher accuracy quadrature, as seen in \cref{fig: comparison across bandwidth different integrals}. However, since the quadratures can be computed cheaply on a classical computer to a very high precision, increasing accuracy of Hamiltonian simulation without affecting the circuit complexity, there seems to be no good reason for preferring low accuracy quadratures.

    \begin{figure}[h]
    \centering
    \iftoggle{chirped_CFTO_result}{
        \resizebox{\textwidth}{!}{\input{fig_chirped_CFTO}}
        }{}
        \caption{{\em (Isotropic Hamiltonian under chirped pulse; fourth-order methods without commutators)} The accuracy of Hamiltonian simulation for isotropic Hamiltonian \eqref{eq: isotropic homonuclear example} under a chirped pulse, \ie, \Cref{ex:isotropic homonuclear} with pulse \ref{case: chirped}. The methods compared are the fourth-order methods based on Magnus expansion \eqref{eq: Theta_2 homo} with Yoshida splitting \eqref{eq: 4th splitting}, the CF42 commutator-free method with two exponents~\eqref{eq: CF42 form}, and the autonomized Yoshida splitting~\eqref{eq: time-orderng}. The integrals in the Magnus expansion and the CF42 method are computed using \texttt{scipy}, while the autonomized Yoshida splitting involves sampling the Hamiltonian at discrete points by design, and does not feature any integrals.}
        \label{fig: Magnus_cf_to}
    \end{figure}
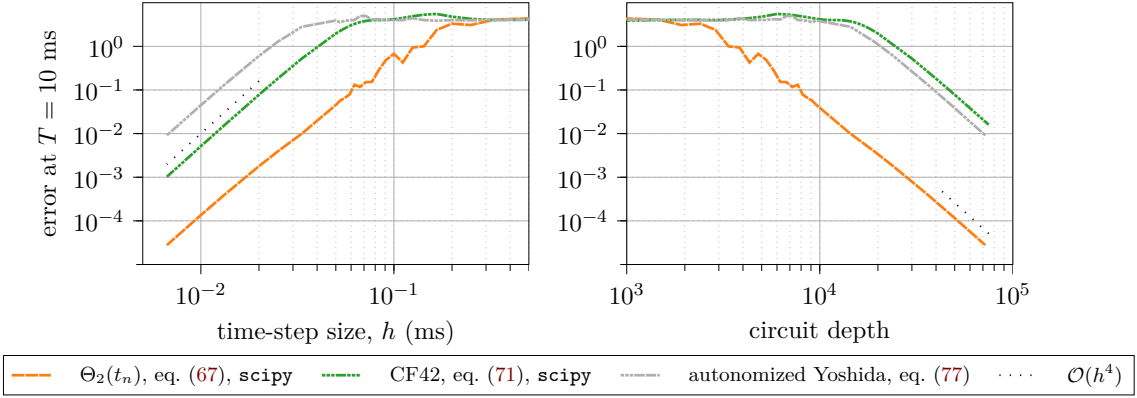

{\em Other methods without commutators.} In \cref{fig: Magnus_cf_to}, we compare the propagator accuracies for three fourth-order methods that are free of commutators: the fourth-order Magnus-based method, \cref{eq: Theta_2 homo}, proposed in this paper, the commutator-free approach CF42, \cref{eq: CF42 form}, and the fourth-order autonomized Yoshida splitting, \cref{eq: time-orderng}. We find that our fourth-order Magnus based method consistently requires a shallower circuit than the CF42 and the autonomized Yoshida splitting. Even for an extremely low accuracy $10^{-1}$, where the circuit depth of our method, \cref{eq: Theta_2 homo}, is $7,876$, CF42 \cref{eq: CF42 form} and the autonomized Yoshida splitting \eqref{eq: time-orderng} require circuit depths of $46,804$ and $38,932$, respectively -- $\ie$, circuit depths that are roughly $4.85$ and $3.94$ times longer, respectively. These observations remain similar when we require higher accuracies.

{\bf Benchmarking the effect of oscillatory pulses}

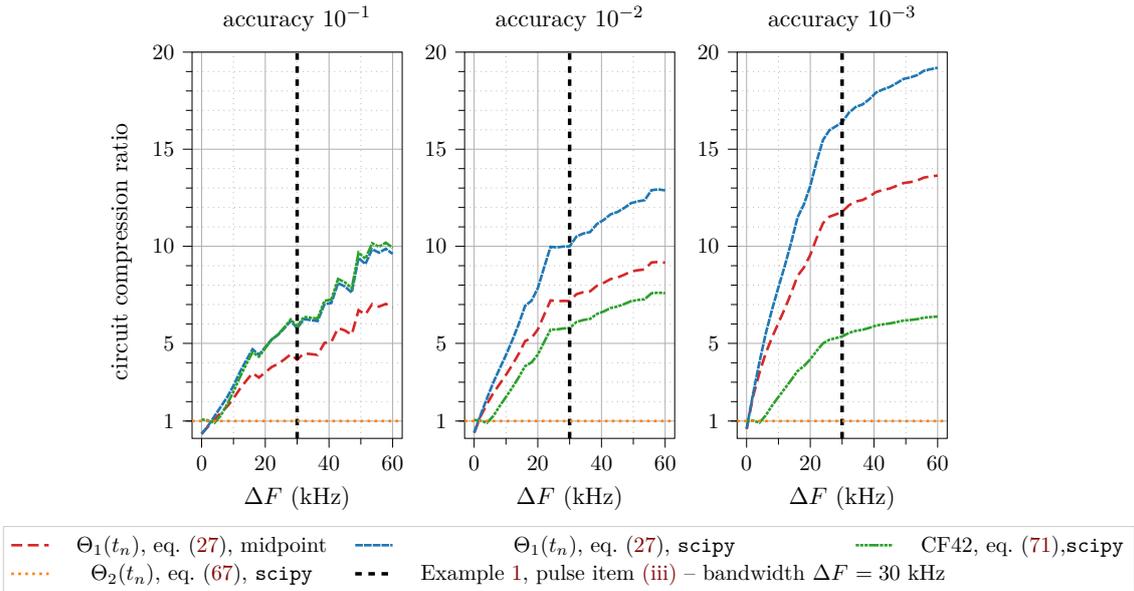
\begin{figure}[h]
    \centering
    \iftoggle{bandwidth_methods}{
        \resizebox{\textwidth}{!}{\input{fig_bandwidth_depth_speedup_different_methods}}
        }{}
        \caption{{\em (Circuit compression ratios for oscillatory pulses; methods of different orders)}
        Comparison of circuit depth across four computational methods at varying bandwidths, $\Delta F$, for different accuracy levels, $10^{-1} $,  $10^{-2} $, and  $10^{-3} $.
        The y-axis shows the circuit compression ratio, indicating each method's efficiency relative to the proposed method, ``$\Theta_2(t_n)$, \cref{eq: Theta_2 homo}, \texttt{scipy}'', with integrals computed using $\texttt{scipy}$. Black dashed vertical lines indicate the reference values used in \Cref{ex:isotropic homonuclear}, pulse \ref{case: chirped}, where the bandwidth is $\Delta F=30$ kHz.
        The horizontal line at $y = 1$ marks the point where the efficiency of the methods being compared is equivalent to that of the proposed method. Higher values signify a less efficient method.
        }
        \label{fig: comparison across bandwidth different methods}
\end{figure}

One of the advantages of our fourth-order Magnus-based method, \cref{eq: Theta_2 homo}, is its efficacy for highly oscillatory pulses. The oscillatory nature of a chirped pulse, \crefrange{eq: pxpy start}{eq: pxpy end}, increases with the bandwidth, $\Delta F$. In \cref{fig: comparison across bandwidth different methods} we compare the circuit depths required for achieving a prescribed propagator accuracy for a wide range of bandwidths, with other parameters of the pulse and the specification of the Hamiltonian remaining the same as in \Cref{ex:isotropic homonuclear}, pulse \ref{case: chirped}. For reference, the bandwidth used in \Cref{ex:isotropic homonuclear}, pulse \ref{case: chirped}, is marked by a black dashed vertical line (the behaviour under these parameters has been studied earlier in this subsection, in \cref{fig: accuracy chirped isotropic homonuclear,fig: Magnus_cf_to}.)

In addition to the proposed method ``$\Theta_2(t_n)$, \cref{eq: Theta_2 homo}, \texttt{scipy}'' we benchmark the methods ``$\Theta_1(t_n)$, \cref{eq: theta_1 final}, midpoint'', ``$\Theta_1(t_n)$, \cref{eq: theta_1 final}, $\texttt{scipy}$'' and  ``CF42, \cref{eq: CF42 form} $\texttt{scipy}$'' -- these methods are chosen for benchmarking since they were found to be the most efficient first-order, second-order and fourth-order alternatives to our approach in \cref{fig: accuracy chirped isotropic homonuclear}.
In \cref{fig: comparison across bandwidth different methods}, the x-axis represents different bandwidths, while the y-axis shows the {\em circuit compression ratio}. This ratio indicates how many times deeper circuit is required for the three comparative methods compared to the proposed method, which is treated as a baseline method and is depicted for reference with a circuit compression ratio of $1$. A higher value on the y-axis signifies a less efficient method. In particular, values higher than $1$ signify a method less efficient than the proposed method, while values lower than $1$ signify a method more efficient than the proposed method.

Each subplot in \cref{fig: accuracy chirped isotropic homonuclear} corresponds to a specific accuracy target, $10^{-1}, 10^{-2}$ and $10^{-3}$, with lower accuracies being of more immediate relevance to near-term quantum devices.
The benchmarking demonstrates that our fourth-order Magnus-based method consistently outperforms the others across all accuracy levels and almost all bandwidths, with a minor exception being a combination of extremely slowly varying pulses (small bandwidth, $\Delta F$) and low accuracies. As bandwidth increases, the relative advantage conferred by our method compared to the other methods increases as well, indicating its increased benefits under the typically hard to simulate case of highly-oscillatory pulses.

As evident in \cref{fig: size depth Theta2 different bandwidth}, this is largely due to a resilience of our method to highly oscillatory pulses, which requires a very mild increase in circuit depths for increasingly oscillatory pulses.
We note that the maximum amplitude of the pulses defined in \crefrange{eq: pxpy start}{eq: pxpy end} grows as $\max_{t\in [0,T]} \|\ve(t)\|_\infty = \order(\sqrt{\Delta F})$, which increases the Hamiltonian spectral radius $\rho$ in \cref{eq: Hamiltonian norm time dependent} as a consequence. We suspect that, along with the increase in $\gamma$ \eqref{eq: control frequency max} due to increased oscillations, this could be the reason for the increase in circuit-depth for higher bandwidths in \cref{fig: size depth Theta2 different bandwidth} seen for the higher-accuracy of $10^{-3}$.

\Cref{fig: comparison across bandwidth different integrals} compares circuit depths required when applying our fourth-order method, \cref{eq: Theta_2 homo}, in combination with different methods for approximating the integrals, namely, $\texttt{scipy}$, GL2 and GL3. Across all levels of accuracy investigated here, $\texttt{scipy}$ method maintains a clear advantage over the GL2 method. However, the advantage of $\texttt{scipy}$ over GL3 diminishes for accuracy levels of $10^{-3}$. As noted earlier, since integrals can be computed cheaply on a classical computer, there is no compelling reason to prefer GL2 or GL3 over \texttt{scipy}.

{\bf Benchmarking the effect of coupling strengths}

We benchmark the effect of coupling strengths on the four different methods considered above, namely ``$\Theta_2(t_n)$, \cref{eq: Theta_2 homo}, \texttt{scipy}'', ``$\Theta_1(t_n)$, \cref{eq: theta_1 final}, midpoint'', ``$\Theta_1(t_n)$, \cref{eq: theta_1 final}, $\texttt{scipy}$'' and  ``CF42, \cref{eq: CF42 form} $\texttt{scipy}$'', using random coupling strengths in \Cref{ex:isotropic homonuclear}, case \ref{case: chirped}, with other parameters kept constant. To quantify the coupling strength, we compute the $\ell^2$ norm of the interaction Hamiltonian, $\|\hi\|$, which is depicted on the x-axis of \cref{fig: comparison across Hin norm different methods}, while the y-axis shows the circuit compression ratio. The mean circuit compression ratio is shown as a function of $\|\hi\|$ with thick lines, while the shadowed areas around it represent a single standard deviation.
As before, the black dashed vertical line indicates the reference setup, $\ie$, \Cref{ex:isotropic homonuclear}, case \ref{case: chirped}, where $\|\hi\| \approx 187.4$. %

It is evident in \cref{fig: comparison across Hin norm different methods} that, while our fourth-order Magnus-based method has the highest comparative advantage under a smaller norm of the interaction Hamiltonian (indicating weaker coupling strengths) and for higher accuracies, it consistently outperforms the other three methods across all levels of accuracy considered here, $10^{-1}$, $10^{-2}$ and $10^{-3}$, and an extremely wide range of interaction strengths. However, as the strength of the interaction, measured by $\|\hi\|$ increases, the circuit compression ratio indicating our method's superiority does decrease. In particular, CF42 becomes as efficient as the proposed approach for $\|\hi\|=10^5$, while the first and second-order methods remain less efficient across the entire range of interaction strengths considered.

\begin{figure}[h]
    \centering
    \iftoggle{Hin_methods}{
        \resizebox{\textwidth}{!}{\input{fig_Hin_norm_depth_speedup_different_methods_log}}
        }{}
        \caption{{\em (Circuit compression ratios for oscillatory pulses with different coupling strength; methods of different orders)}
        Comparison of circuit depth across four computational methods at varying $\|\hi\|$ for different accuracy levels, $10^{-1} $,  $10^{-2} $, and  $10^{-3} $.
        The y-axis shows the circuit compression ratio, indicating each method's efficiency relative to the proposed method, \cref{eq: Theta_2 homo}, with integrals computed using $\texttt{scipy}$. Black dashed vertical lines indicate the reference values used in \Cref{ex:isotropic homonuclear}, pulse \ref{case: chirped}, where $\|\hi\| = 187.4$.
        The horizontal line at $y = 1$ marks the point where the efficiency of the methods being compared is equivalent to that of the proposed method. Higher values signify a less efficient method.
        }{}
        \label{fig: comparison across Hin norm different methods}
\end{figure}

\begin{figure}[htbp]
    \centering
    \iftoggle{Hin_integrals}{
        \resizebox{\textwidth}{!}{\input{fig_Hin_norm_depth_speedup_different_integrals}}
        }{}
        \caption{{\em (Circuit compression ratios for oscillatory pulses with different coupling strength; effect of quadrature on fourth-order Magnus)}
        Comparison of the fourth-order Magnus expansion $\Theta_2$ \cref{eq: Theta_2 homo}, when combined with GL2 and GL3 integration methods for approximating the integrals in \cref{eq: extra commutator rotation gates weight,eq: magnus fourth order r}, against the baseline provided by $\texttt{scipy}$, which allows us to compute the integrals to machine precision.
        The y-axis represents the circuit compression ratio, highlighting the relative efficiency of the GL2 and GL3 integrations in comparison to the standard $\texttt{scipy}$ implementation. Black dashed vertical lines represent \Cref{ex:isotropic homonuclear}, pulse \ref{case: chirped}, where $\|\hi\| = 187.4$. %
        The horizontal line at $y = 1$ marks the point where the efficiency of the methods being compared is equivalent to that of the proposed method. Higher values signify a less efficient method.
        }{}
        \label{fig: comparison across Hin norm different integrals}
\end{figure}

\Cref{fig: comparison across Hin norm different integrals} illustrates a comparison of the circuit depths required when applying our fourth-order method, \cref{eq: Theta_2 homo}, in combination with different integral approximation techniques, \ie, $\texttt{scipy}$, GL2 and GL3.
Across all three levels of accuracy considered here, the use of $\texttt{scipy}$ for approximating the integrals maintains a clear comparative advantage over the GL2 method, except for extremely strong interaction strengths, where GL2 becomes as efficient. However, $\texttt{scipy}$ only outperforms GL3 for relatively smaller values of $\|\hi\|$, and only for accuracy levels of $10^{-1}$. For higher accuracies and interaction strengths, it loses its comparative advantage. However, as noted earlier, since integrals can be computed cheaply on a classical computer, there is no compelling reason to prefer GL2 or GL3 over \texttt{scipy}.

As described by the time-step complexity in \cref{eq: time-step complexity of Trotterization}, the circuit depth required for achieving a prescribed accuracy grows linearly with the Hamiltonian spectral radius $\rho$ \eqref{eq: hamiltonian spectral radius},
\[
\rho = \max_{t \in [0,T]} \| \ha(t)\| \leq  \max_{t \in [0,T]} \| \hs(t)\| + \| \hi\|.
\]
Conversely, the time-step size needs to decrease inversely as $\rho$.

\begin{figure}[htbp]
    \centering
    \iftoggle{couplingstrengthTheta2}{
        \resizebox{\textwidth}{!}{\input{fig_Hin_size_depth_best_method}}
        }{}
        \caption{Plots comparing time-step sizes and circuit depths required to reach accuracies of $10^{-1}$, $10^{-2}$, and $10^{-3}$ against various coupling strengths, using $\Theta_2(t_n)$, \cref{eq: Theta_2 homo}, \texttt{scipy}.}
        \label{fig: size depth Theta2 different coupling strength}
    \end{figure}
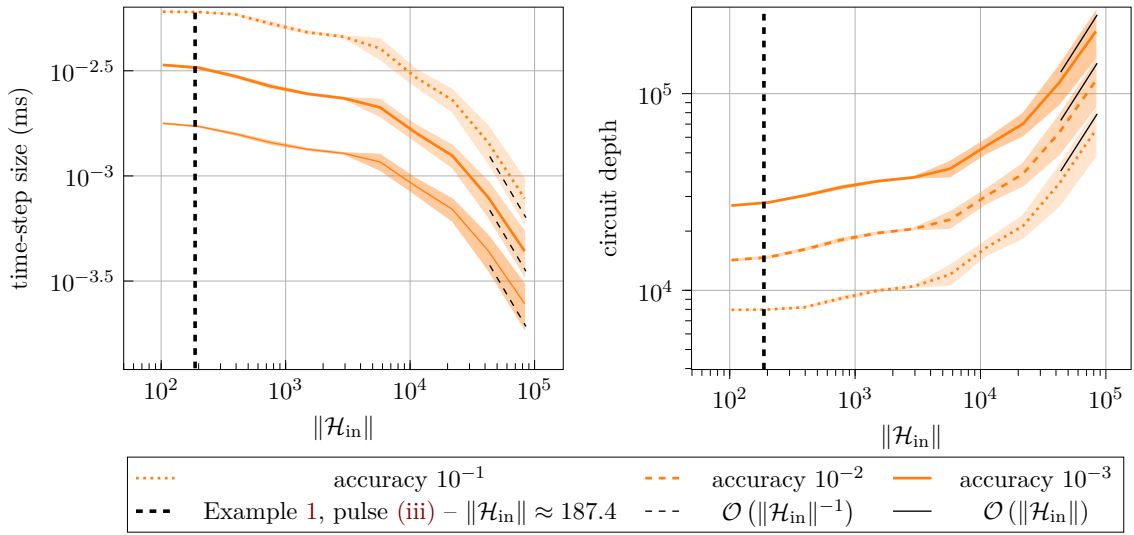

In our baseline case, \Cref{ex:isotropic homonuclear}, pulse~\ref{case: chirped}, $\| \hi\| \approx 187.4$, while $\max_{t \in [0,T]}\|\hs(t)\| = 35511.92$. In \cref{fig: size depth Theta2 different coupling strength}, it takes a while until $\| \hi \|$ dominates and essentially dictates the value of $\rho$, at which point the circuit depth of our fourth-order method, \cref{eq: Theta_2 homo}, almost converges to a linear growth with $\|\hi\|$. Conversely, the time-step size decreases inversely with $\|\hi\|$ at this point, as expected.

{\bf Chirped pulse, general Hamiltonian without mixed coupling} (\Cref{ex:general}, case~\ref{case: no mixed})

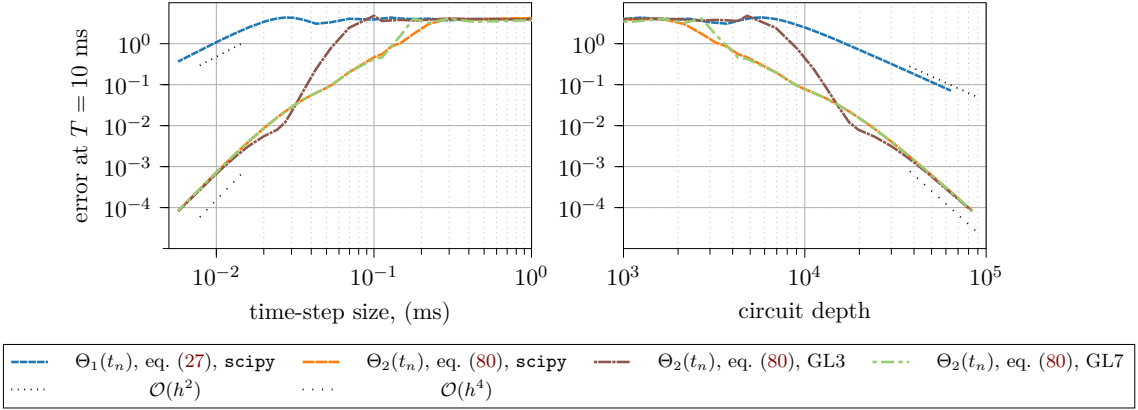
\begin{figure}[htbp]
    \centering
    \iftoggle{general_no_mixed_result}{
        \resizebox{\textwidth}{!}{\input{fig_general_no_mixed}}
        }{}
        \caption{{\em (Non-isotropic Hamiltonian without mixed couplings under multiple chirped pulses; Magnus-based methods)}
        The accuracy of Hamiltonian simulation for non-isotropic Hamiltonian without mixed couplings under a chirped pulse, \ie, \Cref{ex:general}, case \ref{case: no mixed}.
        The methods compared are the second-order Magnus expansion \eqref{eq: theta_1 final} with integrals computed using \texttt{scipy} and combined with the Strang splitting \eqref{eq:strang omega}, and the fourth-order Magnus expansion with commutator eliminated \eqref{eq: Trotter type E} and combined with Yoshida splitting \eqref{eq: 4th splitting}. The integrals in the fourth-order Magnus expansions are either computed to machine precision using the Python package  \texttt{scipy} or approximated with three or seven {\em Gauss--Legendre} quadrature nodes. The corresponding plots are labeled \texttt{scipy}, GL3, and GL7 respectively.}
        \label{fig: accuracy general no mixed coupling}
    \end{figure}

For the case of the general Hamiltonian without mixed couplings described in \Cref{ex:general}, case \ref{case: no mixed}, the fourth-order Magnus expansion requires the elimination of commutators before combination with the Yoshida splitting, and leads to the proposed method described in \cref{eq: Trotter type E}.
As we can see in \cref{fig: accuracy general no mixed coupling}, even for an accuracy of $10^{-1}$, ``$\Theta_1(t_n)$, \cref{eq: theta_1 final}, \texttt{scipy}'' requires a circuit depth of $53,924$, while ``$\Theta_2(t_n)$, \cref{eq: Theta_2 homo}, \texttt{scipy}'' achieves the same accuracy with circuit depth of $8,932$, which is $83.4\%$ shorter. \Cref{fig: accuracy general no mixed coupling} also shows that the choice of integration methods is critical for an accuracy level of $10^{-1}$: for example, $\Theta_2(t_n)$ combining with GL3 requires a circuit depth of $12,820$, which is $43.5\%$ longer than when combined with \texttt{scipy}.

For $10^{-3}$ or higher accuracies, the three integral methods we are examining converge to the same behaviour, all of which require a circuit depth of $42,916$.
The circuit depth required here is $52.3\%$ longer than in \Cref{ex:isotropic homonuclear}, pulse \ref{case: chirped}, where the fourth-order Magnus method with \texttt{scipy} integrals required a circuit depth of $28,180$.

{\bf Chirped pulse, general Hamiltonian with mixed coupling} (\Cref{ex:general}, case~\ref{case: mixed})

\begin{figure}[htbp]
    \centering
    \iftoggle{general_mixed_result}{
        \resizebox{!}{4.8cm}{\input{fig_general_case_with_mix_coupling_vs_timestep}}
        }{}
        \caption{{\em (Non-isotropic Hamiltonian with mixed couplings under multiple chirped pulses; Magnus-based methods)} The accuracy of Hamiltonian simulation accuracy for non-isotropic Hamiltonian with mixed couplings under a chirped pulse, \ie, \Cref{ex:general}, case \ref{case: mixed} .
        The methods compared are the second-order Magnus expansion \eqref{eq: theta_1 final} with integrals computed using \texttt{scipy} and combined with the Strang splitting \eqref{eq:strang omega}, and the fourth-order Magnus expansion with commutator eliminated \eqref{eq:eliminating} and exponentials computed by brute force. The integrals in the fourth-order Magnus expansions are either computed to machine precision using the Python package  \texttt{scipy} or approximated with three or seven {\em Gauss--Legendre} quadrature nodes. The corresponding plots are labeled \texttt{scipy}, GL3, and GL7 respectively.
        }
        \label{fig: accuracy general with mixed coupling}
    \end{figure}
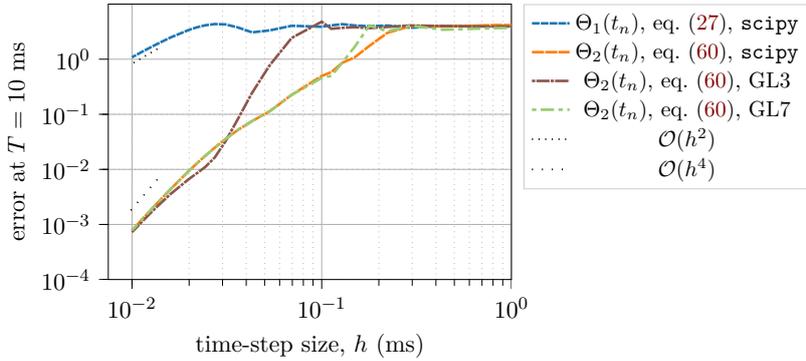

We do not have Trotterized circuits for general Hamiltonians with mixed couplings. Instead, we compute the exponentials in \cref{eq:eliminating} by brute force.
As shown in \cref{fig: accuracy general with mixed coupling}, we observe similar behaviours to previous examples, \ie, the proposed fourth-order method, ``$\Theta_2(t_n)$, \cref{eq:eliminating}, \texttt{scipy}'', allows larger time-step sizes than the second-order method, ``$\Theta_1(t_n)$, \cref{eq: theta_1 final}, \texttt{scipy}''.
Within the context of \cref{eq:eliminating}, using \texttt{scipy} for computing integrals allows larger time-step sizes than GL3 and GL7.

%% file: fig_indep_result.tex
\begin{tikzpicture}
\begin{groupplot}[
    group style={group size=2 by 1,
    horizontal sep=1.5cm,
    y descriptions at=edge left},
    legend pos=south west,
    legend style={
        at={(-0.15,-0.65)},
        anchor=south,
        legend columns=3,
        column sep=10pt,
        font=\footnotesize
      }
]

\nextgroupplot
[
width=0.5\textwidth, height=0.375\textwidth,
log basis x={10},
log basis y={10},
tick align=outside,
tick pos=left,
x grid style={darkgray176},
xlabel={time-step size, $h$ (ms)},
xmin=1e-01*0.25, xmax=14,
xmode=log,
xtick style={color=black, font=\footnotesize},
ymin = 1e-8*0.75,
ymax = 10,
xmajorgrids,
ymajorgrids,
y grid style={darkgray176},
ylabel={error at $T=10$ ms},
ymode=log,
ytick={1, 1e-1, 1e-4, 1e-7},
xminorgrids,
yminorgrids,
minor ytick={1e-8, 1e-7,1e-6,1e-5,1e-4,1e-3,1e-2,1e-1},
minor xtick={2e-2, 3e-2, 4e-2, 5e-2, 6e-2, 7e-2, 8e-2, 9e-2, 1e-1, 2e-1, 3e-1, 4e-1, 5e-1, 6e-1, 7e-1, 8e-1, 9e-1, 1, 2, 3, 4, 5, 6, 7, 8, 9, 10},
minor grid style={lightgray230, dotted, line width=0.1pt}
]

\addplot [very thick, orchid227119194, dash pattern=on 1.5pt off 0.5pt, x filter/.code={\pgfmathparse{\pgfmathresult +3}\pgfmathresult}]
table {%
0.01 1.47172700963905
0.005 0.752804790960832
0.00333333333333333 0.105676563545016
0.0025 0.743617313168305
0.002 0.4472329022845
0.00166666666666667 0.0640381591787306
0.00142857142857143 0.052301148514109
0.00125 0.0323737171852916
0.00111111111111111 0.0311635765396616
0.001 0.0233918631958555
0.000769230769230769 0.0168685182308817
0.000625 0.0133229819504426
0.000526315789473684 0.0110410455546502
0.000454545454545455 0.00942081069526103
0.0004 0.00823672486185121
0.000357142857142857 0.00732324625444222
0.00024390243902439 0.00495598772601405
0.000227272727272727 0.00461165080642324
0.000212765957446809 0.00431309549535773
0.0002 0.00405123317233916
0.000188679245283019 0.0038195576420333
0.000178571428571429 0.003613081260577
0.000169491525423729 0.00342787809631246
0.000161290322580645 0.00326080567156546
0.000153846153846154 0.00310931541878339
0.000147058823529412 0.00297131730288361
0.000138888888888889 0.00280535816475138
0.000133333333333333 0.00269259449051511
0.000128205128205128 0.0025885654301256
0.000123456790123457 0.00249229184965239
0.000119047619047619 0.00240293598870019
0.000114942528735632 0.00231977678002668
0.000111111111111111 0.00224219017694519
0.00010752688172043 0.00216963333866748
0.000104166666666667 0.00210163181684083
0.0001 0.00201733651921818
3.07692307692308e-05 0.000619968289516683
};

\addplot [very thick, steelblue31119180, dash pattern=on 3.5pt off 0.5pt, x filter/.code={\pgfmathparse{\pgfmathresult +3}\pgfmathresult}]
table {%
0.01 1.46169302422396
0.005 0.747981673682183
0.00333333333333333 0.0698411142835281
0.0025 0.742235255574396
0.002 0.442499066623122
0.00166666666666667 0.0496625888252776
0.00142857142857143 0.0294074613524612
0.00125 0.0204206039119126
0.00111111111111111 0.0152671998610361
0.001 0.0119255429642448
0.000769230769230769 0.00667300881602891
0.000625 0.00429649357822476
0.000526315789473684 0.00300603050789549
0.000454545454545455 0.00222394309727399
0.0004 0.00171312214731432
0.000357142857142857 0.00136071792155893
0.00024390243902439 0.000629852393692012
0.000227272727272727 0.00054642161760525
0.000212765957446809 0.000478557286408629
0.0002 0.000422609667909137
0.000188679245283019 0.000375940422780471
0.000178571428571429 0.000336603678534345
0.000169491525423729 0.000303138648118958
0.000161290322580645 0.00027443130991099
0.000153846153846154 0.000249620043602698
0.000147058823529412 0.000228029942252025
0.000138888888888889 0.000203345436650492
0.000133333333333333 0.00018737236705757
0.000128205128205128 0.000173211055804396
0.000123456790123457 0.000160597373125798
0.000119047619047619 0.000149313648235282
0.000114942528735632 0.000139179192680537
0.000111111111111111 0.000130043005650853
0.00010752688172043 0.000121778105213063
0.000104166666666667 0.000114277083431902
0.0001 0.000105308048462654
3.07692307692308e-05 9.96022603199094e-06
};

\addplot [very thick, darkorange25512714, dash pattern=on 5.5pt off 0.5pt, x filter/.code={\pgfmathparse{\pgfmathresult +3}\pgfmathresult}]
table {%
0.01 2.19328812955444
0.005 1.05706692668002
0.00333333333333333 0.0683814565893408
0.0025 1.14337438437574
0.002 0.679641796946404
0.00166666666666667 0.048122632336064
0.00142857142857143 0.0238789333224412
0.00125 0.00873705235009754
0.00111111111111111 0.00967965611656478
0.001 0.00698081318017093
0.000769230769230769 0.004534242170349
0.000625 0.00275188889570346
0.000526315789473684 0.00155908771781287
0.000454545454545455 0.00101335953677387
0.0004 0.000673341284410796
0.000357142857142857 0.000447730623885195
0.00024390243902439 0.000189304859776167
0.000227272727272727 0.000135905494659685
0.000212765957446809 0.000102812340862525
0.0002 7.9981175777038e-05
0.000188679245283019 6.34193987312105e-05
0.000178571428571429 5.10363970873072e-05
0.000169491525423729 4.15778369023525e-05
0.000161290322580645 3.42318420163057e-05
0.000153846153846154 2.84475007927161e-05
0.000147058823529412 2.38386427248768e-05
0.000138888888888889 1.90538767864462e-05
0.000133333333333333 1.62343813520662e-05
0.000128205128205128 1.39175054431149e-05
0.000123456790123457 1.19993142099436e-05
0.000119047619047619 1.04001885824114e-05
0.000114942528735632 9.05849892882379e-06
0.000111111111111111 7.92608696703872e-06
0.00010752688172043 6.96499332902523e-06
0.000104166666666667 6.14505721799401e-06
0.0001 5.23035324436192e-06
3.07692307692308e-05 4.80456187686779e-08
};

\addplot [thick, dash pattern=on 0.5pt off 1pt, x filter/.code={\pgfmathparse{\pgfmathresult +3}\pgfmathresult}]
table {%
8e-05 0.004
3.42465753424658e-05 0.00171232876712329
};

\addplot [thick, dash pattern=on 0.5pt off 2.5pt, x filter/.code={\pgfmathparse{\pgfmathresult+3}\pgfmathresult}]
table {%
8e-05 0.000128
3.42465753424658e-05 2.34565584537437e-05
};

\addplot [thick, dash pattern=on 0.5pt off 4pt, x filter/.code={\pgfmathparse{\pgfmathresult+3}\pgfmathresult}]
table {%
8e-05 8.192e-07
3.42465753424658e-05 2.75105067246947e-08
};

\nextgroupplot
[
width=0.5\textwidth, height=0.375\textwidth,
log basis x={10},
log basis y={10},
tick align=outside,
tick pos=left,
x grid style={darkgray176},
xlabel={circuit depth},
xmin=10, xmax=1e4*3,
xmode=log,
xtick style={color=black},
ymin = 1e-8*0.75,
ymax = 10,
xmajorgrids,
ymajorgrids,
y grid style={darkgray176},
ymode=log,
y grid style={darkgray176},
ymode=log,
ytick={1, 1e-1, 1e-4, 1e-7},
yticklabels={$10^0$, $10^{-1}$, $10^{-4}$, $10^{-7}$},
xminorgrids,
yminorgrids,
minor ytick={1e-8, 1e-7,1e-6,1e-5,1e-4,1e-3,1e-2,1e-1},
minor xtick={2e1, 3e1, 4e1, 5e1, 6e1, 7e1, 8e1, 9e1,2e2, 3e2, 4e2, 5e2, 6e2, 7e2, 8e2, 9e2,2e3, 3e3, 4e3, 5e3, 6e3, 7e3, 8e3, 9e3},
minor grid style={lightgray230, dotted, line width=0.1pt}
]

\addplot [very thick, orchid227119194, dash pattern=on 1.5pt off 0.5pt]
table {%
12 1.47172700963905
24 0.752804790960832
36 0.105676563545016
48 0.743617313168305
60 0.4472329022845
72 0.0640381591787306
84 0.052301148514109
96 0.0323737171852916
108 0.0311635765396616
120 0.0233918631958555
156 0.0168685182308817
192 0.0133229819504426
228 0.0110410455546502
264 0.00942081069526103
300 0.00823672486185121
336 0.00732324625444222
492 0.00495598772601405
528 0.00461165080642324
564 0.00431309549535773
600 0.00405123317233916
636 0.0038195576420333
672 0.003613081260577
708 0.00342787809631246
744 0.00326080567156546
780 0.00310931541878339
816 0.00297131730288361
864 0.00280535816475138
900 0.00269259449051511
936 0.0025885654301256
972 0.00249229184965239
1008 0.00240293598870019
1044 0.00231977678002668
1080 0.00224219017694519
1116 0.00216963333866748
1152 0.00210163181684083
1200 0.00201733651921818
3900 0.000619968289516683
6600 0.000366314799923766
9300 0.000259959654916378
12000 0.000201466946639702
};
\addlegendentry{Trotter, \cref{eq:trotter omega}}
\addplot [very thick, steelblue31119180, dash pattern=on 3.5pt off 0.5pt]
table {%
20 1.46169302422396
36 0.747981673682183
52 0.0698411142835281
68 0.742235255574396
84 0.442499066623122
100 0.0496625888252776
116 0.0294074613524612
132 0.0204206039119126
148 0.0152671998610361
164 0.0119255429642448
212 0.00667300881602891
260 0.00429649357822476
308 0.00300603050789549
356 0.00222394309727399
404 0.00171312214731432
452 0.00136071792155893
500 0.00110747258030821
548 0.0009186220376627
596 0.000774548196447309
660 0.000629852393692012
708 0.00054642161760525
756 0.000478557286408629
804 0.000422609667909137
852 0.000375940422780471
900 0.000336603678534345
948 0.000303138648118958
996 0.00027443130991099
1044 0.000249620043602698
1092 0.000228029942252025
1156 0.000203345436650492
1204 0.00018737236705757
1252 0.000173211055804396
1300 0.000160597373125798
1348 0.000149313648235282
1396 0.000139179192680537
1444 0.000130043005650853
1492 0.000121778105213063
1540 0.000114277083431902
1604 0.000105308048462654
5204 9.96022603199094e-06
8804 3.47761582265349e-06
12404 1.75144245895957e-06
16004 1.05195257464487e-06
};
\addlegendentry{Strang, \cref{eq:strang omega}}
\addplot [very thick, darkorange25512714, dash pattern=on 5.5pt off 0.5pt]
table {%
52 2.19328812955444
100 1.05706692668002
148 0.0683814565893408
196 1.14337438437574
244 0.679641796946404
292 0.048122632336064
340 0.0238789333224412
388 0.00873705235009754
436 0.00967965611656478
484 0.00698081318017093
628 0.004534242170349
772 0.00275188889570346
916 0.00155908771781287
1060 0.00101335953677387
1204 0.000673341284410796
1348 0.000447730623885195
1972 0.000189304859776167
2116 0.000135905494659685
2260 0.000102812340862525
2404 7.9981175777038e-05
2548 6.34193987312105e-05
2692 5.10363970873072e-05
2836 4.15778369023525e-05
2980 3.42318420163057e-05
3124 2.84475007927161e-05
3268 2.38386427248768e-05
3460 1.90538767864462e-05
3604 1.62343813520662e-05
3748 1.39175054431149e-05
3892 1.19993142099436e-05
4036 1.04001885824114e-05
4180 9.05849892882379e-06
4324 7.92608696703872e-06
4468 6.96499332902523e-06
4612 6.14505721799401e-06
4804 5.23035324436192e-06
15604 4.80456187686779e-08
};
\addlegendentry{Yoshida, \cref{eq: 4th splitting}}
\addplot [thick, black, dash pattern=on 0.5pt off 1pt]
table {%
11904.7619047619 0.00042
5952.38095238095 0.00084
4000 0.00125
};
\addlegendentry{$\order\lb h \rb$}
\addplot [thick, black, dash pattern=on 0.5pt off 2.5pt]
table {%
11904.7619047619 3.528e-06
5952.38095238095 1.4112e-05
4000 3.125e-05
};
\addlegendentry{$\order\lb h^2 \rb$}
\addplot [thick, black, dash pattern=on 0.5pt off 4pt]
table {%
11904.7619047619 6.223392e-08
5952.38095238095 9.9574272e-07
4000 4.8828125e-06
};
\addlegendentry{$\order \lb h^4 \rb$}

\end{groupplot}

\end{tikzpicture} 

%% file: fig_chirped_result.tex
\begin{tikzpicture}

\begin{groupplot}[
    group style={group size=2 by 1,
    horizontal sep=1.5cm,
    y descriptions at=edge left},
    width=\textwidth, %
    legend pos= south east,
    legend style={
        at={(-0.15,-0.75)},
        anchor=south,
        legend columns=3,column sep=10pt,
        font=\footnotesize
      }
]

\nextgroupplot
[
width=0.5\textwidth, height=0.375\textwidth,
legend style={draw=none},
log basis x={10},
log basis y={10},
tick align=outside,
tick pos=left,
x grid style={darkgray176},
xlabel={time-step size, $h$ (ms)},
xmin=1e-02*0.5, xmax=0.5,
xmode=log,
xtick style={color=black, font=\footnotesize},
ymin = 1e-5,
ymax = 10,
xmajorgrids,
ymajorgrids,
y grid style={darkgray176},
ylabel={error at $T=10$ ms},
ymode=log,
ytick={1, 1e-1, 1e-2, 1e-3, 1e-4},
ytick style={color=black},
xminorgrids,
yminorgrids,
minor ytick={1e-7,1e-6,1e-5,1e-4,1e-3,1e-2,1e-1},
minor xtick={2e-2, 3e-2, 4e-2, 5e-2, 6e-2, 7e-2, 8e-2, 9e-2, 1e-1, 2e-1, 3e-1, 4e-1, 5e-1, 6e-1, 7e-1, 8e-1, 9e-1, 1, 2, 3, 4, 5, 6, 7, 8, 9, 10},
minor grid style={lightgray230, dotted, line width=0.1pt}
]
\addplot [very thick, orchid227119194, dash pattern=on 1.5pt off 0.5pt, x filter/.code={\pgfmathparse{\pgfmathresult +3}\pgfmathresult}]
table {%
0.001 4.02237706091091
0.0005 3.99568698713503
0.000333333333333333 3.8507338677303
0.00025 4.04797044970421
0.0002 3.75959914671215
0.000166666666666667 3.95470460079529
0.000142857142857143 3.85212843053157
0.000125 3.9330189895024
0.000111111111111111 4.0034254465864
0.0001 4.02691456126561
9.09090909090909e-05 4.33531637366652
8.33333333333333e-05 4.05225762850164
7.69230769230769e-05 3.92417037669378
7.14285714285714e-05 3.99059809235397
6.66666666666667e-05 4.03073773654815
6.25e-05 4.06524146921515
5.88235294117647e-05 4.02175096366423
5.55555555555556e-05 4.42022106924293
5.26315789473684e-05 4.66983928322516
5e-05 4.53234047019901
3.33333333333333e-05 3.81160034352927
2.5e-05 2.77340188245801
2e-05 1.98857603130444
1.33333333333333e-05 1.10569965778719
1e-05 0.775344968141025
6.66666666666667e-06 0.505065556230199
};

\addplot [very thick, crimson2143940, dash pattern=on 5pt off 2.5pt, x filter/.code={\pgfmathparse{\pgfmathresult +3}\pgfmathresult}]
table {%
0.001 4.00332035668664
0.0005 4.01306709118163
0.000333333333333333 3.96582695873658
0.00025 4.06193591826303
0.0002 3.94948700144925
0.000166666666666667 3.96978076277121
0.000142857142857143 4.12131553998926
0.000125 4.2473913733047
0.000111111111111111 4.01472673783758
0.0001 3.83730954622572
9.09090909090909e-05 3.2084389751291
8.33333333333333e-05 4.10594994739305
7.69230769230769e-05 4.00667565035
7.14285714285714e-05 4.00105041056668
6.66666666666667e-05 4.00084764991346
6.25e-05 3.57102453669805
5.88235294117647e-05 3.9785923532205
5.55555555555556e-05 5.06094991684658
5.26315789473684e-05 5.62987940976471
5e-05 5.2531824233838
3.33333333333333e-05 3.73856354938931
2.5e-05 2.53407680770128
2e-05 1.69268404695405
1.33333333333333e-05 0.765223060490001
1e-05 0.430664489801599
6.66666666666667e-06 0.191159074867562
};

\addplot [very thick, steelblue31119180, dash pattern=on 3.5pt off 0.5pt, x filter/.code={\pgfmathparse{\pgfmathresult +3}\pgfmathresult}]
table {%
0.001 4.00149082772768
0.0005 4.1333581261857
0.000333333333333333 4.00272904673226
0.00025 4.0116515591049
0.0002 3.84021485301396
0.000166666666666667 4.35166805909276
0.000142857142857143 3.75927288067588
0.000125 3.82393348622147
0.000111111111111111 4.0155353753298
0.0001 4.01465053532475
9.09090909090909e-05 4.06667498479278
8.33333333333333e-05 4.00492530889255
7.69230769230769e-05 4.36805030195266
7.14285714285714e-05 4.57950199933317
6.66666666666667e-05 4.93927532344454
6.25e-05 5.27393273595648
5.88235294117647e-05 5.57106240651764
5.55555555555556e-05 5.61870183257519
5.26315789473684e-05 5.51270993494116
5e-05 5.29986286090306
3.33333333333333e-05 3.99999616337183
2.5e-05 3.57093552125802
2e-05 2.74672013050798
1.33333333333333e-05 1.39397845042133
1e-05 0.809321630787861
6.66666666666667e-06 0.365922316161506
};

\addplot [very thick, darkorange25512714, dash pattern=on 5.5pt off 0.5pt, x filter/.code={\pgfmathparse{\pgfmathresult +3}\pgfmathresult}]
table {%
0.001 3.76556539536615
0.0005 4.36300290619118
0.000333333333333333 4.12340879919882
0.00025 3.07882585456268
0.0002 3.30818989148196
0.000166666666666667 2.38480318533574
0.000142857142857143 1.00050659459663
0.000125 0.940933213483284
0.000111111111111111 0.426180188768672
0.0001 0.674736254802783
9.09090909090909e-05 0.484874334621512
8.33333333333333e-05 0.282456497532165
7.69230769230769e-05 0.153040787696044
7.14285714285714e-05 0.150827729541816
6.66666666666667e-05 0.117359393299361
6.25e-05 0.131719766688122
5.88235294117647e-05 0.0783275884179961
5.55555555555556e-05 0.0656067517870598
5.26315789473684e-05 0.0569050496770939
5e-05 0.0453464932984029
3.33333333333333e-05 0.00996643819528166
2.5e-05 0.0038948678876822
2e-05 0.00179157052092477
1.33333333333333e-05 0.000406182690750903
1e-05 0.000135440197307964
6.66666666666667e-06 2.77911261242837e-05
};

\addplot [very thick, mediumpurple148103189, dash pattern=on 1.5pt off 0.5pt on 3pt off 0.5pt, x filter/.code={\pgfmathparse{\pgfmathresult +3}\pgfmathresult}]
table {%
0.001 3.95695651852345
0.0005 3.97338940558971
0.000333333333333333 4.00778828767396
0.00025 4.04667676891754
0.0002 4.00078448205719
0.000166666666666667 3.70084429413842
0.000142857142857143 4.02442795515759
0.000125 3.95123061202946
0.000111111111111111 4.90845552794999
0.0001 4.12637467340067
9.09090909090909e-05 4.01683397450332
8.33333333333333e-05 4.0451178949415
7.69230769230769e-05 3.99907212431411
7.14285714285714e-05 3.73496984813179
6.66666666666667e-05 3.67798734531223
6.25e-05 3.23326071654821
5.88235294117647e-05 2.50602706184943
5.55555555555556e-05 1.99819322259551
5.26315789473684e-05 1.63980302709773
5e-05 1.35705892326483
3.33333333333333e-05 0.286553660202277
2.5e-05 0.0919892072157959
2e-05 0.0379021360292505
1.33333333333333e-05 0.00751332674309211
1e-05 0.0023797956509516
6.66666666666667e-06 0.000470405474085901
};

\addplot [very thick, sienna1408675, dash pattern=on 1.5pt off 0.5pt on 6pt off 0.5pt, x filter/.code={\pgfmathparse{\pgfmathresult +3}\pgfmathresult}]
table {%
0.001 4.06348828197809
0.0005 3.9209092507249
0.000333333333333333 4.55759148364618
0.00025 4.02911068587173
0.0002 3.94730890173775
0.000166666666666667 4.01657686671465
0.000142857142857143 4.18347327200844
0.000125 4.1222289916582
0.000111111111111111 3.92294584054728
0.0001 3.22888302057476
9.09090909090909e-05 2.34844034241217
8.33333333333333e-05 1.67240531810144
7.69230769230769e-05 1.20328879183195
7.14285714285714e-05 0.882289086767094
6.66666666666667e-05 0.658584547254092
6.25e-05 0.451957630677245
5.88235294117647e-05 0.339127032201099
5.55555555555556e-05 0.249339732394089
5.26315789473684e-05 0.189322571188004
5e-05 0.142420814660203
3.33333333333333e-05 0.0144199362425502
2.5e-05 0.00354277312803308
2e-05 0.00154866219084941
1.33333333333333e-05 0.000373055806701315
1e-05 0.000128987257259443
6.66666666666667e-06 2.71935564912697e-05
};

\addplot [thick, black, dash pattern=on 0.5pt off 1pt, x filter/.code={\pgfmathparse{\pgfmathresult +3}\pgfmathresult}]
table {%
1e-05 0.95
6.66666666666667e-06 0.633333333333333
};

\addplot [thick, black, dash pattern=on 0.5pt off 2.5pt, x filter/.code={\pgfmathparse{\pgfmathresult +3}\pgfmathresult}]
table {%
1e-05 0.3
6.66666666666667e-06 0.133333333333333
};
\addplot [thick, black, dash pattern=on 0.5pt off 4pt, x filter/.code={\pgfmathparse{\pgfmathresult +3}\pgfmathresult}]
table {%
1e-05 0.0003
6.66666666666667e-06 5.92592592592593e-05
};

\nextgroupplot
[
width=0.5\textwidth, height=0.375\textwidth,
log basis x={10},
log basis y={10},
tick align=outside,
tick pos=left,
x grid style={darkgray176},
xlabel={circuit depth},
xmin=1e3, xmax=1e5,
xmode=log,
xtick style={color=black, font=\footnotesize},
ymin = 1e-5,
ymax = 10,
xmajorgrids,
ymajorgrids,
y grid style={darkgray176},
ymode=log,
ytick={1, 1e-1, 1e-2, 1e-3, 1e-4},
ytick style={color=black},
yticklabels={$10^0$, $10^{-1}$, $10^{-2}$, $10^{-3}$, $10^{-4}$},
xminorgrids,
yminorgrids,
minor ytick={1e-7,1e-6,1e-5,1e-4,1e-3,1e-2,1e-1},
minor xtick={2e3, 3e3, 4e3, 5e3, 6e3, 7e3, 8e3, 9e3, 1e4, 2e4, 3e4, 4e4, 5e4, 6e4, 7e4, 8e4, 9e4},
minor grid style={lightgray230, dotted, line width=0.1pt}
]

\addplot [very thick, orchid227119194, dash pattern=on 1.5pt off 0.5pt]
table {%
120 4.02237706091091
240 3.99568698713503
360 3.8507338677303
480 4.04797044970421
600 3.75959914671215
720 3.95470460079529
840 3.85212843053157
960 3.9330189895024
1080 4.0034254465864
1200 4.02691456126561
1320 4.33531637366652
1440 4.05225762850164
1560 3.92417037669378
1680 3.99059809235397
1800 4.03073773654815
1920 4.06524146921515
2040 4.02175096366423
2160 4.42022106924293
2280 4.66983928322516
2400 4.53234047019901
3600 3.81160034352927
4800 2.77340188245801
6000 1.98857603130444
9000 1.10569965778719
12000 0.775344968141025
18000 0.505065556230199
24000 0.382195375682143
48000 0.198518404373704
60000 0.160530396509942
};
\addlegendentry{$\Theta_1(t_n)$, \cref{eq: theta_1 final}, starting point}
\addplot [very thick, crimson2143940, dash pattern=on 5pt off 2.5pt]
table {%
164 4.00332035668664
324 4.01306709118163
484 3.96582695873658
644 4.06193591826303
804 3.94948700144925
964 3.96978076277121
1124 4.12131553998926
1284 4.2473913733047
1444 4.01472673783758
1604 3.83730954622572
1764 3.2084389751291
1924 4.10594994739305
2084 4.00667565035
2244 4.00105041056668
2404 4.00084764991346
2564 3.57102453669805
2724 3.9785923532205
2884 5.06094991684658
3044 5.62987940976471
3204 5.2531824233838
4804 3.73856354938931
6404 2.53407680770128
8004 1.69268404695405
12004 0.765223060490001
16004 0.430664489801599
24004 0.191159074867562
32004 0.107443619089402
48004 0.0477214327681691
64004 0.0268366183072929
};
\addlegendentry{$\Theta_1(t_n)$, \cref{eq: theta_1 final}, midpoint}
\addplot [very thick, steelblue31119180, dash pattern=on 3.5pt off 0.5pt]
table {%
164 4.00149082772768
324 4.1333581261857
484 4.00272904673226
644 4.0116515591049
804 3.84021485301396
964 4.35166805909276
1124 3.75927288067588
1284 3.82393348622147
1444 4.0155353753298
1604 4.01465053532475
1764 4.06667498479278
1924 4.00492530889255
2084 4.36805030195266
2244 4.57950199933317
2404 4.93927532344454
2564 5.27393273595648
2724 5.57106240651764
2884 5.61870183257519
3044 5.51270993494116
3204 5.29986286090306
4804 3.99999616337183
6404 3.57093552125802
8004 2.74672013050798
12004 1.39397845042133
16004 0.809321630787861
24004 0.365922316161506
32004 0.206835016229288
48004 0.0922120933231947
64004 0.0519217278234325
};
\addlegendentry{$\Theta_1(t_n)$, \cref{eq: theta_1 final}, \texttt{scipy}}
\addplot [very thick, darkorange25512714, dash pattern=on 5.5pt off 0.5pt]
table {%
484 3.76556539536615
964 4.36300290619118
1444 4.12340879919882
1924 3.07882585456268
2404 3.30818989148196
2884 2.38480318533574
3364 1.00050659459663
3844 0.940933213483284
4324 0.426180188768672
4804 0.674736254802783
5284 0.484874334621512
5764 0.282456497532165
6244 0.153040787696044
6724 0.150827729541816
7204 0.117359393299361
7684 0.131719766688122
8164 0.0783275884179961
8644 0.0656067517870598
9124 0.0569050496770939
9604 0.0453464932984029
14404 0.00996643819528166
19204 0.0038948678876822
24004 0.00179157052092477
36004 0.000406182690750903
48004 0.000135440197307964
72004 2.77911261242837e-05
};
\addlegendentry{$\Theta_2(t_n)$, \cref{eq: Theta_2 homo}, \texttt{scipy}}
\addplot [very thick, mediumpurple148103189, dash pattern=on 1.5pt off 0.5pt on 3pt off 0.5pt]
table {%
484 3.95695651852345
964 3.97338940558971
1444 4.00778828767396
1924 4.04667676891754
2404 4.00078448205719
2884 3.70084429413842
3364 4.02442795515759
3844 3.95123061202946
4324 4.90845552794999
4804 4.12637467340067
5284 4.01683397450332
5764 4.0451178949415
6244 3.99907212431411
6724 3.73496984813179
7204 3.67798734531223
7684 3.23326071654821
8164 2.50602706184943
8644 1.99819322259551
9124 1.63980302709773
9604 1.35705892326483
14404 0.286553660202277
19204 0.0919892072157959
24004 0.0379021360292505
36004 0.00751332674309211
48004 0.0023797956509516
72004 0.000470405474085901
};
\addlegendentry{$\Theta_2(t_n)$, \cref{eq: Theta_2 homo}, GL2}
\addplot [very thick, sienna1408675, dash pattern=on 1.5pt off 0.5pt on 6pt off 0.5pt]
table {%
484 4.06348828197809
964 3.9209092507249
1444 4.55759148364618
1924 4.02911068587173
2404 3.94730890173775
2884 4.01657686671465
3364 4.18347327200844
3844 4.1222289916582
4324 3.92294584054728
4804 3.22888302057476
5284 2.34844034241217
5764 1.67240531810144
6244 1.20328879183195
6724 0.882289086767094
7204 0.658584547254092
7684 0.451957630677245
8164 0.339127032201099
8644 0.249339732394089
9124 0.189322571188004
9604 0.142420814660203
14404 0.0144199362425502
19204 0.00354277312803308
24004 0.00154866219084941
36004 0.000373055806701315
48004 0.000128987257259443
72004 2.71935564912697e-05
};
\addlegendentry{$\Theta_2(t_n)$, \cref{eq: Theta_2 homo}, GL3}
\addplot [thick, black, dash pattern=on 0.5pt off 1pt]
table {%
66666.6666666667 0.21
28571.4285714286 0.49
};
\addlegendentry{$\order(h)$}
\addplot [thick, black, dash pattern=on 0.5pt off 2.5pt]
table {%
66666.6666666667 0.018
28571.4285714286 0.098
};
\addlegendentry{$\order(h^2)$}
\addplot [thick, black, dash pattern=on 0.5pt off 4pt]
table {%
66666.6666666667 8.1e-05
28571.4285714286 0.002401
};
\addlegendentry{$\order(h^4)$}

\end{groupplot}

\end{tikzpicture} 

%% file: fig_bandwidth_size_depth_best_method.tex
\begin{tikzpicture}

\begin{groupplot}[group style={group size=2 by 1,
                    horizontal sep=1.5cm,
                    },
                    title style={yshift=5mm},
                    legend pos=south west,
                    legend style={
                            at={(-0.15,-0.45)},
                            anchor=south,
                            legend columns=3,
                            column sep = 10pt,
                          }
                          ]
\nextgroupplot
[
tick pos=left,
x grid style={darkgray176},
xlabel={$\Delta F$ (kHz)},
xmin=-2.9895, xmax=62.9995,
xtick style={color=black},
y grid style={darkgray176},
ylabel={time-step size (ms)},
ymode = log,
log basis y={10},
ymin=-0.00931841302136318, ymax=1,
ytick style={color=black},
xmajorgrids,
ymajorgrids,
xminorgrids,
yminorgrids,
minor grid style={lightgray230, dotted, line width=0.1pt}
]
\addplot [very thick, darkorange25512714, dotted]
table {%
0.01 0.4
2.009 0.0775193798449612
4.008 0.0675675675675676
6.008 0.0680272108843537
8.007 0.0709219858156028
10.006 0.0763358778625954
12.006 0.0826446280991736
14.005 0.0854700854700855
16.004 0.0869565217391304
18.004 0.0735294117647059
20.003 0.072992700729927
22.002 0.072463768115942
24.002 0.0699300699300699
26.001 0.0689655172413793
28 0.0684931506849315
30 0.0602409638554217
32.142 0.0609756097560976
34.285 0.0568181818181818
36.428 0.0531914893617021
38.571 0.0574712643678161
40.714 0.054945054945055
42.857 0.0598802395209581
45 0.0558659217877095
47.142 0.0512820512820513
49.285 0.0606060606060606
51.428 0.0561797752808989
53.571 0.0584795321637427
55.714 0.0552486187845304
57.857 0.0543478260869565
60 0.0510204081632653
};
\addplot [very thick, darkorange25512714, dashed]
table {%
0.01 0.144927536231884
2.009 0.0423728813559322
4.008 0.0371747211895911
6.008 0.0367647058823529
8.007 0.0371747211895911
10.006 0.037593984962406
12.006 0.0383141762452107
14.005 0.0392156862745098
16.004 0.0404858299595142
18.004 0.0378787878787879
20.003 0.037593984962406
22.002 0.0392156862745098
24.002 0.0404858299595142
26.001 0.0374531835205993
28 0.0350877192982456
30 0.0328947368421053
32.142 0.0323624595469256
34.285 0.0308641975308642
36.428 0.0293255131964809
38.571 0.0288184438040346
40.714 0.0278551532033426
42.857 0.0271739130434783
45 0.0261780104712042
47.142 0.0254452926208651
49.285 0.0248756218905473
51.428 0.0240384615384615
53.571 0.0232018561484919
55.714 0.0232558139534884
57.857 0.0224719101123596
60 0.0215982721382289
};
\addplot [very thick, darkorange25512714]
table {%
0.01 0.0699300699300699
2.009 0.0239808153477218
4.008 0.0210970464135021
6.008 0.0207900207900208
8.007 0.0208333333333333
10.006 0.0208333333333333
12.006 0.0208768267223382
14.005 0.0209643605870021
16.004 0.0211416490486258
18.004 0.0202020202020202
20.003 0.0198412698412698
22.002 0.0199600798403194
24.002 0.0198412698412698
26.001 0.0190114068441065
28 0.0179533213644524
30 0.0170357751277683
32.142 0.0164473684210526
34.285 0.0157232704402516
36.428 0.0149476831091181
38.571 0.014367816091954
40.714 0.0138888888888889
42.857 0.0133333333333333
45 0.0128040973111396
47.142 0.0123609394313968
49.285 0.0119760479041916
51.428 0.0115340253748558
53.571 0.0111482720178372
55.714 0.01085776330076
57.857 0.0105042016806723
60 0.0101729399796541
};%
\draw [dashed, ultra thick] (axis cs:30,0.001) -- (axis cs:30,1);

\nextgroupplot
[
tick align=outside,
tick pos=left,
x grid style={darkgray176},
xlabel={$\Delta F$ (kHz)},
xmin=-2.9895, xmax=62.9995,
xtick style={color=black},
y grid style={darkgray176},
ylabel={circuit depth},
ymin=-1134.05, ymax=51029.05,
ytick style={color=black},
xmajorgrids,
ymajorgrids,
xminorgrids,
yminorgrids,
minor grid style={lightgray230, dotted, line width=0.1pt}
]
\addplot [very thick, darkorange25512714, dotted]
table {%
0.01 1237
2.009 6385
4.008 7325.5
6.008 7276
8.007 6979
10.006 6484
12.006 5989
14.005 5791
16.004 5692
18.004 6731.5
20.003 6781
22.002 6830.5
24.002 7078
26.001 7177
28 7226.5
30 8216.5
32.142 8117.5
34.285 8711.5
36.428 9305.5
38.571 8612.5
40.714 9008.5
42.857 8266
45 8860
47.142 9652
49.285 8167
51.428 8810.5
53.571 8464
55.714 8959
57.857 9107.5
60 9701.5
};
\addlegendentry{accuracy $10^{-1}$}
\addplot [very thick, darkorange25512714, dashed]
table {%
0.01 3415
2.009 11681.5
4.008 13315
6.008 13463.5
8.007 13315
10.006 13166.5
12.006 12919
14.005 12622
16.004 12226
18.004 13067.5
20.003 13166.5
22.002 12622
24.002 12226
26.001 13216
28 14107
30 15047.5
32.142 15295
34.285 16037.5
36.428 16879
38.571 17176
40.714 17770
42.857 18215.5
45 18908.5
47.142 19453
49.285 19898.5
51.428 20591.5
53.571 21334
55.714 21284.5
57.857 22027
60 22918
};
\addlegendentry{accuracy $10^{-2}$}
\addplot [very thick, darkorange25512714]
table {%
0.01 7078
2.009 20641
4.008 23462.5
6.008 23809
8.007 23759.5
10.006 23759.5
12.006 23710
14.005 23611
16.004 23413
18.004 24502
20.003 24947.5
22.002 24799
24.002 24947.5
26.001 26036.5
28 27571
30 29056
32.142 30095.5
34.285 31481.5
36.428 33115
38.571 34451.5
40.714 35639.5
42.857 37124.5
45 38659
47.142 40045
49.285 41332
51.428 42916
53.571 44401
55.714 45589
57.857 47123.5
60 48658
};
\addlegendentry{accuracy $10^{-3}$}
\draw [dashed, ultra thick] (axis cs:30,\pgfkeysvalueof{/pgfplots/ymin}) -- (axis cs:30,\pgfkeysvalueof{/pgfplots/ymax});
\addlegendimage{black, dashed, ultra thick}
\addlegendentry{\Cref{ex:isotropic homonuclear}, pulse~\ref{case: chirped} -- bandwidth $\Delta F$ $= 30$ kHz \hspace*{-5cm}}
\end{groupplot}

\end{tikzpicture}

%% file: fig_chirped_CFTO.tex
\begin{tikzpicture}

\begin{groupplot}[
    group style={group size=2 by 1,
    horizontal sep=1.5cm,
    y descriptions at=edge left},
    legend pos=south west,
    legend style={
        at={(-0.15,-0.5)},
        anchor=south,
        legend columns=4,
        column sep=10pt,
        font=\footnotesize
      }
]

\nextgroupplot
[
width=0.5\textwidth, height=0.375\textwidth,
log basis x={10},
log basis y={10},
tick align=outside,
tick pos=left,
x grid style={darkgray176},
xlabel={time-step size, $h$ (ms)},
xmin=1e-2*0.5, xmax=0.5,
xmode=log,
xtick style={color=black, font=\footnotesize},
ymin = 1e-5,
ymax = 10,
xmajorgrids,
ymajorgrids,
y grid style={darkgray176},
ylabel={error at $T=10$ ms},
ymode=log,
ytick={1, 1e-1, 1e-2, 1e-3, 1e-4},
ytick style={color=black},
xminorgrids,
yminorgrids,
minor ytick={1e-7,1e-6,1e-5,1e-4,1e-3,1e-2,1e-1},
minor xtick={2e-2, 3e-2, 4e-2, 5e-2, 6e-2, 7e-2, 8e-2, 9e-2, 1e-1, 2e-1, 3e-1, 4e-1, 5e-1, 6e-1, 7e-1, 8e-1, 9e-1, 1, 2, 3, 4, 5, 6, 7, 8, 9, 10},
minor grid style={lightgray230, dotted, line width=0.1pt}
]

\addplot [very thick, darkorange25512714, dash pattern=on 5.5pt off 0.5pt, x filter/.code={\pgfmathparse{\pgfmathresult +3}\pgfmathresult}]
table {%
0.001 3.76556539536615
0.0005 4.36300290619118
0.000333333333333333 4.12340879919882
0.00025 3.07882585456268
0.0002 3.30818989148196
0.000166666666666667 2.38480318533574
0.000142857142857143 1.00050659459663
0.000125 0.940933213483284
0.000111111111111111 0.426180188768672
0.0001 0.674736254802783
9.09090909090909e-05 0.484874334621512
8.33333333333333e-05 0.282456497532165
7.69230769230769e-05 0.153040787696044
7.14285714285714e-05 0.150827729541816
6.66666666666667e-05 0.117359393299361
6.25e-05 0.131719766688122
5.88235294117647e-05 0.0783275884179961
5.55555555555556e-05 0.0656067517870598
5.26315789473684e-05 0.0569050496770939
5e-05 0.0453464932984029
3.33333333333333e-05 0.00996643819528166
2.5e-05 0.0038948678876822
2e-05 0.00179157052092477
1.33333333333333e-05 0.000406182690750903
1e-05 0.000135440197307964
6.66666666666667e-06 2.77911261242837e-05
};

\addplot [very thick, forestgreen4416044, dash pattern=on 1.5pt off 0.5pt on 1.5pt off 0.5pt on 1.5pt off 0.5pt on 4.5pt off 0.5pt, x filter/.code={\pgfmathparse{\pgfmathresult +3}\pgfmathresult}]
table {%
0.001 3.91026506896007
0.0005 4.06284802694207
0.000333333333333333 4.00523101826701
0.00025 4.25985662632848
0.0002 4.7826222016767
0.000166666666666667 5.52383024148232
0.000142857142857143 5.31524114103071
0.000125 4.90123478161349
0.000111111111111111 4.49123916734154
0.0001 4.15780672015274
9.09090909090909e-05 4.0177009029124
8.33333333333333e-05 4.00049475434984
7.69230769230769e-05 3.96819817791166
7.14285714285714e-05 3.83101102065978
6.66666666666667e-05 3.58616527702362
6.25e-05 3.27041837683571
5.88235294117647e-05 2.92287872781456
5.55555555555556e-05 2.57905660113611
5.26315789473684e-05 2.25733407086867
5e-05 1.96685390326883
3.33333333333333e-05 0.524611490544049
2.5e-05 0.1820340951719
2e-05 0.0777484062538406
1.33333333333333e-05 0.0160038941142163
1e-05 0.0051373375245193
6.66666666666667e-06 0.00102531150522364
};

\addplot [very thick, darkgray176, dash pattern=on 1.5pt off 0.5pt on 1.5pt off 0.5pt on 1.5pt off 0.5pt on 4.5pt off 0.5pt on 4.5pt off 0.5pt on 4.5pt off 0.5pt, x filter/.code={\pgfmathparse{\pgfmathresult +3}\pgfmathresult}]
table {%
0.001 4.51477423760619
0.0005 4.09313456678381
0.000333333333333333 4.10406336825042
0.00025 3.94678954066106
0.0002 3.98139950682402
0.000166666666666667 3.86973892540263
0.000142857142857143 4.00249966585883
0.000125 4.42682876831882
0.000111111111111111 3.99529568909024
0.0001 4.02970410872492
9.09090909090909e-05 4.01395065549515
8.33333333333333e-05 4.2065838785905
7.69230769230769e-05 4.11219835972693
7.14285714285714e-05 4.97268179949666
6.66666666666667e-05 4.98204134471097
6.25e-05 4.02066515188326
5.88235294117647e-05 3.9354592852699
5.55555555555556e-05 3.85165223649933
5.26315789473684e-05 3.57099096971804
5e-05 3.90070249036844
3.33333333333333e-05 2.80874673709684
2.5e-05 1.26755079219161
2e-05 0.603779999668127
1.33333333333333e-05 0.135575250726158
1e-05 0.0448350899409418
6.66666666666667e-06 0.00913858358657821
};

\addplot [thick, black, dash pattern=on 0.5pt off 4pt, x filter/.code={\pgfmathparse{\pgfmathresult +3}\pgfmathresult}]
table {%
2e-05 0.16
6.66666666666667e-06 0.00197530864197531
};

\nextgroupplot
[
width=0.5\textwidth, height=0.375\textwidth,
log basis x={10},
log basis y={10},
tick align=outside,
tick pos=left,
x grid style={darkgray176},
xlabel={circuit depth},
xmin=1e3, xmax=1e5,
xmode=log,
xtick style={color=black, font=\footnotesize},
ymin = 1e-5,
ymax = 10,
xmajorgrids,
ymajorgrids,
y grid style={darkgray176},
ymode=log,
ytick={1, 1e-1, 1e-2, 1e-3, 1e-4},
yticklabels={$10^0$, $10^{-1}$, $10^{-2}$, $10^{-3}$, $10^{-4}$},
ytick style={color=black},
xminorgrids,
yminorgrids,
minor ytick={1e-7,1e-6,1e-5,1e-4,1e-3,1e-2,1e-1,1},
minor xtick={2e3, 3e3, 4e3, 5e3, 6e3, 7e3, 8e3, 9e3, 1e4, 2e4, 3e4, 4e4, 5e4, 6e4, 7e4, 8e4, 9e4},
minor grid style={lightgray230, dotted, line width=0.1pt}
]

\addplot [very thick, darkorange25512714, dash pattern=on 5.5pt off 0.5pt]
table {%
484 3.76556539536615
964 4.36300290619118
1444 4.12340879919882
1924 3.07882585456268
2404 3.30818989148196
2884 2.38480318533574
3364 1.00050659459663
3844 0.940933213483284
4324 0.426180188768672
4804 0.674736254802783
5284 0.484874334621512
5764 0.282456497532165
6244 0.153040787696044
6724 0.150827729541816
7204 0.117359393299361
7684 0.131719766688122
8164 0.0783275884179961
8644 0.0656067517870598
9124 0.0569050496770939
9604 0.0453464932984029
14404 0.00996643819528166
19204 0.0038948678876822
24004 0.00179157052092477
36004 0.000406182690750903
48004 0.000135440197307964
72004 2.77911261242837e-05
};
\addlegendentry{$\Theta_2(t_n)$, \cref{eq: Theta_2 homo}, \texttt{scipy}}

\addplot [very thick, forestgreen4416044, dash pattern=on 1.5pt off 0.5pt on 1.5pt off 0.5pt on 1.5pt off 0.5pt on 4.5pt off 0.5pt]
table {%
1004 3.91026506896007
2004 4.06284802694207
3004 4.00523101826701
4004 4.25985662632848
5004 4.7826222016767
6004 5.52383024148232
7004 5.31524114103071
8004 4.90123478161349
9004 4.49123916734154
10004 4.15780672015274
11004 4.0177009029124
12004 4.00049475434984
13004 3.96819817791166
14004 3.83101102065978
15004 3.58616527702362
16004 3.27041837683571
17004 2.92287872781456
18004 2.57905660113611
19004 2.25733407086867
20004 1.96685390326883
30004 0.524611490544049
40004 0.1820340951719
50004 0.0777484062538406
75004 0.0160038941142163
};
\addlegendentry{CF42, \cref{eq: CF42 form}, \texttt{scipy}}

\addplot [very thick, darkgray176, dash pattern=on 1.5pt off 0.5pt on 1.5pt off 0.5pt on 1.5pt off 0.5pt on 4.5pt off 0.5pt on 4.5pt off 0.5pt on 4.5pt off 0.5pt]
table {%
484 4.51477423760619
964 4.09313456678381
1444 4.10406336825042
1924 3.94678954066106
2404 3.98139950682402
2884 3.86973892540263
3364 4.00249966585883
3844 4.42682876831882
4324 3.99529568909024
4804 4.02970410872492
5284 4.01395065549515
5764 4.2065838785905
6244 4.11219835972693
6724 4.97268179949666
7204 4.98204134471097
7684 4.02066515188326
8164 3.9354592852699
8644 3.85165223649933
9124 3.57099096971804
9604 3.90070249036844
14404 2.80874673709684
19204 1.26755079219161
24004 0.603779999668127
36004 0.135575250726158
48004 0.0448350899409418
72004 0.00913858358657821
};
\addlegendentry{autonomized Yoshida, \cref{eq: time-orderng}}

\addplot [thick, black, dash pattern=on 0.5pt off 4pt]
table {%
75000 5.12e-05
42857.1428571429 0.0004802
};
\addlegendentry{$\order(h^4)$}

\end{groupplot}

\end{tikzpicture} 

%% file: fig_bandwidth_depth_speedup_different_methods.tex
\begin{tikzpicture}

\begin{groupplot}[
    group style={group size=3 by 1, y descriptions at=edge left},
    width=0.33\textwidth, %
    height=0.33\textheight, %
    xlabel={$\Delta F$ (kHz)},
    xtick style={color=black},
    y grid style={darkgray176},
    ytick style={color=black},
    yticklabel style={font=\small},
    xticklabel style={font=\small},
    legend pos=south west,
    legend style={
        at={(-0.8,-0.4)},
        anchor=south,
        legend columns=3,column sep=10pt,
        font=\footnotesize
      }
]

\nextgroupplot[
tick align=outside,
tick pos=left,
title={accuracy $10^{-1}$},
x grid style={darkgray176},
xlabel={$\Delta F$ (kHz)},
xmin=-1/1000*2989.5, xmax=1/1000*62999.5,
xtick style={color=black},
y grid style={darkgray176},
ymin=0.107047887361378, ymax=20,
ytick style={color=black},
ytick={1,5,10,15,20},
xmajorgrids,
ymajorgrids,
xminorgrids,
yminorgrids,
minor ytick={1,2,3,4,6,7, 8,9,11,12,13,14,16,17,18,19},
minor xtick={1e1,3e1,5e1},
minor grid style={lightgray230, dotted, line width=0.1pt}
]
\addplot [very thick, crimson2143940, dash pattern=on 5pt off 2.5pt, x filter/.code={\pgfmathparse{\pgfmathresult / 1000}\pgfmathresult}]
table {%
10 0.335548172757475
2009 0.71078114912847
4008 1.06978052898143
6008 1.41926345609065
8007 1.74660366213822
10006 2.16211061665607
12006 2.65450791465933
14005 3.07259786476868
16004 3.47067342505431
18004 3.2363747703613
20003 3.50942249240122
22002 3.7833433916717
24002 3.94234129295282
26001 4.17748420448018
28000 4.43867655447804
30000 4.16106372303061
32142 4.48806500761808
34285 4.44723142451491
36428 4.40629153743908
38571 5.02489229296314
40714 5.06407322654462
42857 5.79201995012469
45000 5.66449511400651
47142 5.44254592054677
49285 6.71024734982332
51428 6.48245203556387
53571 7.02630297126157
55714 6.89047399907961
57857 7.03349932095971
60000 6.84275393115172
};
\addplot [very thick, steelblue31119180, dash pattern=on 3.5pt off 0.5pt, x filter/.code={\pgfmathparse{\pgfmathresult / 1000}\pgfmathresult}]
table {%
10 0.335548172757475
2009 0.749515816655907
4008 1.23635340461452
6008 1.7342776203966
8007 2.21913762551683
10006 2.815638906548
12006 3.51617343427392
14005 4.12313167259786
16004 4.69876900796524
18004 4.40722596448255
20003 4.80790273556231
22002 5.20277610138805
24002 5.44030285381479
26001 5.78116025272832
28000 6.15687393040502
30000 5.78073256397391
32142 6.24733367191468
34285 6.19640321817321
36428 6.14842711564023
38571 7.02010531354715
40714 7.07597254004577
42857 8.10224438902743
45000 7.92787342950209
47142 7.61939342161469
49285 9.40181726400808
51428 9.08610201216659
53571 9.84948855333658
55714 9.67004141739531
57857 9.8727931190584
60000 9.60858478538037
};
\addplot [very thick, forestgreen4416044, dash pattern=on 1.5pt off 0.5pt on 1.5pt off 0.5pt on 1.5pt off 0.5pt on 4.5pt off 0.5pt, x filter/.code={\pgfmathparse{\pgfmathresult / 1000}\pgfmathresult}]
table {%
10 1.08305647840532
2009 1.03357004519045
4008 0.900956668542487
6008 1.27535410764873
8007 1.86119314825753
10006 2.51176096630642
12006 3.25258086717137
14005 3.9153024911032
16004 4.54453294713975
18004 4.31781996325781
20003 4.75744680851064
22002 5.19070609535305
24002 5.46068724519511
26001 5.83055715106261
28000 6.23274386765545
30000 5.87104867034621
32142 6.36160487557136
34285 6.31850449597728
36428 6.28090385467435
38571 7.19291527046434
40714 7.25446224256293
42857 8.31720698254364
45000 8.14378780828292
47142 7.83895771038018
49285 9.67995961635538
51428 9.35938231165185
53571 10.1563565513882
55714 9.97514956281638
57857 10.1860570393843
60000 9.92392690182746
};
\draw [dashed, ultra thick] (axis cs:30,\pgfkeysvalueof{/pgfplots/ymin}) -- (axis cs:30,\pgfkeysvalueof{/pgfplots/ymax});
\addplot [very thick, darkorange25512714, dotted]
table {%
-2.9895 1
62.9995 1
};

\nextgroupplot[
tick align=outside,
tick pos=left,
title={accuracy $10^{-2}$},
x grid style={darkgray176},
xlabel={$\Delta F$ (kHz)},
xmin=-1/1000*2989.5, xmax=1/1000*62999.5,
xtick style={color=black},
y grid style={darkgray176},
ymin=0.107047887361378, ymax=20,
ytick style={color=black},
ytick={1,5,10,15,20},
yticklabels={1, 5, 10, 15, 20},
yticklabels={1, 5, 10, 15, 20},
xmajorgrids,
ymajorgrids,
xminorgrids,
yminorgrids,
minor ytick={1,2,3,4,6,7, 8,9,11,12,13,14,16,17,18,19},
minor xtick={1e1,3e1,5e1},
minor grid style={lightgray230, dotted, line width=0.1pt}
]
\addplot [very thick, crimson2143940, dash pattern=on 5pt off 2.5pt, x filter/.code={\pgfmathparse{\pgfmathresult / 1000}\pgfmathresult}]
table {%
10 0.392038600723764
2009 1.23296858453936
4008 1.86838030349954
6008 2.4333843797856
8007 2.90523381851967
10006 3.3764484810523
12006 3.89562719438238
14005 4.46292061417837
16004 5.11467116357504
18004 5.27264121173872
20003 5.71531475101785
22002 6.47141457040183
24002 7.21382799325464
26001 7.17160686427457
28000 7.18766442560655
30000 7.18141956700466
32142 7.53221892693448
34285 7.63718179480586
36428 7.68067432201319
38571 7.96662665066026
40714 8.11626827570202
42857 8.3117500565995
45000 8.39018538713195
47142 8.53360186559254
49285 8.7040414507772
51428 8.7652713799319
53571 8.80746182099362
55714 9.1681844603759
57857 9.19322224302565
60000 9.15691920100774
};
\addplot [very thick, steelblue31119180, dash pattern=on 3.5pt off 0.5pt, x filter/.code={\pgfmathparse{\pgfmathresult / 1000}\pgfmathresult}]
table {%
10 0.406513872135103
2009 1.30215319449347
4008 2.15949210281821
6008 2.97733537519142
8007 3.69433261071539
10006 4.40119010335108
12006 5.16597510373444
14005 5.99444625939236
16004 6.9318718381113
18004 7.19248974439886
20003 7.84246789852803
22002 8.91898072525318
24002 9.97537942664418
26001 9.94227769110764
28000 9.98801520023385
30000 9.99643738010414
32142 10.5022917228363
34285 10.6590383131911
36428 10.7336916687027
38571 11.1464585834334
40714 11.3597122302158
42857 11.645234321938
45000 11.764667393675
47142 11.9730761076956
49285 12.2215544041451
51428 12.3142399359103
53571 12.3759907210516
55714 12.8899438093393
57857 12.9288522748549
60000 12.8805110671225
};
\addplot [very thick, forestgreen4416044, dash pattern=on 1.5pt off 0.5pt on 1.5pt off 0.5pt on 1.5pt off 0.5pt on 4.5pt off 0.5pt, x filter/.code={\pgfmathparse{\pgfmathresult / 1000}\pgfmathresult}]
table {%
10 1.05669481302774
2009 1.02400282386163
4008 0.906162898730257
6008 1.25604900459418
8007 1.76556209352741
10006 2.23958659567805
12006 2.72933290775614
14005 3.25089839921594
16004 3.8283305227656
18004 4.01577784790155
20003 4.42405261509552
22002 5.07219862789938
24002 5.70860033726813
26001 5.71794071762871
28000 5.76615024846536
30000 5.78953137845985
32142 6.10029657589647
34285 6.20365132424788
36428 6.26093330075739
38571 6.51284513805522
40714 6.64330471106985
42857 6.82046637989586
45000 6.89770992366412
47142 7.02268390926436
49285 7.17637305699482
51428 7.23532946124574
53571 7.27353566595786
55714 7.58108893625266
57857 7.61112151282531
60000 7.58520784596005
};
\draw [dashed, ultra thick] (axis cs:30,\pgfkeysvalueof{/pgfplots/ymin}) -- (axis cs:30,\pgfkeysvalueof{/pgfplots/ymax});
\addplot [very thick, darkorange25512714, dotted]
table {%
-2.9895 1
62.9995 1
};

\nextgroupplot[
tick align=outside,
tick pos=left,
title={accuracy $10^{-3}$},
x grid style={darkgray176},
xlabel={$\Delta F$ (kHz)},
xmin=-1/1000*2989.5, xmax=1/1000*62999.5,
xtick style={color=black},
y grid style={darkgray176},
ymin=0.107047887361378, ymax=20,
ytick style={color=black},
ytick={1,5,10,15,20},
yticklabels={1, 5, 10, 15, 20},
xmajorgrids,
ymajorgrids,
xminorgrids,
yminorgrids,
minor ytick={1,2,3,4,6,7, 8,9,11,12,13,14,16,17,18,19},
minor xtick={1e1,3e1,5e1},
minor grid style={lightgray230, dotted, line width=0.1pt},
legend style={fill opacity=0.8, draw opacity=1, text opacity=1, draw=lightgray204, font=\small}
]
\addplot [very thick, crimson2143940, dash pattern=on 5pt off 2.5pt, x filter/.code={\pgfmathparse{\pgfmathresult / 1000}\pgfmathresult}]
table {%
10 0.585323238206174
2009 2.26113886113886
4008 3.47073299349622
6008 4.49696864715053
8007 5.30966846033675
10006 6.0560666550946
12006 6.78744129413811
14005 7.5907423580786
16004 8.47577946098291
18004 8.91247264770241
20003 9.55810877831046
22002 10.4322301679694
24002 11.1940816663911
26001 11.5236812925709
28000 11.6411368735976
30000 11.7707594038325
32142 12.1119638207483
34285 12.3082667365387
36428 12.3837339643791
38571 12.5642284209266
40714 12.7995602360838
42857 12.8969003444062
45000 12.9782353568761
47142 13.1157688742404
49285 13.2570601736354
51428 13.3048534358482
53571 13.3886669763121
55714 13.5421152628246
57857 13.5944857768053
60000 13.6438925150462
};
\addlegendentry{$\Theta_1(t_n)$, \cref{eq: theta_1 final}, midpoint}
\addplot [very thick, steelblue31119180, dash pattern=on 3.5pt off 0.5pt, x filter/.code={\pgfmathparse{\pgfmathresult / 1000}\pgfmathresult}]
table {%
10 0.615608619685498
2009 2.38741258741259
4008 4.014238003164
6008 5.50857439805993
8007 6.75733379621593
10006 7.89810796736678
12006 9.00278309271178
14005 10.1968558951965
16004 11.4900475603312
18004 12.1610839925938
20003 13.1124152752521
22002 14.375020788292
24002 15.4777649198215
26001 15.9760811024869
28000 16.1754674644727
30000 16.3850958126331
32142 16.8832396875428
34285 17.1755535176209
36428 17.3058911446008
38571 17.5779959296061
40714 17.9239671334336
42857 18.0807688034663
45000 18.2085778299371
47142 18.4044700793079
49285 18.6190000997904
51428 18.6957232099952
53571 18.8158848118904
55714 19.0425223921107
57857 19.1220131291028
60000 19.1964906332118
};
\addlegendentry{$\Theta_1(t_n)$, \cref{eq: theta_1 final}, \texttt{scipy}}
\addplot [very thick, forestgreen4416044, dash pattern=on 1.5pt off 0.5pt on 1.5pt off 0.5pt on 1.5pt off 0.5pt on 4.5pt off 0.5pt, x filter/.code={\pgfmathparse{\pgfmathresult / 1000}\pgfmathresult}]
table {%
10 1.04892253931275
2009 1.02917082917083
4008 0.909825979961329
6008 1.29066343322363
8007 1.79673667766013
10006 2.24370768963722
12006 2.6745520960167
14005 3.11371179039301
16004 3.57600845516998
18004 3.83369803063457
20003 4.17854190775335
22002 4.61516713786795
24002 5.00099189948752
26001 5.1878663076192
28000 5.26940912490651
30000 5.35500354861604
32142 5.54008496642456
34285 5.64666579326608
36428 5.7013326690746
38571 5.8004309828804
40714 5.9195694942715
42857 5.98000222197534
45000 6.02805931932146
47142 6.09753836646411
49285 6.17463327013272
51428 6.2038443056223
53571 6.24718996748723
55714 6.32642721433095
57857 6.35676148796499
60000 6.38306349071798
};
\addlegendentry{CF42, \cref{eq: CF42 form},\texttt{scipy}}
\draw [dashed, ultra thick] (axis cs:30,\pgfkeysvalueof{/pgfplots/ymin}) -- (axis cs:30,\pgfkeysvalueof{/pgfplots/ymax});

\addlegendimage{very thick, darkorange25512714, dotted}
\addlegendentry{$\Theta_2(t_n)$, \cref{eq: Theta_2 homo}, \texttt{scipy}}

\addlegendimage{black, dashed, ultra thick}
\addlegendentry{\Cref{ex:isotropic homonuclear}, pulse \cref{case: chirped} -- bandwidth $\Delta F$ $= 30$ kHz \hspace*{-2cm}}

\addplot [very thick, darkorange25512714, dotted]
table {%
-2.9895 1
62.9995 1
};

\end{groupplot}

\node[anchor=south, rotate=90, xshift= -10ex, yshift= 5ex] at ($(group c1r1.west)!0.5!(group c1r1.north west)$) {circuit compression ratio};

\end{tikzpicture} 

%% file: fig_bandwidth_depth_speedup_different_integrals.tex
\begin{tikzpicture}

\begin{groupplot}[
    group style={group size=3 by 1, y descriptions at=edge left},
    width=0.33\textwidth,
    height=0.28\textheight,
    xlabel={$\Delta F$ (kHz)},
    xtick style={color=black},
    y grid style={darkgray176},
    ytick style={color=black},
    yticklabel style={font=\small},
    xticklabel style={font=\small},
    legend style={
        at={(-0.8,-0.5)},
        anchor=south,
        legend columns=3,column sep=10pt,
        font=\footnotesize
      }
]

\nextgroupplot[
tick align=outside,
tick pos=left,
title={accuracy $10^{-1}$},
x grid style={darkgray176},
xlabel={$\Delta F$ (kHz)},
xtick style={color=black},
y grid style={darkgray176},
ymin=0.507047887361378, ymax=5,
ytick style={color=black},
ytick={1,5},
xmajorgrids,
ymajorgrids,
xminorgrids,
yminorgrids,
minor ytick={1,2,3,4,6,7, 8,9,11,12,13,14,16,17,18,19},
minor xtick={1e1,3e1,5e1},
minor grid style={lightgray230, dotted, line width=0.1pt}
]
\addplot [very thick, mediumpurple148103189, dash pattern=on 1.5pt off 0.5pt on 3pt off 0.5pt, x filter/.code={\pgfmathparse{\pgfmathresult / 1000}\pgfmathresult}]
table {%
10 1
2009 1
4008 1
6008 1
8007 0.950354609929078
10006 0.870229007633588
12006 1.1404958677686
14005 1.43589743589744
16004 1.7304347826087
18004 1.67647058823529
20003 1.86861313868613
22002 2.05797101449275
24002 2.17482517482518
26001 2.33103448275862
28000 2.5
30000 2.35542168674699
32142 2.5609756097561
34285 2.54545454545455
36428 2.53723404255319
38571 2.90229885057471
40714 2.92857142857143
42857 3.36526946107784
45000 3.29608938547486
47142 3.16923076923077
49285 3.92121212121212
51428 3.79213483146067
53571 4.11111111111111
55714 4.04419889502762
57857 4.1304347826087
60000 4.02040816326531
};
\addplot [very thick, sienna1408675, dash pattern=on 1.5pt off 0.5pt on 6pt off 0.5pt, x filter/.code={\pgfmathparse{\pgfmathresult / 1000}\pgfmathresult}]
table {%
10 1
2009 1
4008 1
6008 1
8007 1
10006 1
12006 0.950413223140496
14005 0.965811965811966
16004 1.00869565217391
18004 1.03676470588235
20003 1.12408759124088
22002 1.16666666666667
24002 1.21678321678322
26001 1.28275862068966
28000 1.37671232876712
30000 1.28313253012048
32142 1.39024390243902
34285 1.375
36428 1.3563829787234
38571 1.55172413793103
40714 1.56593406593407
42857 1.78443113772455
45000 1.74301675977654
47142 1.67692307692308
49285 2.06666666666667
51428 1.99438202247191
53571 2.16374269005848
55714 2.11602209944751
57857 2.16847826086957
60000 2.10204081632653
};
\draw [dashed, ultra thick] (axis cs:30,\pgfkeysvalueof{/pgfplots/ymin}) -- (axis cs:30,\pgfkeysvalueof{/pgfplots/ymax});
\addplot [very thick, darkorange25512714, dotted]
table {%
-2.9895 1
62.9995 1
};

\nextgroupplot[
tick align=outside,
tick pos=left,
title={accuracy $10^{-2}$},
x grid style={darkgray176},
xlabel={$\Delta F$ (kHz)},
xtick style={color=black},
y grid style={darkgray176},
ymin=0.507047887361378, ymax=5,
ytick style={color=black},
ytick={1,5},
yticklabels={1, 5},
xmajorgrids,
ymajorgrids,
xminorgrids,
yminorgrids,
minor ytick={1,2,3,4,6,7, 8,9,11,12,13,14,16,17,18,19},
minor xtick={1e1,3e1,5e1},
minor grid style={lightgray230, dotted, line width=0.1pt}
]
\addplot [very thick, mediumpurple148103189, dash pattern=on 1.5pt off 0.5pt on 3pt off 0.5pt, x filter/.code={\pgfmathparse{\pgfmathresult / 1000}\pgfmathresult}]
table {%
10 1
2009 1
4008 1.00371747211896
6008 1
8007 0.977695167286245
10006 0.902255639097744
12006 0.881226053639847
14005 1.12156862745098
16004 1.40080971659919
18004 1.51893939393939
20003 1.70300751879699
22002 1.97254901960784
24002 2.23886639676113
26001 2.25093632958801
28000 2.28070175438596
30000 2.29605263157895
32142 2.42394822006472
34285 2.46913580246914
36428 2.49560117302053
38571 2.59942363112392
40714 2.65459610027855
42857 2.72826086956522
45000 2.75916230366492
47142 2.8117048346056
49285 2.87562189054726
51428 2.90144230769231
53571 2.91647331786543
55714 3.04186046511628
57857 3.05393258426966
60000 3.04535637149028
};
\addplot [very thick, sienna1408675, dash pattern=on 1.5pt off 0.5pt on 6pt off 0.5pt, x filter/.code={\pgfmathparse{\pgfmathresult / 1000}\pgfmathresult}]
table {%
10 1
2009 1
4008 1
6008 1
8007 1
10006 1
12006 0.996168582375479
14005 0.988235294117647
16004 0.979757085020243
18004 0.977272727272727
20003 0.969924812030075
22002 0.976470588235294
24002 1.02834008097166
26001 1.04494382022472
28000 1.05964912280702
30000 1.0625
32142 1.10032362459547
34285 1.1141975308642
36428 1.12023460410557
38571 1.14985590778098
40714 1.16991643454039
42857 1.1929347826087
45000 1.19633507853403
47142 1.22137404580153
49285 1.24129353233831
51428 1.25
53571 1.25522041763341
55714 1.30232558139535
57857 1.30561797752809
60000 1.29805615550756
};
\draw [dashed, ultra thick] (axis cs:30,\pgfkeysvalueof{/pgfplots/ymin}) -- (axis cs:30,\pgfkeysvalueof{/pgfplots/ymax});
\addplot [very thick, darkorange25512714, dotted]
table {%
-2.9895 1
62.9995 1
};

\nextgroupplot[
tick align=outside,
tick pos=left,
title={accuracy $10^{-3}$},
x grid style={darkgray176},
xlabel={$\Delta F$ (kHz)},
xtick style={color=black},
y grid style={darkgray176},
ymin=0.507047887361378, ymax=5,
ytick style={color=black},
ytick={1,5},
yticklabels={1, 5},
xmajorgrids,
ymajorgrids,
xminorgrids,
yminorgrids,
minor ytick={1,2,3,4,6,7, 8,9,11,12,13,14,16,17,18,19},
minor xtick={1e1,3e1,5e1},
minor grid style={lightgray230, dotted, line width=0.1pt},
legend style={fill opacity=0.8, draw opacity=1, text opacity=1, draw=lightgray204, font=\small}
]
\addplot [very thick, mediumpurple148103189, dash pattern=on 1.5pt off 0.5pt on 3pt off 0.5pt, x filter/.code={\pgfmathparse{\pgfmathresult / 1000}\pgfmathresult}]
table {%
10 1
2009 1.00239808153477
4008 1.00210970464135
6008 1.002079002079
8007 0.979166666666667
10006 0.920833333333333
12006 0.855949895615866
14005 1.0482180293501
16004 1.28752642706131
18004 1.43636363636364
20003 1.59722222222222
22002 1.78443113772455
24002 1.95039682539683
26001 2.03422053231939
28000 2.0754039497307
30000 2.11584327086882
32142 2.19243421052632
34285 2.24056603773585
36428 2.2660687593423
38571 2.30747126436782
40714 2.35694444444444
42857 2.384
45000 2.40460947503201
47142 2.43386897404203
49285 2.46586826347305
51428 2.479815455594
53571 2.49721293199554
55714 2.52985884907709
57857 2.54306722689076
60000 2.55442522889115
};
\addlegendentry{$\Theta_2(t_n)$, \cref{eq: Theta_2 homo}, GL2}
\addplot [very thick, sienna1408675, dash pattern=on 1.5pt off 0.5pt on 6pt off 0.5pt, x filter/.code={\pgfmathparse{\pgfmathresult / 1000}\pgfmathresult}]
table {%
10 1
2009 1
4008 1
6008 1
8007 1
10006 1
12006 1
14005 0.9979035639413
16004 0.995771670190275
18004 0.993939393939394
20003 0.992063492063492
22002 0.978043912175649
24002 0.968253968253968
26001 0.963878326996198
28000 0.96588868940754
30000 0.964224872231687
32142 0.952302631578947
34285 0.952830188679245
36428 0.95067264573991
38571 0.941091954022989
40714 0.934722222222222
42857 0.929333333333333
45000 0.921895006402049
47142 0.919653893695921
49285 0.908982035928144
51428 0.901960784313726
53571 0.907469342251951
55714 0.894679695982628
57857 0.892857142857143
60000 0.899287894201424
};
\addlegendentry{$\Theta_2(t_n)$, \cref{eq: Theta_2 homo}, GL3}
\draw [dashed, ultra thick] (axis cs:30,\pgfkeysvalueof{/pgfplots/ymin}) -- (axis cs:30,\pgfkeysvalueof{/pgfplots/ymax});
\addplot [very thick, darkorange25512714, dotted]
table {%
-2.9895 1
62.9995 1
};
\addlegendentry{$\Theta_2(t_n)$, \cref{eq: Theta_2 homo}, \texttt{scipy}}

\addlegendimage{black, dashed, ultra thick}
\addlegendentry{\Cref{ex:isotropic homonuclear}, pulse (ii) -- bandwidth $\Delta F$ $= 30$ kHz \hspace*{-2cm}}

\end{groupplot}

\node[anchor=south, rotate=90, xshift= -12ex, yshift= 5ex] at ($(group c1r1.west)!0.5!(group c1r1.north west)$) {circuit compression ratio};

\end{tikzpicture}

%% file: fig_Hin_norm_depth_speedup_different_methods_log.tex
\begin{tikzpicture}

\begin{groupplot}[
    group style={group size=3 by 1, y descriptions at=edge left},
    width=0.33\textwidth, %
    height=0.3\textheight, %
    xlabel={$\|\hi\|$},
    xtick style={color=black},
    xmode = log,
    log basis x={10},
    y grid style={darkgray176},
    ytick style={color=black},
    yticklabel style={font=\small},
    xticklabel style={font=\small},
    legend style={
        at={(-0.8,-0.5)},
        anchor=south,
        legend columns=3,column sep=10pt,
        font=\footnotesize
      }
]

\nextgroupplot[
tick align=outside,
tick pos=left,
title={accuracy $10^{-1}$},
x grid style={darkgray176},
xmin=-2989.5, xmax=156414.464418882,
xtick style={color=black},
y grid style={darkgray176},
ymin=0, ymax=20,
ytick style={color=black},
ytick={1,5,10,15,20},
xmajorgrids,
ymajorgrids,
xminorgrids,
yminorgrids,
minor ytick={1,2,3,4,6,7, 8,9,11,12,13,14,16,17,18,19},
minor grid style={lightgray230, dotted, line width=0.1pt}
]
\path [draw=crimson2143940, fill=crimson2143940, opacity=0.3]
(axis cs:103.300315108378,4.18134781485098)
--(axis cs:103.300315108378,4.15758352307507)
--(axis cs:201.705394079863,4.14405389135252)
--(axis cs:393.852293269657,4.01382883068842)
--(axis cs:769.040558490716,3.57057657630911)
--(axis cs:1501.63751921786,3.27517509428746)
--(axis cs:2932.11484651493,3.13234872417927)
--(axis cs:5725.28147647193,2.46729032558157)
--(axis cs:11179.2510528,1.82699577664189)
--(axis cs:21828.735340807,1.53794270163852)
--(axis cs:42623.0419487404,1.4807220941837)
--(axis cs:83226.2463491358,1.67157521000668)
--(axis cs:83226.2463491358,2.25055475030105)
--(axis cs:83226.2463491358,2.25055475030105)
--(axis cs:42623.0419487404,1.97338135455967)
--(axis cs:21828.735340807,1.8681597509432)
--(axis cs:11179.2510528,2.23031465129696)
--(axis cs:5725.28147647193,3.11704393979938)
--(axis cs:2932.11484651493,3.20944814717527)
--(axis cs:1501.63751921786,3.38172411279564)
--(axis cs:769.040558490716,3.7298428949199)
--(axis cs:393.852293269657,4.08463402587083)
--(axis cs:201.705394079863,4.16562149369214)
--(axis cs:103.300315108378,4.18134781485098)
--cycle;
\path [draw=steelblue31119180, fill=steelblue31119180, opacity=0.3]
(axis cs:103.300315108378,5.81092892393073)
--(axis cs:103.300315108378,5.7779030632389)
--(axis cs:201.705394079863,5.75910057854068)
--(axis cs:393.852293269657,5.57618738193191)
--(axis cs:769.040558490716,4.96040187334587)
--(axis cs:1501.63751921786,4.5500171543815)
--(axis cs:2932.11484651493,4.35100868339956)
--(axis cs:5725.28147647193,3.41973562385897)
--(axis cs:11179.2510528,2.50870177561362)
--(axis cs:21828.735340807,2.00730488557429)
--(axis cs:42623.0419487404,1.64615215819967)
--(axis cs:83226.2463491358,1.69446747686676)
--(axis cs:83226.2463491358,2.2635791170334)
--(axis cs:83226.2463491358,2.2635791170334)
--(axis cs:42623.0419487404,2.13276974227697)
--(axis cs:21828.735340807,2.50915537253279)
--(axis cs:11179.2510528,3.08101565332814)
--(axis cs:5725.28147647193,4.32656582825846)
--(axis cs:2932.11484651493,4.4581041215132)
--(axis cs:1501.63751921786,4.69803973272864)
--(axis cs:769.040558490716,5.18166164142918)
--(axis cs:393.852293269657,5.67455307030722)
--(axis cs:201.705394079863,5.7890736422045)
--(axis cs:103.300315108378,5.81092892393073)
--cycle;

\path [draw=forestgreen4416044, fill=forestgreen4416044, opacity=0.3]
(axis cs:103.300315108378,5.89966848927665)
--(axis cs:103.300315108378,5.86613828572307)
--(axis cs:201.705394079863,5.84704866546677)
--(axis cs:393.852293269657,5.66330774724288)
--(axis cs:769.040558490716,5.03790142522524)
--(axis cs:1501.63751921786,4.62110500159864)
--(axis cs:2932.11484651493,4.41745358823932)
--(axis cs:5725.28147647193,3.47116068071873)
--(axis cs:11179.2510528,2.54135336306211)
--(axis cs:21828.735340807,1.93311039223016)
--(axis cs:42623.0419487404,1.15468348295474)
--(axis cs:83226.2463491358,1.02745796967053)
--(axis cs:83226.2463491358,1.07608453622991)
--(axis cs:83226.2463491358,1.07608453622991)
--(axis cs:42623.0419487404,1.64206498398017)
--(axis cs:21828.735340807,2.48927108412543)
--(axis cs:11179.2510528,3.12048244306756)
--(axis cs:5725.28147647193,4.38720536310492)
--(axis cs:2932.11484651493,4.52618449681787)
--(axis cs:1501.63751921786,4.77144023198141)
--(axis cs:769.040558490716,5.26261807710813)
--(axis cs:393.852293269657,5.76321026609364)
--(axis cs:201.705394079863,5.877479452272)
--(axis cs:103.300315108378,5.89966848927665)
--cycle;
\addplot [very thick, darkorange25512714, dotted]
table {%
103.300315108378 1
201.705394079863 1
393.852293269657 1
769.040558490716 1
1501.63751921786 1
2932.11484651493 1
5725.28147647193 1
11179.2510528 1
21828.735340807 1
42623.0419487404 1
83226.2463491358 1
};
\addplot [very thick, crimson2143940, dash pattern=on 5pt off 2.5pt]
table {%
103.300315108378 4.16946566896303
201.705394079863 4.15483769252233
393.852293269657 4.04923142827963
769.040558490716 3.65020973561451
1501.63751921786 3.32844960354155
2932.11484651493 3.17089843567727
5725.28147647193 2.79216713269047
11179.2510528 2.02865521396942
21828.735340807 1.70305122629086
42623.0419487404 1.72705172437168
83226.2463491358 1.96106498015387
};
\addplot [very thick, steelblue31119180, dash pattern=on 3.5pt off 0.5pt]
table {%
103.300315108378 5.79441599358482
201.705394079863 5.77408711037259
393.852293269657 5.62537022611957
769.040558490716 5.07103175738752
1501.63751921786 4.62402844355507
2932.11484651493 4.40455640245638
5725.28147647193 3.87315072605872
11179.2510528 2.79485871447088
21828.735340807 2.25823012905354
42623.0419487404 1.88946095023832
83226.2463491358 1.97902329695008
};
\addplot [very thick, forestgreen4416044, dash pattern=on 1.5pt off 0.5pt on 1.5pt off 0.5pt on 1.5pt off 0.5pt on 4.5pt off 0.5pt]
table {%
103.300315108378 5.88290338749986
201.705394079863 5.86226405886938
393.852293269657 5.71325900666826
769.040558490716 5.15025975116669
1501.63751921786 4.69627261679002
2932.11484651493 4.4718190425286
5725.28147647193 3.92918302191183
11179.2510528 2.83091790306484
21828.735340807 2.2111907381778
42623.0419487404 1.39837423346746
83226.2463491358 1.05177125295022
};
\draw [dashed, ultra thick] (axis cs:187.39277589884816,\pgfkeysvalueof{/pgfplots/ymin}) -- (axis cs:187.39277589884816,\pgfkeysvalueof{/pgfplots/ymax});

\nextgroupplot[
tick align=outside,
tick pos=left,
title={accuracy $10^{-2}$},
x grid style={darkgray176},
xmin=-2989.5, xmax=156414.464418882,
xtick style={color=black},
y grid style={darkgray176},
ymin=0, ymax=20,
ytick style={color=black},
ytick={1,5,10,15,20},
yticklabels={1,5,10,15,20},
xmajorgrids,
ymajorgrids,
xminorgrids,
yminorgrids,
minor ytick={1,2,3,4,6,7, 8,9,11,12,13,14,16,17,18,19},
minor grid style={lightgray230, dotted, line width=0.1pt}
]
\path [draw=crimson2143940, fill=crimson2143940, opacity=0.3]
(axis cs:103.300315108378,7.45618877834394)
--(axis cs:103.300315108378,7.33227209875749)
--(axis cs:201.705394079863,7.04785674134584)
--(axis cs:393.852293269657,6.39931703175752)
--(axis cs:769.040558490716,5.69108723161034)
--(axis cs:1501.63751921786,5.29613874893011)
--(axis cs:2932.11484651493,5.05739862528978)
--(axis cs:5725.28147647193,4.18547071079419)
--(axis cs:11179.2510528,3.17000978458524)
--(axis cs:21828.735340807,2.6487517902394)
--(axis cs:42623.0419487404,2.54700679513927)
--(axis cs:83226.2463491358,2.93413920417135)
--(axis cs:83226.2463491358,4.04523242813788)
--(axis cs:83226.2463491358,4.04523242813788)
--(axis cs:42623.0419487404,3.46717660898192)
--(axis cs:21828.735340807,3.20075422115022)
--(axis cs:11179.2510528,3.8266477590644)
--(axis cs:5725.28147647193,5.06689689566177)
--(axis cs:2932.11484651493,5.16773154328504)
--(axis cs:1501.63751921786,5.43103828313905)
--(axis cs:769.040558490716,5.93813750621926)
--(axis cs:393.852293269657,6.60841206150531)
--(axis cs:201.705394079863,7.24980140521781)
--(axis cs:103.300315108378,7.45618877834394)
--cycle;
\path [draw=steelblue31119180, fill=steelblue31119180, opacity=0.3]
(axis cs:103.300315108378,10.3768251001827)
--(axis cs:103.300315108378,10.2051663730231)
--(axis cs:201.705394079863,9.80944072746511)
--(axis cs:393.852293269657,8.90622592836091)
--(axis cs:769.040558490716,7.9218564126046)
--(axis cs:1501.63751921786,7.37161469549741)
--(axis cs:2932.11484651493,7.03687540526319)
--(axis cs:5725.28147647193,5.81133638697323)
--(axis cs:11179.2510528,4.3599435682004)
--(axis cs:21828.735340807,3.45667543168424)
--(axis cs:42623.0419487404,2.84868591290309)
--(axis cs:83226.2463491358,2.97432836247938)
--(axis cs:83226.2463491358,4.06864425218415)
--(axis cs:83226.2463491358,4.06864425218415)
--(axis cs:42623.0419487404,3.72769670956742)
--(axis cs:21828.735340807,4.31153306278947)
--(axis cs:11179.2510528,5.29465675818737)
--(axis cs:5725.28147647193,7.04435043444986)
--(axis cs:2932.11484651493,7.18980343395571)
--(axis cs:1501.63751921786,7.55937071279253)
--(axis cs:769.040558490716,8.26744300535651)
--(axis cs:393.852293269657,9.19717244209849)
--(axis cs:201.705394079863,10.0899488561848)
--(axis cs:103.300315108378,10.3768251001827)
--cycle;

\path [draw=forestgreen4416044, fill=forestgreen4416044, opacity=0.3]
(axis cs:103.300315108378,5.9968951942805)
--(axis cs:103.300315108378,5.89723096066271)
--(axis cs:201.705394079863,5.67186693104536)
--(axis cs:393.852293269657,5.15599000930502)
--(axis cs:769.040558490716,4.5875768523549)
--(axis cs:1501.63751921786,4.2688723460727)
--(axis cs:2932.11484651493,4.0699080722969)
--(axis cs:5725.28147647193,3.36475786247514)
--(axis cs:11179.2510528,2.51973397131807)
--(axis cs:21828.735340807,1.89812353709876)
--(axis cs:42623.0419487404,1.15529134076825)
--(axis cs:83226.2463491358,1.02608794162512)
--(axis cs:83226.2463491358,1.07382219420681)
--(axis cs:83226.2463491358,1.07382219420681)
--(axis cs:42623.0419487404,1.61129914281812)
--(axis cs:21828.735340807,2.44276704669808)
--(axis cs:11179.2510528,3.05876214091165)
--(axis cs:5725.28147647193,4.07193759905447)
--(axis cs:2932.11484651493,4.15786713031274)
--(axis cs:1501.63751921786,4.37760109861174)
--(axis cs:769.040558490716,4.78722867214476)
--(axis cs:393.852293269657,5.32323016359175)
--(axis cs:201.705394079863,5.83458306337526)
--(axis cs:103.300315108378,5.9968951942805)
--cycle;
\addlegendimage{area legend, draw=forestgreen4416044, fill=forestgreen4416044, opacity=0.3}
\addplot [very thick, crimson2143940, dash pattern=on 5pt off 2.5pt]
table {%
103.300315108378 7.39423043855071
201.705394079863 7.14882907328183
393.852293269657 6.50386454663141
769.040558490716 5.8146123689148
1501.63751921786 5.36358851603458
2932.11484651493 5.11256508428741
5725.28147647193 4.62618380322798
11179.2510528 3.49832877182482
21828.735340807 2.92475300569481
42623.0419487404 3.00709170206059
83226.2463491358 3.48968581615462
};
\addplot [very thick, steelblue31119180, dash pattern=on 3.5pt off 0.5pt]
table {%
103.300315108378 10.2909957366029
201.705394079863 9.94969479182497
393.852293269657 9.0516991852297
769.040558490716 8.09464970898055
1501.63751921786 7.46549270414497
2932.11484651493 7.11333941960945
5725.28147647193 6.42784341071154
11179.2510528 4.82730016319389
21828.735340807 3.88410424723685
42623.0419487404 3.28819131123525
83226.2463491358 3.52148630733176
};
\addplot [very thick, forestgreen4416044, dash pattern=on 1.5pt off 0.5pt on 1.5pt off 0.5pt on 1.5pt off 0.5pt on 4.5pt off 0.5pt]
table {%
103.300315108378 5.94706307747161
201.705394079863 5.75322499721031
393.852293269657 5.23961008644839
769.040558490716 4.68740276224983
1501.63751921786 4.32323672234222
2932.11484651493 4.11388760130482
5725.28147647193 3.71834773076481
11179.2510528 2.78924805611486
21828.735340807 2.17044529189842
42623.0419487404 1.38329524179319
83226.2463491358 1.04995506791596
};
\addplot [very thick, darkorange25512714, dotted]
table {%
103.300315108378 1
201.705394079863 1
393.852293269657 1
769.040558490716 1
1501.63751921786 1
2932.11484651493 1
5725.28147647193 1
11179.2510528 1
21828.735340807 1
42623.0419487404 1
83226.2463491358 1
};
\draw [dashed, ultra thick] (axis cs:187.39277589884816,\pgfkeysvalueof{/pgfplots/ymin}) -- (axis cs:187.39277589884816,\pgfkeysvalueof{/pgfplots/ymax});

\nextgroupplot[
tick align=outside,
tick pos=left,
title={accuracy $10^{-3}$},
x grid style={darkgray176},
xmin=-2989.5, xmax=156414.464418882,
xtick style={color=black},
y grid style={darkgray176},
ymin=0, ymax=20,
ytick style={color=black},
ytick={1,5,10,15,20},
yticklabels={1,5,10,15,20},
xmajorgrids,
ymajorgrids,
xminorgrids,
yminorgrids,
minor ytick={1,2,3,4,6,7, 8,9,11,12,13,14,16,17,18,19},
minor grid style={lightgray230, dotted, line width=0.1pt}
]

\path [draw=crimson2143940, fill=crimson2143940, opacity=0.3]
(axis cs:103.300315108378,13.0044772803136)
--(axis cs:103.300315108378,12.8388785634307)
--(axis cs:201.705394079863,12.0452213238435)
--(axis cs:393.852293269657,10.8385704567861)
--(axis cs:769.040558490716,9.74756838222786)
--(axis cs:1501.63751921786,9.13906168578943)
--(axis cs:2932.11484651493,8.76147315638736)
--(axis cs:5725.28147647193,7.33823758271122)
--(axis cs:11179.2510528,5.59487102903913)
--(axis cs:21828.735340807,4.66594661648133)
--(axis cs:42623.0419487404,4.478458330988)
--(axis cs:83226.2463491358,5.19775244755956)
--(axis cs:83226.2463491358,7.21700339978759)
--(axis cs:83226.2463491358,7.21700339978759)
--(axis cs:42623.0419487404,6.12457656980324)
--(axis cs:21828.735340807,5.62954849596089)
--(axis cs:11179.2510528,6.72898550416587)
--(axis cs:5725.28147647193,8.77076641278895)
--(axis cs:2932.11484651493,8.9282008536357)
--(axis cs:1501.63751921786,9.35706440391476)
--(axis cs:769.040558490716,10.1402279752821)
--(axis cs:393.852293269657,11.1605387667186)
--(axis cs:201.705394079863,12.3176222673967)
--(axis cs:103.300315108378,13.0044772803136)
--cycle;
\path [draw=steelblue31119180, fill=steelblue31119180, opacity=0.3]
(axis cs:103.300315108378,18.1069901493944)
--(axis cs:103.300315108378,17.8760642045714)
--(axis cs:201.705394079863,16.767079654168)
--(axis cs:393.852293269657,15.0899435872238)
--(axis cs:769.040558490716,13.5668292495466)
--(axis cs:1501.63751921786,12.7202456479158)
--(axis cs:2932.11484651493,12.192353197196)
--(axis cs:5725.28147647193,10.1906154243963)
--(axis cs:11179.2510528,7.69592608473975)
--(axis cs:21828.735340807,6.08586345576223)
--(axis cs:42623.0419487404,5.01730246989763)
--(axis cs:83226.2463491358,5.26936724523194)
--(axis cs:83226.2463491358,7.26116088569421)
--(axis cs:83226.2463491358,7.26116088569421)
--(axis cs:42623.0419487404,6.57632451602139)
--(axis cs:21828.735340807,7.58665535182373)
--(axis cs:11179.2510528,9.31164820401481)
--(axis cs:5725.28147647193,12.195246837042)
--(axis cs:2932.11484651493,12.4221790223593)
--(axis cs:1501.63751921786,13.0234132937387)
--(axis cs:769.040558490716,14.1108146810863)
--(axis cs:393.852293269657,15.5396598228604)
--(axis cs:201.705394079863,17.1445946834244)
--(axis cs:103.300315108378,18.1069901493944)
--cycle;
\path [draw=forestgreen4416044, fill=forestgreen4416044, opacity=0.3]
(axis cs:103.300315108378,5.77194948276053)
--(axis cs:103.300315108378,5.69905702495589)
--(axis cs:201.705394079863,5.42172299021847)
--(axis cs:393.852293269657,4.92316693276004)
--(axis cs:769.040558490716,4.43601238898997)
--(axis cs:1501.63751921786,4.15795304696319)
--(axis cs:2932.11484651493,3.98569819828905)
--(axis cs:5725.28147647193,3.33330951031442)
--(axis cs:11179.2510528,2.51276158245517)
--(axis cs:21828.735340807,1.88754218230515)
--(axis cs:42623.0419487404,1.15403333557134)
--(axis cs:83226.2463491358,1.02541388312987)
--(axis cs:83226.2463491358,1.07294900378542)
--(axis cs:83226.2463491358,1.07294900378542)
--(axis cs:42623.0419487404,1.59876677919107)
--(axis cs:21828.735340807,2.42925063832573)
--(axis cs:11179.2510528,3.03885520876709)
--(axis cs:5725.28147647193,3.98310675736569)
--(axis cs:2932.11484651493,4.06081475373609)
--(axis cs:1501.63751921786,4.25726686183592)
--(axis cs:769.040558490716,4.61259120257824)
--(axis cs:393.852293269657,5.06419740861365)
--(axis cs:201.705394079863,5.52304520298373)
--(axis cs:103.300315108378,5.77194948276053)
--cycle;
\addplot [very thick, crimson2143940, dash pattern=on 5pt off 2.5pt]
table {%
103.300315108378 12.9216779218722
201.705394079863 12.1814217956201
393.852293269657 10.9995546117524
769.040558490716 9.94389817875496
1501.63751921786 9.24806304485209
2932.11484651493 8.84483700501153
5725.28147647193 8.05450199775008
11179.2510528 6.1619282666025
21828.735340807 5.14774755622111
42623.0419487404 5.30151745039562
83226.2463491358 6.20737792367358
};
\addlegendentry{$\Theta_1(t_n)$, \cref{eq: theta_1 final}, midpoint}
\addplot [very thick, steelblue31119180, dash pattern=on 3.5pt off 0.5pt]
table {%
103.300315108378 17.9915271769829
201.705394079863 16.9558371687962
393.852293269657 15.3148017050421
769.040558490716 13.8388219653164
1501.63751921786 12.8718294708273
2932.11484651493 12.3072661097777
5725.28147647193 11.1929311307191
11179.2510528 8.50378714437728
21828.735340807 6.83625940379298
42623.0419487404 5.79681349295951
83226.2463491358 6.26526406546307
};
\addlegendentry{$\Theta_1(t_n)$, \cref{eq: theta_1 final}, \texttt{scipy}}
\addplot [very thick, forestgreen4416044, dash pattern=on 1.5pt off 0.5pt on 1.5pt off 0.5pt on 1.5pt off 0.5pt on 4.5pt off 0.5pt]
table {%
103.300315108378 5.73550325385821
201.705394079863 5.4723840966011
393.852293269657 4.99368217068685
769.040558490716 4.5243017957841
1501.63751921786 4.20760995439955
2932.11484651493 4.02325647601257
5725.28147647193 3.65820813384006
11179.2510528 2.77580839561113
21828.735340807 2.15839641031544
42623.0419487404 1.37640005738121
83226.2463491358 1.04918144345764
};
\addlegendentry{CF42, \cref{eq: CF42 form}, \texttt{scipy}}
\addplot [very thick, darkorange25512714, dotted]
table {%
103.300315108378 1
201.705394079863 1
393.852293269657 1
769.040558490716 1
1501.63751921786 1
2932.11484651493 1
5725.28147647193 1
11179.2510528 1
21828.735340807 1
42623.0419487404 1
83226.2463491358 1
};
\addlegendentry{$\Theta_2(t_n)$, \cref{eq: Theta_2 homo}, \texttt{scipy}}
\draw [dashed, ultra thick] (axis cs:187.39277589884816,\pgfkeysvalueof{/pgfplots/ymin}) -- (axis cs:187.39277589884816,\pgfkeysvalueof{/pgfplots/ymax});
\addlegendimage{black, dashed, ultra thick}
\addlegendentry{\Cref{ex:isotropic homonuclear}, pulse~\ref{case: chirped} -- $\|\hi\| \approx 187.4$\hspace*{-2cm}}
\end{groupplot}

\node[anchor=south, rotate=90, xshift= -10ex, yshift= 5ex] at ($(group c1r1.west)!0.5!(group c1r1.north west)$) {circuit compression ratio};
\end{tikzpicture}

%% file: fig_Hin_norm_depth_speedup_different_integrals.tex
\begin{tikzpicture}

\begin{groupplot}[
    group style={group size=3 by 1, y descriptions at=edge left},
    width=0.33\textwidth, %
    height=0.3\textheight, %
    xlabel={$\|\hi\|$},
    xtick style={color=black},
    xmode = log,
    log basis x={10},
    y grid style={darkgray176},
    ytick style={color=black},
    yticklabel style={font=\small},
    xticklabel style={font=\small},
    legend style={
        at={(-0.8,-0.5)},
        anchor=south,
        legend columns=3,column sep=10pt,
        font=\footnotesize
      }
]

\nextgroupplot[
tick align=outside,
tick pos=left,
title={accuracy $10^{-1}$},
x grid style={darkgray176},
xmin=-2989.5, xmax=156414.464418882,
xtick style={color=black},
y grid style={darkgray176},
ymin=0.8, ymax=3,
ytick style={color=black},
ytick={1,3},
xmajorgrids,
ymajorgrids,
xminorgrids,
yminorgrids,
minor ytick={1,2,3,4,6,7, 8,9,11,12,13,14,16,17,18,19},
minor grid style={lightgray230, dotted, line width=0.1pt}
]
\path [draw=mediumpurple148103189, fill=mediumpurple148103189, opacity=0.3]
(axis cs:103.300315108378,2.36690954731517)
--(axis cs:103.300315108378,2.35345068147879)
--(axis cs:201.705394079863,2.34578825602627)
--(axis cs:393.852293269657,2.27203722084569)
--(axis cs:769.040558490716,2.02102502834044)
--(axis cs:1501.63751921786,1.85644348317705)
--(axis cs:2932.11484651493,1.77112975483684)
--(axis cs:5725.28147647193,1.38494383423672)
--(axis cs:11179.2510528,1.0395076040128)
--(axis cs:21828.735340807,0.962338326100934)
--(axis cs:42623.0419487404,0.978680012382784)
--(axis cs:83226.2463491358,0.998510292845163)
--(axis cs:83226.2463491358,1.00008281397041)
--(axis cs:83226.2463491358,1.00008281397041)
--(axis cs:42623.0419487404,0.998514013676091)
--(axis cs:21828.735340807,1.01638010129976)
--(axis cs:11179.2510528,1.2427678204006)
--(axis cs:5725.28147647193,1.75118779650526)
--(axis cs:2932.11484651493,1.81286791664493)
--(axis cs:1501.63751921786,1.91326787594544)
--(axis cs:769.040558490716,2.11121407457148)
--(axis cs:393.852293269657,2.31213636197524)
--(axis cs:201.705394079863,2.35800295672895)
--(axis cs:103.300315108378,2.36690954731517)
--cycle;

\path [draw=sienna1408675, fill=sienna1408675, opacity=0.3]
(axis cs:103.300315108378,1.28939062296197)
--(axis cs:103.300315108378,1.28205881113806)
--(axis cs:201.705394079863,1.28388410943637)
--(axis cs:393.852293269657,1.25614195367168)
--(axis cs:769.040558490716,1.14651492621466)
--(axis cs:1501.63751921786,1.07982638893636)
--(axis cs:2932.11484651493,1.0498942570643)
--(axis cs:5725.28147647193,0.980686674958072)
--(axis cs:11179.2510528,0.989672484245996)
--(axis cs:21828.735340807,0.996429870887626)
--(axis cs:42623.0419487404,0.99939019634876)
--(axis cs:83226.2463491358,0.999882395414803)
--(axis cs:83226.2463491358,1.0000963466878)
--(axis cs:83226.2463491358,1.0000963466878)
--(axis cs:42623.0419487404,1.00038581292721)
--(axis cs:21828.735340807,0.999377413306293)
--(axis cs:11179.2510528,0.999458993241546)
--(axis cs:5725.28147647193,1.02786359860196)
--(axis cs:2932.11484651493,1.06023087914249)
--(axis cs:1501.63751921786,1.10555489986312)
--(axis cs:769.040558490716,1.18891171749008)
--(axis cs:393.852293269657,1.27626385191753)
--(axis cs:201.705394079863,1.29056939319692)
--(axis cs:103.300315108378,1.28939062296197)
--cycle;

\addplot [very thick, mediumpurple148103189, dash pattern=on 1.5pt off 0.5pt on 3pt off 0.5pt]
table {%
103.300315108378 2.36018011439698
201.705394079863 2.35189560637761
393.852293269657 2.29208679141047
769.040558490716 2.06611955145596
1501.63751921786 1.88485567956124
2932.11484651493 1.79199883574088
5725.28147647193 1.56806581537099
11179.2510528 1.1411377122067
21828.735340807 0.989359213700346
42623.0419487404 0.988597013029437
83226.2463491358 0.999296553407788
};
\addplot [very thick, sienna1408675, dash pattern=on 1.5pt off 0.5pt on 6pt off 0.5pt]
table {%
103.300315108378 1.28572471705002
201.705394079863 1.28722675131664
393.852293269657 1.2662029027946
769.040558490716 1.16771332185237
1501.63751921786 1.09269064439974
2932.11484651493 1.0550625681034
5725.28147647193 1.00427513678001
11179.2510528 0.994565738743771
21828.735340807 0.997903642096959
42623.0419487404 0.999888004637986
83226.2463491358 0.999989371051299
};
\addplot [very thick, darkorange25512714, dotted]
table {%
103.300315108378 1
201.705394079863 1
393.852293269657 1
769.040558490716 1
1501.63751921786 1
2932.11484651493 1
5725.28147647193 1
11179.2510528 1
21828.735340807 1
42623.0419487404 1
83226.2463491358 1
};
\draw [dashed, ultra thick] (axis cs:187.39277589884816,\pgfkeysvalueof{/pgfplots/ymin}) -- (axis cs:187.39277589884816,\pgfkeysvalueof{/pgfplots/ymax});

\nextgroupplot[
tick align=outside,
tick pos=left,
title={accuracy $10^{-2}$},
x grid style={darkgray176},
xmin=-2989.5, xmax=156414.464418882,
xtick style={color=black},
y grid style={darkgray176},
ymin=0.8, ymax=3,
ytick style={color=black},
ytick={1,3},
yticklabels={1,3},
xmajorgrids,
ymajorgrids,
xminorgrids,
yminorgrids,
minor ytick={1,2,3,4,6,7, 8,9,11,12,13,14,16,17,18,19},
minor grid style={lightgray230, dotted, line width=0.1pt}
]
\path [draw=mediumpurple148103189, fill=mediumpurple148103189, opacity=0.3]
(axis cs:103.300315108378,2.37609585004291)
--(axis cs:103.300315108378,2.33659582122019)
--(axis cs:201.705394079863,2.24963404476169)
--(axis cs:393.852293269657,2.0447369987346)
--(axis cs:769.040558490716,1.8192695271313)
--(axis cs:1501.63751921786,1.69423645436888)
--(axis cs:2932.11484651493,1.61530863016165)
--(axis cs:5725.28147647193,1.32896023049868)
--(axis cs:11179.2510528,1.02904295481607)
--(axis cs:21828.735340807,0.96842328150063)
--(axis cs:42623.0419487404,0.981157936817225)
--(axis cs:83226.2463491358,0.998260485832195)
--(axis cs:83226.2463491358,1.00011640720462)
--(axis cs:83226.2463491358,1.00011640720462)
--(axis cs:42623.0419487404,0.997460802706571)
--(axis cs:21828.735340807,1.00953362594972)
--(axis cs:11179.2510528,1.20829250812398)
--(axis cs:5725.28147647193,1.60657812709296)
--(axis cs:2932.11484651493,1.64849926800479)
--(axis cs:1501.63751921786,1.73556342990752)
--(axis cs:769.040558490716,1.89846243754701)
--(axis cs:393.852293269657,2.1110770957932)
--(axis cs:201.705394079863,2.31452205092447)
--(axis cs:103.300315108378,2.37609585004291)
--cycle;
\path [draw=sienna1408675, fill=sienna1408675, opacity=0.3]
(axis cs:103.300315108378,1.08888641245254)
--(axis cs:103.300315108378,1.07304347662021)
--(axis cs:201.705394079863,1.0487467119688)
--(axis cs:393.852293269657,1.01068278771321)
--(axis cs:769.040558490716,0.999621722774088)
--(axis cs:1501.63751921786,0.994826953047106)
--(axis cs:2932.11484651493,0.991454060446535)
--(axis cs:5725.28147647193,0.989464293548096)
--(axis cs:11179.2510528,0.997043776295859)
--(axis cs:21828.735340807,0.998927942626103)
--(axis cs:42623.0419487404,0.999749158087776)
--(axis cs:83226.2463491358,0.999882564222806)
--(axis cs:83226.2463491358,1.00010570564945)
--(axis cs:83226.2463491358,1.00010570564945)
--(axis cs:42623.0419487404,1.00018281131977)
--(axis cs:21828.735340807,1.00018265578417)
--(axis cs:11179.2510528,1.00046655855621)
--(axis cs:5725.28147647193,0.996666859867356)
--(axis cs:2932.11484651493,0.994486530048903)
--(axis cs:1501.63751921786,0.997292742927461)
--(axis cs:769.040558490716,1.00220476124418)
--(axis cs:393.852293269657,1.02162866709895)
--(axis cs:201.705394079863,1.08066832433512)
--(axis cs:103.300315108378,1.08888641245254)
--cycle;

\addplot [very thick, mediumpurple148103189, dash pattern=on 1.5pt off 0.5pt on 3pt off 0.5pt]
table {%
103.300315108378 2.35634583563155
201.705394079863 2.28207804784308
393.852293269657 2.0779070472639
769.040558490716 1.85886598233916
1501.63751921786 1.7148999421382
2932.11484651493 1.63190394908322
5725.28147647193 1.46776917879582
11179.2510528 1.11866773147002
21828.735340807 0.988978453725174
42623.0419487404 0.989309369761898
83226.2463491358 0.99918844651841
};
\addplot [very thick, sienna1408675, dash pattern=on 1.5pt off 0.5pt on 6pt off 0.5pt]
table {%
103.300315108378 1.08096494453637
201.705394079863 1.06470751815196
393.852293269657 1.01615572740608
769.040558490716 1.00091324200913
1501.63751921786 0.996059847987283
2932.11484651493 0.992970295247719
5725.28147647193 0.993065576707726
11179.2510528 0.998755167426035
21828.735340807 0.999555299205139
42623.0419487404 0.999965984703775
83226.2463491358 0.99999413493613
};
\addplot [very thick, darkorange25512714, dotted]
table {%
103.300315108378 1
201.705394079863 1
393.852293269657 1
769.040558490716 1
1501.63751921786 1
2932.11484651493 1
5725.28147647193 1
11179.2510528 1
21828.735340807 1
42623.0419487404 1
83226.2463491358 1
};
\draw [dashed, ultra thick] (axis cs:187.39277589884816,\pgfkeysvalueof{/pgfplots/ymin}) -- (axis cs:187.39277589884816,\pgfkeysvalueof{/pgfplots/ymax});

\nextgroupplot[
tick align=outside,
tick pos=left,
title={accuracy $10^{-3}$},
x grid style={darkgray176},
xmin=-2989.5, xmax=156414.464418882,
xtick style={color=black},
y grid style={darkgray176},
ymin=0.8, ymax=3,
ytick style={color=black},
ytick={1,3},
yticklabels={1,3},
xmajorgrids,
ymajorgrids,
xminorgrids,
yminorgrids,
minor ytick={1,2,3,4,6,7, 8,9,11,12,13,14,16,17,18,19},
minor grid style={lightgray230, dotted, line width=0.1pt}
]
\path [draw=mediumpurple148103189, fill=mediumpurple148103189, opacity=0.3]
(axis cs:103.300315108378,2.28026813348645)
--(axis cs:103.300315108378,2.25018128606269)
--(axis cs:201.705394079863,2.14108538312997)
--(axis cs:393.852293269657,1.94482587716893)
--(axis cs:769.040558490716,1.75307919272029)
--(axis cs:1501.63751921786,1.64489272173997)
--(axis cs:2932.11484651493,1.57504471842454)
--(axis cs:5725.28147647193,1.31268241237759)
--(axis cs:11179.2510528,1.02591168556173)
--(axis cs:21828.735340807,0.970244014864574)
--(axis cs:42623.0419487404,0.982487657973597)
--(axis cs:83226.2463491358,0.998162926020076)
--(axis cs:83226.2463491358,1.00013865519489)
--(axis cs:83226.2463491358,1.00013865519489)
--(axis cs:42623.0419487404,0.997292299930709)
--(axis cs:21828.735340807,1.00841465217565)
--(axis cs:11179.2510528,1.19780514489854)
--(axis cs:5725.28147647193,1.56553058015421)
--(axis cs:2932.11484651493,1.60206896533968)
--(axis cs:1501.63751921786,1.68318804497835)
--(axis cs:769.040558490716,1.82178227940286)
--(axis cs:393.852293269657,2.00017915318656)
--(axis cs:201.705394079863,2.18036203706119)
--(axis cs:103.300315108378,2.28026813348645)
--cycle;

\path [draw=sienna1408675, fill=sienna1408675, opacity=0.3]
(axis cs:103.300315108378,0.964145697797337)
--(axis cs:103.300315108378,0.95632635402618)
--(axis cs:201.705394079863,0.966835401301875)
--(axis cs:393.852293269657,0.980371358474207)
--(axis cs:769.040558490716,0.991474363973022)
--(axis cs:1501.63751921786,0.995677997145084)
--(axis cs:2932.11484651493,0.996665681095312)
--(axis cs:5725.28147647193,0.996121242831036)
--(axis cs:11179.2510528,0.998992007675856)
--(axis cs:21828.735340807,0.999545685203258)
--(axis cs:42623.0419487404,0.999899539320092)
--(axis cs:83226.2463491358,0.999983460492706)
--(axis cs:83226.2463491358,1.00001962964666)
--(axis cs:83226.2463491358,1.00001962964666)
--(axis cs:42623.0419487404,1.00006815616458)
--(axis cs:21828.735340807,1.00014083675024)
--(axis cs:11179.2510528,1.00012531618951)
--(axis cs:5725.28147647193,0.999133821441552)
--(axis cs:2932.11484651493,0.997701375802128)
--(axis cs:1501.63751921786,0.996814451544206)
--(axis cs:769.040558490716,0.994101345252558)
--(axis cs:393.852293269657,0.985989659107701)
--(axis cs:201.705394079863,0.968746142315715)
--(axis cs:103.300315108378,0.964145697797337)
--cycle;

\addplot [very thick, mediumpurple148103189, dash pattern=on 1.5pt off 0.5pt on 3pt off 0.5pt]
table {%
103.300315108378 2.26522470977457
201.705394079863 2.16072371009558
393.852293269657 1.97250251517774
769.040558490716 1.78743073606158
1501.63751921786 1.66404038335916
2932.11484651493 1.58855684188211
5725.28147647193 1.4391064962659
11179.2510528 1.11185841523013
21828.735340807 0.989329333520114
42623.0419487404 0.989889978952153
83226.2463491358 0.999150790607484
};
\addlegendentry{$\Theta_2(t_n)$, \cref{eq: Theta_2 homo}, GL2}
\addplot [very thick, sienna1408675, dash pattern=on 1.5pt off 0.5pt on 6pt off 0.5pt]
table {%
103.300315108378 0.960236025911758
201.705394079863 0.967790771808795
393.852293269657 0.983180508790954
769.040558490716 0.99278785461279
1501.63751921786 0.996246224344645
2932.11484651493 0.99718352844872
5725.28147647193 0.997627532136294
11179.2510528 0.999558661932681
21828.735340807 0.99984326097675
42623.0419487404 0.999983847742335
83226.2463491358 1.00000154506968
};
\addlegendentry{$\Theta_2(t_n)$, \cref{eq: Theta_2 homo}, GL3}
\addplot [very thick, darkorange25512714, dotted]
table {%
103.300315108378 1
201.705394079863 1
393.852293269657 1
769.040558490716 1
1501.63751921786 1
2932.11484651493 1
5725.28147647193 1
11179.2510528 1
21828.735340807 1
42623.0419487404 1
83226.2463491358 1
};
\addlegendentry{$\Theta_2(t_n)$, \cref{eq: Theta_2 homo}, \texttt{scipy}}
\draw [dashed, ultra thick] (axis cs:187.39277589884816,\pgfkeysvalueof{/pgfplots/ymin}) -- (axis cs:187.39277589884816,\pgfkeysvalueof{/pgfplots/ymax});
\addlegendimage{black, dashed, ultra thick}
\addlegendentry{\Cref{ex:isotropic homonuclear}, pulse~\ref{case: chirped} -- $\|\hi\| \approx 187.4$}
\end{groupplot}

\node[anchor=south, rotate=90, xshift= -10ex, yshift= 5ex] at ($(group c1r1.west)!0.5!(group c1r1.north west)$) {circuit compression ratio};
\end{tikzpicture} 

%% file: fig_Hin_size_depth_best_method.tex
\begin{tikzpicture}

\begin{groupplot}[group style={group size=2 by 1,
                    horizontal sep=2cm,
                    },
                    title style={yshift=5mm},
                    legend pos=south west,
                    legend style={
                            at={(-0.15,-0.45)},
                            anchor=south,
                            legend columns=3,
                            column sep = 10pt,
                          }
                          ]
\nextgroupplot
[
tick pos=left,
x grid style={darkgray176},
xlabel={$\|\hi\|$},
xtick style={color=black},
y grid style={darkgray176},
ylabel={time-step size (ms)},
xmode = log,
ymode = log,
xmajorgrids,
ymajorgrids,
xmin=50, xmax=170000,
xmode=log,
ymin=0.00012, ymax=0.00634680718424548,
ytick style={color=black}
]
\path [draw=darkorange25512714, fill=darkorange25512714, opacity=0.2]
(axis cs:103.300315108378,0.00605347710310786)
--(axis cs:103.300315108378,0.0060190554513524)
--(axis cs:201.705394079863,0.00599945845531014)
--(axis cs:393.852293269657,0.00581083688195829)
--(axis cs:769.040558490716,0.00516886196506506)
--(axis cs:1501.63751921786,0.00474106273865758)
--(axis cs:2932.11484651493,0.00453204564483022)
--(axis cs:5725.28147647193,0.00357661269464331)
--(axis cs:11179.2510528,0.00264305211606743)
--(axis cs:21828.735340807,0.0020135041336245)
--(axis cs:42623.0419487404,0.00114366983548709)
--(axis cs:83226.2463491358,0.000594028126444931)
--(axis cs:83226.2463491358,0.000967905720167082)
--(axis cs:83226.2463491358,0.000967905720167082)
--(axis cs:42623.0419487404,0.00172867102717035)
--(axis cs:21828.735340807,0.00257807628131842)
--(axis cs:11179.2510528,0.00322108373519143)
--(axis cs:5725.28147647193,0.0045044722761407)
--(axis cs:2932.11484651493,0.00464363977456458)
--(axis cs:1501.63751921786,0.00489536232590503)
--(axis cs:769.040558490716,0.00539952448739509)
--(axis cs:393.852293269657,0.00591339223011571)
--(axis cs:201.705394079863,0.00603069809905101)
--(axis cs:103.300315108378,0.00605347710310786)
--cycle;

\path [draw=darkorange25512714, fill=darkorange25512714, opacity=0.3]
(axis cs:103.300315108378,0.0033992787554262)
--(axis cs:103.300315108378,0.0033427694151934)
--(axis cs:201.705394079863,0.00322082289837681)
--(axis cs:393.852293269657,0.00292942263429027)
--(axis cs:769.040558490716,0.002606403333999)
--(axis cs:1501.63751921786,0.00242529694390441)
--(axis cs:2932.11484651493,0.00231497649251607)
--(axis cs:5725.28147647193,0.00191963012277517)
--(axis cs:11179.2510528,0.00145077683891542)
--(axis cs:21828.735340807,0.00109402878810401)
--(axis cs:42623.0419487404,0.000635369200894102)
--(axis cs:83226.2463491358,0.000332602132704274)
--(axis cs:83226.2463491358,0.000545064636805783)
--(axis cs:83226.2463491358,0.000545064636805783)
--(axis cs:42623.0419487404,0.000940836091545919)
--(axis cs:21828.735340807,0.00140089119233119)
--(axis cs:11179.2510528,0.00174816868878701)
--(axis cs:5725.28147647193,0.00231601437924446)
--(axis cs:2932.11484651493,0.00236501770691599)
--(axis cs:1501.63751921786,0.00248708221320518)
--(axis cs:769.040558490716,0.00271986022571205)
--(axis cs:393.852293269657,0.00302446575328539)
--(axis cs:201.705394079863,0.00331102804973186)
--(axis cs:103.300315108378,0.0033992787554262)
--cycle;

\path [draw=darkorange25512714, fill=darkorange25512714, opacity=0.4]
(axis cs:103.300315108378,0.00179125540729493)
--(axis cs:103.300315108378,0.0017676208060194)
--(axis cs:201.705394079863,0.00170539862947724)
--(axis cs:393.852293269657,0.00156234587051126)
--(axis cs:769.040558490716,0.00141036137789243)
--(axis cs:1501.63751921786,0.0013227200655708)
--(axis cs:2932.11484651493,0.0012679168744304)
--(axis cs:5725.28147647193,0.00106426596715897)
--(axis cs:11179.2510528,0.000809632705541915)
--(axis cs:21828.735340807,0.000608772177162116)
--(axis cs:42623.0419487404,0.000355391220380849)
--(axis cs:83226.2463491358,0.000186875480355494)
--(axis cs:83226.2463491358,0.000306650741092767)
--(axis cs:83226.2463491358,0.000306650741092767)
--(axis cs:42623.0419487404,0.000522627525505279)
--(axis cs:21828.735340807,0.000779657177899538)
--(axis cs:11179.2510528,0.000972023745107545)
--(axis cs:5725.28147647193,0.00126755230083224)
--(axis cs:2932.11484651493,0.00129181526734982)
--(axis cs:1501.63751921786,0.00135431710780723)
--(axis cs:769.040558490716,0.00146563337039651)
--(axis cs:393.852293269657,0.00160477991069856)
--(axis cs:201.705394079863,0.00172972520299879)
--(axis cs:103.300315108378,0.00179125540729493)
--cycle;

\addplot [very thick, darkorange25512714, dotted]
table {%
103.300315108378 0.00603626627723013
201.705394079863 0.00601507827718058
393.852293269657 0.005862114556037
769.040558490716 0.00528419322623008
1501.63751921786 0.0048182125322813
2932.11484651493 0.0045878427096974
5725.28147647193 0.00404054248539201
11179.2510528 0.00293206792562943
21828.735340807 0.00229579020747146
42623.0419487404 0.00143617043132872
83226.2463491358 0.000780966923306007
};
\addplot [very thick, darkorange25512714]
table {%
103.300315108378 0.0033710240853098
201.705394079863 0.00326592547405434
393.852293269657 0.00297694419378783
769.040558490716 0.00266313177985552
1501.63751921786 0.00245618957855479
2932.11484651493 0.00233999709971603
5725.28147647193 0.00211782225100982
11179.2510528 0.00159947276385121
21828.735340807 0.0012474599902176
42623.0419487404 0.000788102646220011
83226.2463491358 0.000438833384755029
};
\addplot [semithick, darkorange25512714]
table {%
103.300315108378 0.00177943810665717
201.705394079863 0.00171756191623801
393.852293269657 0.00158356289060491
769.040558490716 0.00143799737414447
1501.63751921786 0.00133851858668902
2932.11484651493 0.00127986607089011
5725.28147647193 0.0011659091339956
11179.2510528 0.00089082822532473
21828.735340807 0.000694214677530827
42623.0419487404 0.000439009372943064
83226.2463491358 0.000246763110724131
};

\addplot [black, dashed, ultra thick]
table {%
187.4 0.00001
187.400000000000001 0.1
};

\addplot [semithick, dashed]
table {%
43657.470681614 0.000375651629468006
85246.0838974809 0.000192384204061773
};
\addplot [semithick, dashed]
table {%
43657.470681614 0.000687167614880499
85246.0838974809 0.000351922324503244
};
\addplot [semithick, dashed]
table {%
43657.470681614 0.0012369017067849
85246.0838974809 0.000633460184105839
};

\draw [dashed, ultra thick] (axis cs:187.39277589884816,\pgfkeysvalueof{/pgfplots/ymin}) -- (axis cs:187.39277589884816,\pgfkeysvalueof{/pgfplots/ymax});

\nextgroupplot
[
tick align=outside,
tick pos=left,
x grid style={darkgray176},
xlabel={$\|\hi\|$},
xmin=-2.9895, xmax=158966,
xtick style={color=black},
y grid style={darkgray176},
ylabel={circuit depth},
ymin= 3956, ymax=275419.696959967,
ytick style={color=black},
xmajorgrids,
ymajorgrids,
xmode = log,
log basis x={10},
ymode = log,
minor grid style={lightgray230, dotted, line width=0.1pt}
]
\path [draw=darkorange25512714, fill=darkorange25512714, opacity=0.2]
(axis cs:103.300315108378,7978.62741699797)
--(axis cs:103.300315108378,7933.37258300203)
--(axis cs:201.705394079863,7963.21539030917)
--(axis cs:393.852293269657,8120.96017817394)
--(axis cs:769.040558490716,8894.73875190499)
--(axis cs:1501.63751921786,9806.44926855724)
--(axis cs:2932.11484651493,10339.202484496)
--(axis cs:5725.28147647193,10633.9221369511)
--(axis cs:11179.2510528,15024.5773195992)
--(axis cs:21828.735340807,18420.0973301897)
--(axis cs:42623.0419487404,26769.7458865252)
--(axis cs:83226.2463491358,48144.6833171312)
--(axis cs:83226.2463491358,82686.2732046079)
--(axis cs:83226.2463491358,82686.2732046079)
--(axis cs:42623.0419487404,43283.111256332)
--(axis cs:21828.735340807,24092.8626698103)
--(axis cs:11179.2510528,18017.7083946865)
--(axis cs:5725.28147647193,13454.919968312)
--(axis cs:2932.11484651493,10596.797515504)
--(axis cs:1501.63751921786,10131.1507314428)
--(axis cs:769.040558490716,9289.26124809501)
--(axis cs:393.852293269657,8264.63982182606)
--(axis cs:201.705394079863,8004.78460969083)
--(axis cs:103.300315108378,7978.62741699797)
--cycle;

\path [draw=darkorange25512714, fill=darkorange25512714, opacity=0.3]
(axis cs:103.300315108378,14363.7330363768)
--(axis cs:103.300315108378,14124.2669636232)
--(axis cs:201.705394079863,14502.1287539049)
--(axis cs:393.852293269657,15876.2000781861)
--(axis cs:769.040558490716,17655.3479278922)
--(axis cs:1501.63751921786,19300.3695042737)
--(axis cs:2932.11484651493,20299.0269771293)
--(axis cs:5725.28147647193,20663.9174931068)
--(axis cs:11179.2510528,27651.9457939017)
--(axis cs:21828.735340807,33951.7258332526)
--(axis cs:42623.0419487404,49054.3361378427)
--(axis cs:83226.2463491358,85503.4842884275)
--(axis cs:83226.2463491358,147625.211363746)
--(axis cs:83226.2463491358,147625.211363746)
--(axis cs:42623.0419487404,78143.3781478716)
--(axis cs:21828.735340807,44267.4741667474)
--(axis cs:11179.2510528,32856.6256346698)
--(axis cs:5725.28147647193,25085.5561911037)
--(axis cs:2932.11484651493,20739.3730228707)
--(axis cs:1501.63751921786,19798.8304957263)
--(axis cs:769.040558490716,18416.6520721078)
--(axis cs:393.852293269657,16387.7999218139)
--(axis cs:201.705394079863,14905.8712460951)
--(axis cs:103.300315108378,14363.7330363768)
--cycle;

\path [draw=darkorange25512714, fill=darkorange25512714, opacity=0.4]
(axis cs:103.300315108378,27159.5995545651)
--(axis cs:103.300315108378,26800.4004454349)
--(axis cs:201.705394079863,27753.0025125787)
--(axis cs:393.852293269657,29917.3714933225)
--(axis cs:769.040558490716,32758.8061519443)
--(axis cs:1501.63751921786,35440.9194196141)
--(axis cs:2932.11484651493,37160.3071958561)
--(axis cs:5725.28147647193,37754.1095773697)
--(axis cs:11179.2510528,49712.1711899041)
--(axis cs:21828.735340807,61037.0815602138)
--(axis cs:42623.0419487404,88286.8567019449)
--(axis cs:83226.2463491358,152008.877492766)
--(axis cs:83226.2463491358,262682.252942017)
--(axis cs:83226.2463491358,262682.252942017)
--(axis cs:42623.0419487404,139757.143298055)
--(axis cs:21828.735340807,79505.3184397862)
--(axis cs:11179.2510528,58892.4002386673)
--(axis cs:5725.28147647193,45243.3641068408)
--(axis cs:2932.11484651493,37862.0928041439)
--(axis cs:1501.63751921786,36298.2805803859)
--(axis cs:769.040558490716,34033.1938480557)
--(axis cs:393.852293269657,30724.2285066775)
--(axis cs:201.705394079863,28150.9974874213)
--(axis cs:103.300315108378,27159.5995545651)
--cycle;

\addplot [very thick, darkorange25512714, dotted]
table {%
103.300315108378 7956
201.705394079863 7984
393.852293269657 8192.8
769.040558490716 9092
1501.63751921786 9968.8
2932.11484651493 10468
5725.28147647193 12044.4210526316
11179.2510528 16521.1428571429
21828.735340807 21256.48
42623.0419487404 35026.4285714286
83226.2463491358 65415.4782608696
};
\addlegendentry{accuracy $10^{-1}$}
\addplot [very thick, darkorange25512714, dashed]
table {%
103.300315108378 14244
201.705394079863 14704
393.852293269657 16132
769.040558490716 18036
1501.63751921786 19549.6
2932.11484651493 20519.2
5725.28147647193 22874.7368421053
11179.2510528 30254.2857142857
21828.735340807 39109.6
42623.0419487404 63598.8571428571
83226.2463491358 116564.347826087
};
\addlegendentry{accuracy $10^{-2}$}
\addplot [very thick, darkorange25512714]
table {%
103.300315108378 26980
201.705394079863 27952
393.852293269657 30320.8
769.040558490716 33396
1501.63751921786 35869.6
2932.11484651493 37511.2
5725.28147647193 41498.7368421053
11179.2510528 54302.2857142857
21828.735340807 70271.2
42623.0419487404 114022
83226.2463491358 207345.565217391
};
\addlegendentry{accuracy $10^{-3}$}

\addplot [black, dashed, ultra thick]
table {%
187.4 3956
187.400000000000001 275419
};
\addlegendentry{\Cref{ex:isotropic homonuclear}, pulse~\ref{case: chirped} -- $\|\hi\| \approx 187.4$}

\addlegendimage{black, semithick, dashed}
\addlegendentry{$\order\lb \|\hi\|^{-1} \rb$}

\addplot [semithick]
table {%
43657.470681614 128789.538510761
85246.0838974809 251475.947497569
};
\addplot [semithick]
table {%
43657.470681614 73126.2633917035
85246.0838974809 142787.19052828
};
\addplot [semithick]
table {%
43657.470681614 40383.160380493
85246.0838974809 78852.6276051698
};
\addlegendentry{$\order\lb \|\hi\| \rb$}

\end{groupplot}

\end{tikzpicture}

%% file: fig_general_no_mixed.tex
\begin{tikzpicture}

\begin{groupplot}[
    group style={group size=2 by 1,
    horizontal sep=1.5cm,
    y descriptions at=edge left},
    legend pos=south west,
    legend style={
        at={(-0.15,-0.65)},
        anchor=south,
        legend columns=4,column sep=10pt,
        font=\footnotesize
      }
]

\nextgroupplot
[
width=0.5\textwidth, height=0.375\textwidth,
log basis x={10},
log basis y={10},
tick align=outside,
tick pos=left,
x grid style={darkgray176},
xlabel={time-step size, (ms)},
xmin=1e-2*0.5, xmax=1,
xmode=log,
xtick style={color=black, font=\footnotesize},
ymin = 1e-5,
ymax = 10,
xmajorgrids,
ymajorgrids,
y grid style={darkgray176},
ylabel={error at $T=10$ ms},
ymode=log,
ytick={1, 1e-1, 1e-2, 1e-3, 1e-4},
ytick style={color=black},
xminorgrids,
yminorgrids,
minor ytick={1e-7,1e-6,1e-5,1e-4,1e-3,1e-2,1e-1},
minor xtick={
1e-2, 2e-2, 3e-2, 4e-2, 5e-2, 6e-2, 7e-2, 8e-2, 9e-2,2e-1, 3e-1, 4e-1, 5e-1, 6e-1, 7e-1, 8e-1, 9e-1},
minor grid style={lightgray230, dotted, line width=0.1pt}
]

\addplot [very thick, steelblue31119180, dash pattern=on 3.5pt off 0.5pt, x filter/.code={\pgfmathparse{\pgfmathresult +3}\pgfmathresult}]
table {%
0.01 3.99984140566706
0.000833333333333333 4.00051385361553
0.000434782608695652 4.00137614433442
0.000294117647058824 3.71980315847756
0.000222222222222222 4.0909685949355
0.000178571428571429 3.97361158211439
0.000149253731343284 4.09894079316434
0.000128205128205128 4.33183549369391
0.000112359550561798 4.11222088491636
0.0001 3.88170968890877
6.94444444444444e-05 4.04710241031181
5.31914893617021e-05 3.32949679599795
4.29184549356223e-05 3.0699461858252
3.6101083032491e-05 3.77848159448635
3.1055900621118e-05 4.28660372632185
2.73224043715847e-05 4.34905246724168
2.4330900243309e-05 4.13148996005064
2.1978021978022e-05 3.79540888238209
2e-05 3.41852933986323
1.6e-05 2.49014569924267
1.33333333333333e-05 1.8355984853272
1.14285714285714e-05 1.39200426051452
1e-05 1.08580417314065
5.71428571428571e-06 0.367231276738675
};
\addplot [very thick, darkorange25512714, dash pattern=on 5.5pt off 0.5pt, x filter/.code={\pgfmathparse{\pgfmathresult +3}\pgfmathresult}, x filter/.code={\pgfmathparse{\pgfmathresult +3}\pgfmathresult}]
table {%
0.01 3.99963857690243
0.000833333333333333 4.17129694434699
0.000434782608695652 3.92606933877013
0.000294117647058824 3.98736318118574
0.000222222222222222 3.01983848078937
0.000178571428571429 1.74650105106496
0.000149253731343284 1.06835263950141
0.000128205128205128 0.839800209119785
0.000112359550561798 0.554521758842203
0.0001 0.4644807498354
6.94444444444444e-05 0.200479532343661
5.31914893617021e-05 0.0946941598687099
4.29184549356223e-05 0.0630796279187407
3.6101083032491e-05 0.0448360562137721
3.1055900621118e-05 0.0310655041029287
2.73224043715847e-05 0.0220960413480539
2.4330900243309e-05 0.0158525408869988
2.1978021978022e-05 0.0116399669587566
2e-05 0.00862427526165561
1.6e-05 0.00406201743869333
1.33333333333333e-05 0.00211927908856364
1.14285714285714e-05 0.00120061707140145
1e-05 0.000726482898840302
5.71428571428571e-06 8.31439600073649e-05
};
\addplot [very thick, sienna1408675, dash pattern=on 1.5pt off 0.5pt on 6pt off 0.5pt, x filter/.code={\pgfmathparse{\pgfmathresult +3}\pgfmathresult}]
table {%
0.01 4.001417227508
0.000833333333333333 3.91778902205307
0.000434782608695652 3.98787469492035
0.000294117647058824 4.10878491277743
0.000222222222222222 3.8553049867521
0.000178571428571429 3.66431001040099
0.000149253731343284 3.80828768158469
0.000128205128205128 3.71569450565557
0.000112359550561798 3.56993943668817
0.0001 4.80118117387089
6.94444444444444e-05 2.43260986251056
5.31914893617021e-05 0.773624447302828
4.29184549356223e-05 0.241325114818697
3.6101083032491e-05 0.0770151078475029
3.1055900621118e-05 0.027232454956086
2.73224043715847e-05 0.0120076818268364
2.4330900243309e-05 0.0079561767682249
2.1978021978022e-05 0.00657213996535847
2e-05 0.00549989033853373
1.6e-05 0.00316022956895147
1.33333333333333e-05 0.00180531642738028
1.14285714285714e-05 0.00107351309743473
1e-05 0.000668720169481089
5.71428571428571e-06 8.10807445471776e-05
};
\addplot [very thick, yellowgreen152204112, dash pattern=on 2pt off 2pt on 6pt off 2pt, x filter/.code={\pgfmathparse{\pgfmathresult +3}\pgfmathresult}]
table {%
0.01 3.99531498299411
0.000833333333333333 3.68004593187773
0.000434782608695652 3.41968724838927
0.000294117647058824 4.06243821473565
0.000222222222222222 3.20676229337643
0.000178571428571429 4.10513072717499
0.000149253731343284 1.67409421427503
0.000128205128205128 0.843813887867536
0.000112359550561798 0.462445302098505
0.0001 0.422668155844383
6.94444444444444e-05 0.19959375677531
5.31914893617021e-05 0.0946556978027295
4.29184549356223e-05 0.0630772271835768
3.6101083032491e-05 0.0448358134753509
3.1055900621118e-05 0.0310654708547427
2.73224043715847e-05 0.0220960353740835
2.4330900243309e-05 0.0158525396447548
2.1978021978022e-05 0.0116399666485647
2e-05 0.00862427517653552
1.6e-05 0.00406201743474259
1.33333333333333e-05 0.00211927908826051
1.14285714285714e-05 0.00120061707137478
1e-05 0.000726482898870986
5.71428571428571e-06 8.31439599876547e-05
};
\addplot [thick, black, dash pattern=on 0.5pt off 2.5pt, x filter/.code={\pgfmathparse{\pgfmathresult +3}\pgfmathresult}]
table {%
1.42857142857143e-05 0.979591836734694
1e-05 0.48
7.69230769230769e-06 0.284023668639053
};
\addplot [thick, black, dash pattern=on 0.5pt off 4pt, x filter/.code={\pgfmathparse{\pgfmathresult +3}\pgfmathresult}]
table {%
1.42857142857143e-05 0.000639733444398168
1e-05 0.0001536
7.69230769230769e-06 5.37796295647912e-05
};

\nextgroupplot
[
width=0.5\textwidth, height=0.375\textwidth,
log basis x={10},
log basis y={10},
tick align=outside,
tick pos=left,
x grid style={darkgray176},
xlabel={circuit depth},
xmin=1e3, xmax=1e5,
xmode=log,
xtick={1e3, 1e4, 1e5},
xtick style={color=black, font=\footnotesize},
ymin = 1e-5,
ymax = 10,
xmajorgrids,
ymajorgrids,
y grid style={darkgray176},
ymode=log,
ytick={1, 1e-1, 1e-2, 1e-3, 1e-4},
ytick style={color=black},
xminorgrids,
yminorgrids,
minor ytick={1,1e-1,1e-2,1e-3,1e-4},
yticklabels = {$10^0$, $10^{-1}$, $10^{-2}$, $10^{-3}$, $10^{-4}$},
minor xtick={2e3, 3e3,4e3,5e3,6e3,7e3,8e3,9e3,2e4,3e4,4e4,5e4,6e4,7e4,8e4,9e4},
minor grid style={lightgray230, dotted, line width=0.1pt}
]

\addplot [very thick, steelblue31119180, dash pattern=on 3.5pt off 0.5pt]
table {%
20 3.99984140566706
196 4.00051385361553
372 4.00137614433442
548 3.71980315847756
724 4.0909685949355
900 3.97361158211439
1076 4.09894079316434
1252 4.33183549369391
1428 4.11222088491636
1604 3.88170968890877
2308 4.04710241031181
3012 3.32949679599795
3732 3.0699461858252
4436 3.77848159448635
5156 4.28660372632185
5860 4.34905246724168
6580 4.13148996005064
7284 3.79540888238209
8004 3.41852933986323
10004 2.49014569924267
12004 1.8355984853272
14004 1.39200426051452
16004 1.08580417314065
28004 0.367231276738675
40004 0.181270778629294
52004 0.107553928198986
64004 0.0710955844258802
};
\addlegendentry{$\Theta_1(t_n)$, \cref{eq: theta_1 final}, \texttt{scipy}}
\addplot [very thick, darkorange25512714, dash pattern=on 5.5pt off 0.5pt]
table {%
52 3.99963857690243
580 4.17129694434699
1108 3.92606933877013
1636 3.98736318118574
2164 3.01983848078937
2692 1.74650105106496
3220 1.06835263950141
3748 0.839800209119785
4276 0.554521758842203
4804 0.4644807498354
6916 0.200479532343661
9028 0.0946941598687099
11188 0.0630796279187407
13300 0.0448360562137721
15460 0.0310655041029287
17572 0.0220960413480539
19732 0.0158525408869988
21844 0.0116399669587566
24004 0.00862427526165561
30004 0.00406201743869333
36004 0.00211927908856364
42004 0.00120061707140145
48004 0.000726482898840302
84004 8.31439600073649e-05
};
\addlegendentry{$\Theta_2(t_n)$, \cref{eq: Trotter type E}, \texttt{scipy}}
\addplot [very thick, sienna1408675, dash pattern=on 1.5pt off 0.5pt on 6pt off 0.5pt]
table {%
52 4.001417227508
580 3.91778902205307
1108 3.98787469492035
1636 4.10878491277743
2164 3.8553049867521
2692 3.66431001040099
3220 3.80828768158469
3748 3.71569450565557
4276 3.56993943668817
4804 4.80118117387089
6916 2.43260986251056
9028 0.773624447302828
11188 0.241325114818697
13300 0.0770151078475029
15460 0.027232454956086
17572 0.0120076818268364
19732 0.0079561767682249
21844 0.00657213996535847
24004 0.00549989033853373
30004 0.00316022956895147
36004 0.00180531642738028
42004 0.00107351309743473
48004 0.000668720169481089
84004 8.10807445471776e-05
};
\addlegendentry{$\Theta_2(t_n)$, \cref{eq: Trotter type E}, GL3}
\addplot [very thick, yellowgreen152204112, dash pattern=on 2pt off 2pt on 6pt off 2pt]
table {%
52 3.99531498299411
580 3.68004593187773
1108 3.41968724838927
1636 4.06243821473565
2164 3.20676229337643
2692 4.10513072717499
3220 1.67409421427503
3748 0.843813887867536
4276 0.462445302098505
4804 0.422668155844383
6916 0.19959375677531
9028 0.0946556978027295
11188 0.0630772271835768
13300 0.0448358134753509
15460 0.0310654708547427
17572 0.0220960353740835
19732 0.0158525396447548
21844 0.0116399666485647
24004 0.00862427517653552
30004 0.00406201743474259
36004 0.00211927908826051
42004 0.00120061707137478
48004 0.000726482898870986
84004 8.31439599876547e-05
};
\addlegendentry{$\Theta_2(t_n)$, \cref{eq: Trotter type E}, GL7}
\addplot [thick, black, dash pattern=on 0.5pt off 2.5pt]
table {%
86956.5217391304 0.0529
35714.2857142857 0.3136
};
\addlegendentry{$\order(h^2)$}
\addplot [thick, black, dash pattern=on 0.5pt off 4pt]
table {%
86956.5217391304 2.79841e-05
35714.2857142857 0.0009834496
};
\addlegendentry{$\order(h^4)$}

\end{groupplot}

\end{tikzpicture} 

%% file: fig_general_case_with_mix_coupling_vs_timestep.tex
\begin{tikzpicture}

\begin{axis}[
width=8cm, height=6cm,
legend pos=outer north east,
legend style={fill opacity=0.8, draw opacity=1, text opacity=1, draw=lightgray204, font=\small},
log basis x={10},
log basis y={10},
tick align=outside,
tick pos=left,
x grid style={darkgray176},
xlabel={time-step size, $h$ (ms)},
xmin=1e-2*0.75, xmax=1,
xmode=log,
xtick style={color=black, font=\footnotesize},
ymin = 1e-4,
ymax = 10,
xmajorgrids,
ymajorgrids,
y grid style={darkgray176},
ylabel={error at $T=10$ ms},
ymode=log,
ytick={1, 1e-1, 1e-2, 1e-3, 1e-4},
ytick style={color=black},
xminorgrids,
yminorgrids,
minor ytick={1e-7,1e-6,1e-5,1e-4,1e-3,1e-2,1e-1},
minor xtick={
1e-2, 2e-2, 3e-2, 4e-2, 5e-2, 6e-2, 7e-2, 8e-2, 9e-2,2e-1, 3e-1, 4e-1, 5e-1, 6e-1, 7e-1, 8e-1, 9e-1},
minor grid style={dotted, color=lightgray230, line width=0.1pt}
]

\addplot [very thick, steelblue31119180, dash pattern=on 3.5pt off 0.5pt, x filter/.code={\pgfmathparse{\pgfmathresult +3}\pgfmathresult}]
table {%
0.01 4.00020905513137
0.000833333333333333 4.00079613481729
0.000434782608695652 4.00037078130547
0.000294117647058824 3.72041471130889
0.000222222222222222 4.09096838120513
0.000178571428571429 3.97067723791519
0.000149253731343284 4.09708379695766
0.000128205128205128 4.32993410698598
0.000112359550561798 4.11272435005926
0.0001 3.883018235851
6.94444444444444e-05 4.04587643549667
5.31914893617021e-05 3.33285173447692
4.29184549356223e-05 3.07380641844302
3.6101083032491e-05 3.78009847173548
3.1055900621118e-05 4.28656275256742
2.73224043715847e-05 4.34816442684306
2.4330900243309e-05 4.13022623280637
2.1978021978022e-05 3.79403390048376
2e-05 3.4171703479497
1.6e-05 2.4890392375013
1.33333333333333e-05 1.83474947378388
1.14285714285714e-05 1.39134848658417
1e-05 1.0852876520199
};
\addlegendentry{$\Theta_1(t_n)$, \cref{eq: theta_1 final}, \texttt{scipy}}
\addplot [very thick, darkorange25512714, dash pattern=on 5.5pt off 0.5pt, x filter/.code={\pgfmathparse{\pgfmathresult +3}\pgfmathresult}]
table {%
0.01 3.9987640945023
0.000833333333333333 4.17059485880854
0.000434782608695652 3.92612752928309
0.000294117647058824 3.98515578152807
0.000222222222222222 3.01945892536569
0.000178571428571429 1.75364272012308
0.000149253731343284 1.08284946637959
0.000128205128205128 0.858532020868137
0.000112359550561798 0.584258361367963
0.0001 0.495642712709592
6.94444444444444e-05 0.225629282008543
5.31914893617021e-05 0.113603098187083
4.29184549356223e-05 0.0750489896988686
3.6101083032491e-05 0.0518534673990815
3.1055900621118e-05 0.0353477488230454
2.73224043715847e-05 0.0248055806063061
2.4330900243309e-05 0.0176076721364193
2.1978021978022e-05 0.0128274250795723
2e-05 0.0094471124267055
1.6e-05 0.00440341989837477
1.33333333333333e-05 0.00228466203260771
1.14285714285714e-05 0.00129005586395484
1e-05 0.00077895785762547
};
\addlegendentry{$\Theta_2(t_n)$, \cref{eq:eliminating}, \texttt{scipy}}
\addplot [very thick, sienna1408675, dash pattern=on 1.5pt off 0.5pt on 6pt off 0.5pt, x filter/.code={\pgfmathparse{\pgfmathresult +3}\pgfmathresult}]
table {%
0.01 3.9963832568881
0.000833333333333333 3.91272940829263
0.000434782608695652 3.99218091672427
0.000294117647058824 4.10502714417224
0.000222222222222222 3.85634752189705
0.000178571428571429 3.66348719994019
0.000149253731343284 3.80891868412351
0.000128205128205128 3.71189957884924
0.000112359550561798 3.56810315251227
0.0001 4.80225302718346
6.94444444444444e-05 2.43332783298343
5.31914893617021e-05 0.775482887230652
4.29184549356223e-05 0.244241095879559
3.6101083032491e-05 0.0812454059647657
3.1055900621118e-05 0.0320268985219242
2.73224043715847e-05 0.0164849430337494
2.4330900243309e-05 0.0110613129647992
2.1978021978022e-05 0.0085099854138776
2e-05 0.00672345124075854
1.6e-05 0.00359034880530022
1.33333333333333e-05 0.00199752525451785
1.14285714285714e-05 0.00117295005775687
1e-05 0.000725505685233528
};
\addlegendentry{$\Theta_2(t_n)$, \cref{eq:eliminating}, GL3}
\addplot [very thick, yellowgreen152204112, dash pattern=on 2pt off 2pt on 6pt off 2pt, x filter/.code={\pgfmathparse{\pgfmathresult +3}\pgfmathresult}]
table {%
0.01 3.99318919781973
0.000833333333333333 3.67628865031212
0.000434782608695652 3.42514293099782
0.000294117647058824 4.06002929832479
0.000222222222222222 3.20334552501557
0.000178571428571429 4.1057822911791
0.000149253731343284 1.68256479481038
0.000128205128205128 0.862843164035711
0.000112359550561798 0.498117691640308
0.0001 0.456800665913333
6.94444444444444e-05 0.22484369660505
5.31914893617021e-05 0.113571184684022
4.29184549356223e-05 0.075046983186639
3.6101083032491e-05 0.0518532586114598
3.1055900621118e-05 0.0353477197455028
2.73224043715847e-05 0.0248055753092602
2.4330900243309e-05 0.0176076710228495
2.1978021978022e-05 0.0128274247992676
2e-05 0.00944711234931475
1.6e-05 0.00440341989474011
1.33333333333333e-05 0.00228466203232493
1.14285714285714e-05 0.00129005586392987
1e-05 0.000778957857653613
};
\addlegendentry{$\Theta_2(t_n)$, \cref{eq:eliminating}, GL7}
\addplot [thick, black, dash pattern=on 0.5pt off 2.5pt, x filter/.code={\pgfmathparse{\pgfmathresult +3}\pgfmathresult}]
table {%
1.36e-05 1.536
1.13333333333333e-05 1.06666666666667
9.71428571428572e-06 0.783673469387755
};
\addlegendentry{$\order(h^2)$}
\addplot [thick, black, dash pattern=on 0.5pt off 4pt, x filter/.code={\pgfmathparse{\pgfmathresult +3}\pgfmathresult}]
table {%
1.36e-05 0.0065536
1.13333333333333e-05 0.00316049382716049
9.71428571428572e-06 0.00170595585172845
};
\addlegendentry{$\order(h^4)$}
\end{axis}

\end{tikzpicture} 

%% file: 6_conclusion.tex
\section{Conclusions}
\label{sec: conclusions}

In this paper we consider the efficacy of various Trotterized algorithms for Hamiltonian simulation of coupled many-body two-level systems, with a particular focus on producing shorter depth circuits for applications on near-term quantum devices.

\para{Time-independent Hamiltonians.} While
high-order numerical methods are typically thought to have relevance only for high  accuracies, and some recent attempts at Hamiltonian simulation on existing quantum devices have utilized the first-order Trotter splitting
\cite{IBM23nisq}, our numerical experiments (cf. \Cref{fig: accuracy time independent}) suggest that the second-order Strang splitting produces circuits that are shorter than those produced by the first-order Trotter splitting, even for relatively low accuracies ($\sim10^{-2}$ -- $10^{-5}$) for a three-body problem. Consequently, Strang splitting should be the method of choice for time-independent Hamiltonian simulation for demonstrating {\em quantum advantage} on near-term devices, where high accuracies are neither achievable nor essential.

\para{Time-dependent Hamiltonians.} For the simulation of time-dependent many-body two-level Hamiltonians, we compare a range of first-order, second-order and fourth-order algorithms that produce Trotterized circuits which are reasonable candidates for implementation on near-term devices. In contrast to the time-independent case, we find that fourth-order methods are required for the simulation of time-dependent Hamiltonians, even for extremely low levels of accuracy $\sim 10^{-1}$ (cf. \Cref{fig: accuracy chirped isotropic homonuclear,fig: accuracy general no mixed coupling,fig: accuracy general with mixed coupling}) for a three-body problem. In fact, all the fourth-order methods considered here produce circuits that are shorter than the first and second-order methods for propagator errors of $1$ or lower. We expect that for large spin systems, where any quantum advantage is likely to be demonstrable, the increased spectral radius will make the need for higher-order methods more pronounced.

{\em Circuit compression ratio.} The fourth-order Magnus expansion-based method developed in this paper produces circuits that are consistently found to be shorter in depth than the first-order, second-order and other fourth-order methods considered here, for a very wide range of driving pulse frequencies and interaction strengths, and across all levels of accuracy ~(cf. \Cref{fig: comparison across bandwidth different methods,fig: comparison across Hin norm different methods}). Moreover, the {\em circuit compression ratios} (which measure how much shorter the circuit produced by the proposed method is, as compared to the other methods) increases when we require higher accuracy and with the frequency of the controls. The latter stems from an ability to take time-steps that seem almost frequency-independent and can be larger than the wavelength (cf. \Cref{fig: accuracy chirped isotropic homonuclear,fig: size depth Theta2 different bandwidth}). This is achievable due to the decoupling of the time-step of the time-stepping procedure \eqref{eq: propagator approximation} from the resolution of the driving pulses via the computation of the integrals, \eqref{eq: second-order Magnus parameters without new notation},\eqref{eq: extra commutator rotation gates weight} and \eqref{eq: magnus fourth order r}, appearing in the Magnus expansion.

{\em Circuit structure.} Moreover, the proposed approach results in circuits that are obtained as minor modifications \eqref{eq: Trotter type E} of fourth-order Trotterized circuits (cf. \Cref{fig: trotter circuit,eq: yoshida for Omega}) used for simulating time-independent Hamiltonians, making their implementation straightforward. Specifically, the modifications involve (i) three additional single-qubit gate layers ($\ie$, $3M$ single-qubit gates) per time-step on average (an overhead that disappears entirely in the special case of isotropic coupling with identical spins, which appears frequently in liquid-state NMR applications), and (ii) tweaking the parameters of existing gates in the fourth-order Trotterized circuit for time-independent Hamiltonians.
Computing the parameters for the gates requires the approximation of multiple one-dimensional and two-dimensional integrals appearing in the Magnus expansion, which can be computed efficiently to arbitrary accuracy on classical computers.

The straightforward circuit structure and consistent out-performance across various accuracy levels, control frequencies and coupling strengths highlights the robustness and broad applicability of our fourth-order Magnus-based method, and we recommend it as the method of choice for circuit design for simulation of time-dependent Hamiltonians on near-term devices, even when a very low level of accuracy is achievable or desired.

\newpage

A fourth-order Magnus-based classical algorithm based on \cite{goodacrethesis} is implemented in the open source Python package \texttt{magpy}, which is available on \texttt{github} \cite{magpygithub} and can be installed from the Python Package Index (\texttt{PyPI}) as \texttt{pip install magpy}. Further extensions to this classical algorithm, based on some of the techniques presented in this manuscript, are currently under development and are also expected to be made available in \texttt{magpy}.

%% file: 7_acknowledgements.tex
\section{Acknowledgements}
GC acknowledges support from EPSRC (EP/S022945/1).
CB acknowledges support from EPSRC (EP/V026259/1).

%% file: A_appendix.tex
\appendix

\section[Appendix A: Proof of Theorem 1]{Proof of \Cref{thm: general Theta_2}}\label{sec: proof of 4th Magnus}

\begin{proof}

\normalfont
The computation of the fourth-order Magnus expansion in \cref{eq: Theta_2 general with commutator} begins by establishing some fundamental properties.

Consider $2\times 2$ matrices $A$, $B$, and their commutator $C=[A,B]$. Then
\begin{equation}
   [A_j, B_k] = \delta_{j,k} C_k, \label{eq: commutator of Kronecker products}
\end{equation}
where $\delta_{j,k}$ is the Kronecker delta, so that
\begin{equation}
    [\va ^\top\vca, \vb ^\top\vcb] = \sum_{j=1}^{M} a_j b_j C_j = (\va \odot \vb) ^\top \vcc,\label{eq: commutator of vector multiplications}
\end{equation}
where $\vca = (A_1,\cdots,A_M)$, $\vcb = (B_1,\cdots,B_M)$ and $\vcc = (C_1,\cdots,C_M)$.
\begin{Lemma}
\begin{equation*}
    \lbb\va ^\top \vcs^X, \vb ^\top \vcs^X \rbb = \lbb\va ^\top \vcs^Y, \vb ^\top \vcs^Y \rbb = \lbb\va ^\top \vcs^Z, \vb ^\top \vcs^Z \rbb = 0.
\end{equation*}\label{lemma: commutator of same weighted summed single operations}
\end{Lemma}

\begin{Property}
\normalfont
The Pauli matrices $\alpha\in\{X,Y,Z\}$ satisfy
\begin{equation}
    [\px,\py] = 2\ii \pz,\ [\py,\pz] = 2\ii\px,\ [\pz,\px] = 2\ii\py.
\end{equation}\label{property: single Pauli matrices commutation}
\end{Property}
\Cref{property: single Pauli matrices commutation} immediately leads to \Cref{lemma: commutator of different weighted summed single operations}.

\begin{Lemma}
\normalfont
\begin{align*}
    \lbb \va ^\top \vcs^X, \vb^\top  \vcs^Y \rbb &= 2 \ii \lb\va\odot\vb\rb^\top  \vcs^Z ,\\
    \lbb \va^\top \vcs^Y ,\vb^\top  \vcs^Z\rbb &= 2 \ii \lb\va\odot\vb\rb^\top  \vcs^X ,\\
    \lbb \va ^\top \vcs^Z, \vb^\top  \vcs^X \rbb &= 2 \ii \lb\va\odot\vb\rb^\top  \vcs^Y.
\end{align*} \label{lemma: commutator of different weighted summed single operations}
\end{Lemma}

Our investigation now turns to the commutator $\lbb \A(t_n+\xi), \A(t_n+\zeta) \rbb$ in Magnus expansion, \cref{eq: Theta_2 general with commutator}, which can be expressed as
\begin{align}
&\lbb \A(t_n+\xi), \A(t_n+\zeta) \rbb \nonumber \\
=& - \lbb \hs(t_n+\xi) + \hi, \hs(t_n+\zeta) + \hi \rbb \nonumber\\
=& - \lbb \hs(t_n+\xi), \hs(t_n+\zeta) \rbb  - \lbb \hs(t_n+\xi) , \hi \rbb + \lbb \hs(t_n+\zeta) , \hi \rbb.
\label{eq: general commutatorA}
\end{align}

The first commutator in \cref{eq: general commutatorA}, $\lbb \hs(t_n+\xi), \hs(t_n+\zeta) \rbb$, is represented as

\begin{align*}
\lbb \hs(t_n+\xi), \hs(t_n+\zeta) \rbb
= 2\ii \left((\ve \wedge \ve)(t_n+\xi, t_n+\zeta)\right)^\top \vcs  \qquad \text{by \Cref{lemma: commutator of same weighted summed single operations} and \Cref{lemma: commutator of different weighted summed single operations}},
\end{align*}

The integral of the above should lead to terms of the same form, i.e. $\mathbf{a}^\top \vcs$ for some parameters $\mathbf{a} \in \CC^{3M}$. Structurally, this is not problematic since it fits the common structure \eqref{eq: common structure}.

The second commutator in \cref{eq: general commutatorA}, $\lbb \hs(t_n+\xi) , \hi \rbb$, can be expressed as
\begin{align}
&\lbb \hs(t_n+\xi) , \hi \rbb\nonumber\\
=&\lbb \vex(t_n+\xi)^\top\vcs^X + \vey(t_n+\xi)^\top\vcs^Y + \vez(t_n+\xi)^\top\vcs^Z, \frac12 \vcs^\top \jg \vcs \rbb\nonumber\\
=& \frac12 \lbb  \vex(t_n+\xi)^\top\vcs^X ,  \vcs^\top \jg \vcs \rbb + \frac12 \lbb  \vey(t_n+\xi)^\top\vcs^Y , \vcs^\top \jg \vcs \rbb \nonumber\\
&+ \frac12 \lbb  \vez(t_n+\xi)^\top\vcs^Z , \vcs^\top \jg \vcs \rbb\label{eq: no cancelout1}
\end{align}
The last commutator in \cref{eq: general commutatorA} has similar structure as the second commutator \cref{eq: no cancelout1}, except that the time variable is replaced by $\zeta$. Collectively, \cref{eq: general commutatorA}, $\ie$, $\lbb \A(t_n+\xi), \A(t_n+\zeta) \rbb$, can be summarized as

\begin{align*}
&[ \A(t_n+\xi), \A(t_n+\zeta) ] \\
=& 2\ii \left((\ve \wedge \ve)(t_n+\xi, t_n+\zeta)\right)^\top \vcs- \frac12 \lbb  \lb\vex(t_n+\xi) - \vex(t_n+\zeta)\rb^\top\vcs^X ,  \vcs^\top \jg \vcs \rbb\\
&+ \frac12 \lbb \lb\vey(t_n+\xi) - \vey(t_n+\zeta)\rb^\top\vcs^Y ,  \vcs^\top \jg \vcs \rbb \\
&+ \frac12 \lbb\lb\vez(t_n+\xi) - \vez(t_n+\zeta)\rb^\top\vcs^Z ,  \vcs^\top \jg \vcs \rbb.
\end{align*}
Subsequently, integrating $[ \A(t_n+\xi), \A(t_n+\zeta) ]$ leads to the expression of the fourth-order Magnus expansion.
\end{proof}

\section[Appendix B: Derivation of eq. (63)]{Derivation of \cref{eq: u and r commutator} }
\label{sec: u and r commutator}
\begin{align*}
 &\lbb \vu(t_n)^\top\vcs, \vr(t_n)^\top\vcs \rbb \\
 =&
 \lbb
  \vu^X(t_n)^\top \vcs^X +  \vu^Y(t_n)^\top \vcs^Y +  \vu^Z(t_n)^\top \vcs^Z
 ,
 \vr^X(t_n)^\top\vcs^X +  \vr^Y(t_n)^\top\vcs^Y +  \vr^Z(t_n)^\top\vcs^Z
 \rbb
 \\
 =&  \lb \vu^Y(t_n)^\top \odot \vr^X(t_n)\rb^\top \lb-2\ii\vcs^Z\rb
 +
 \lb \vu^Z(t_n)^\top \odot \vr^X(t_n)\rb^\top \lb 2\ii\vcs^Y\rb\\
 &
 +
 \lb \vu^X(t_n)^\top \odot \vr^Y(t_n)\rb^\top \lb 2\ii\vcs^Z\rb
 +
 \lb \vu^Z(t_n)^\top \odot \vr^Y(t_n)\rb^\top \lb -2\ii\vcs^X\rb\\
 &  \lb \vu^X(t_n)^\top \odot \vr^Z(t_n)\rb^\top \lb-2\ii\vcs^Y\rb
+
 \lb \vu^Y(t_n)^\top \odot \vr^Z(t_n)\rb^\top \lb 2\ii\vcs^X\rb\\
 =&
 -2\ii
 \lb \vu^Z(t_n)^\top \odot \vr^Y(t_n)\rb^\top \vcs^X
 + 2\ii
 \lb \vu^Y(t_n)^\top \odot \vr^Z(t_n)\rb^\top \vcs^X
 \\
 &
2\ii
 \lb \vu^Z(t_n)^\top \odot \vr^X(t_n)\rb^\top \vcs^Y
 -2\ii
 \lb \vu^X(t_n)^\top \odot \vr^Z(t_n)\rb^\top \vcs^Y
 \\
 &
 2\ii
 \lb \vu^X(t_n)^\top \odot \vr^Y(t_n)\rb^\top \vcs^Z
 -2\ii
 \lb \vu^Y(t_n)^\top \odot \vr^X(t_n)\rb^\top \vcs^Z
 \\
 =& 2\ii \lb \vu(t_n) \times \vr(t_n)\rb ^\top \vcs.
\end{align*}

\section[Appendix C: Proof of Theorem 3]{Proof of \Cref{thm:single spin}}\label{sec: Trotter type E parameters derivation}
\begin{align*}
\ee^{\ii a Z} \ee^{\ii b Y} \ee^{\ii c X} =&\lb \lb I \cos(a) + \ii Z\sin(a) \rb \lb I \cos(b) + \ii Y\sin(b) \rb \rb \ee^{\ii c X} \lb I \cos(c) + \ii X\sin(c) \rb\\
=& I \cos(a)\cos(b)\cos(c) + 2\ii X \sin(a)\sin(b)\cos(c)
+ \ii X\cos(a)\cos(b)\sin(c) \\
&+ \ii Y\cos(a)\sin(b)\cos(c)
- 2\ii Y\sin(a)\cos(b)\sin(c) \\
&+ \ii Z\sin(a)\cos(b)\cos(c)
+  2\ii Z\cos(a)\sin(b)\sin(c)\\
=& \ee^{\ii d (k_1X+ k_2Y+ k_3Z)}.
\end{align*}
We want to express the product of the above three exponentials in the form
\[\ee^{\ii d (k_1X+ k_2Y+ k_3Z)}
= I\cos(d) + \ii (k_1X+ k_2Y+ k_3Z)\sin(d).
\]
Matching the corresponding terms, we find
\begin{equation}
\begin{aligned}
d&=\arccos(\cos(a)\cos(b)\cos(c)),\\
k_1 &= p\lbb\cos(a)\cos(b)\cos(c) - \sin(a)\sin(b)\sin(c))\rbb\lb \sin(a)\sin(b) \cos(c) + \sin(c)\cos(a)\cos(b)
\rb
,\\
k_2 &= p\lbb\cos(a)\cos(b)\cos(c) - \sin(a)\sin(b)\sin(c))\rbb\lb\sin(b)\cos(a)\cos(c) - \sin(a)\cos(b)\sin(c)
\rb
,\\
k_3 &= p\lbb\cos(a)\cos(b)\cos(c) - \sin(a)\sin(b)\sin(c))\rbb\lb\sin(a)\cos(b)\cos(c) + \sin(b)\cos(a)\sin(c)
\rb,
\end{aligned}
\label{eq: generic group multiplication}
\end{equation}
where $p[x] = \frac{\arccos(x)}{\sin\lb\arccos(x)\rb}$. Substituting the parameters defining $\eli$ into \cref{eq: generic group multiplication} gives us \cref{eq: Trotter type E parameters}.

\section[Appendix D: Proof of Lemma 2]{Proof of \Cref{lem: homonuclear reduction}}\label{sec: proof of homonuclear reduction theorem}
\begin{proof}[Proof of \Cref{lem: homonuclear reduction}]
\begin{align*}
& \lbb  \oneb^\top \vcs^{\alpha} , \sum_{\beta \in\{ X,Y,Z \}} \lb\vcs^{\beta}\rb^\top C \vcs^{\beta} \rbb \\
=& \lbb  \oneb^\top \vcs^{\alpha}  , \sum_{j=1}^M \sum_{k=j+1}^M C_{j,k}  \px_j\px_k\rbb + \lbb  \oneb^\top \vcs^{\alpha}  , \sum_{j=1}^M \sum_{k=j+1}^M C_{j,k}  \py_j\py_k \rbb + \lbb  \oneb^\top \vcs^{\alpha}  , \sum_{j=1}^M \sum_{k=j+1}^M C_{j,k}  \pz_j\pz_k\rbb\\
=&
\begin{cases}
      \lbb  \oneb^\top \vcs^X , \sum_{j=1}^M \sum_{k=j+1}^M C_{j,k}  \py_j\py_k \rbb + \lbb  \oneb^\top \vcs^X , \sum_{j=1}^M \sum_{k=j+1}^M C_{j,k}  \pz_j\pz_k\rbb  & \text{if $\alpha = X$}\\
      \lbb  \oneb^\top \vcs^Y , \sum_{j=1}^M \sum_{k=j+1}^M C_{j,k}  \px_j\px_k \rbb + \lbb  \oneb^\top \vcs^Y , \sum_{j=1}^M \sum_{k=j+1}^M C_{j,k}  \pz_j\pz_k\rbb & \text{if $\alpha = Y$}\\
      \lbb  \oneb^\top \vcs^Z , \sum_{j=1}^M \sum_{k=j+1}^M C_{j,k}  \px_j\px_k \rbb + \lbb  \oneb^\top \vcs^Z , \sum_{j=1}^M \sum_{k=j+1}^M C_{j,k}  \py_j\py_k\rbb & \text{if $\alpha = Z$}
    \end{cases}
    \tag*{by \Cref{lemma: AA commutators}} \\
=&
\begin{cases}
      2\ii\sum_{j=1}^M \sum_{k=j+1}^M C_{j,k}  (\pz_j\py_k+\py_j\pz_k) -2\ii\sum_{j=1}^M \sum_{k=j+1}^M C_{j,k}  (\py_j\pz_k + \pz_j\py_k)  & \text{if $\alpha = X$}\\
      -2\ii\sum_{j=1}^M \sum_{k=j+1}^M C_{j,k}  (\pz_j\px_k+\px_j\pz_k) +2\ii\sum_{j=1}^M \sum_{k=j+1}^M C_{j,k}  (\px_j\pz_k + \pz_j\px_k) & \text{if $\alpha = Y$}\\
      2\ii\sum_{j=1}^M \sum_{k=j+1}^M C_{j,k}  (\py_j\px_k+\px_j\py_k) -2\ii\sum_{j=1}^M \sum_{k=j+1}^M C_{j,k}  (\px_j\py_k + \py_j\px_k) & \text{if $\alpha = Z$}
    \end{cases}
    \tag*{by \Cref{property: single Pauli matrices commutation}}    \\
=& 0.
\end{align*}
\end{proof}